\documentclass[utf8]{frontiersSCNS} 
\usepackage{url,hyperref,lineno,microtype,subcaption,epstopdf,aas_macros}
\usepackage[onehalfspacing]{setspace}

\usepackage{graphics,graphicx}
\usepackage{helvet}
\usepackage[utf8]{inputenc}
\usepackage[T1]{fontenc}
\usepackage{soul}
\usepackage{amsmath, amssymb, gensymb}
\usepackage{multirow}
\usepackage{morefloats}
\usepackage{rotating}
\usepackage{xspace}
\usepackage{color}
\usepackage{xspace}
\usepackage{color}
\usepackage{xspace}
\usepackage{wrapfig}

\def\aj{{AJ}}                   
\def\araa{{ARA\&A}}             
\def\apj{{ApJ}}                 
\def\apjl{{ApJ}}                
\def\apjs{{ApJS}}               
\def\ao{{Appl.~Opt.}}           
\def\apss{{Ap\&SS}}             
\def\aap{{A\&A}}                

\def\mnras{{MNRAS}}             
\def\pasp{{PASP}}               
\def\pasj{{PASJ}}               
\def\qjras{{QJRAS}}             
\def\nat{{Nature}}              

\def\procspie{{Proc.~SPIE}}   

\def \arcsec{$^{\prime\prime}$\xspace}   

\def\keyFont{\fontsize{8}{11}\helveticabold }
\def\firstAuthorLast{Hull \& Zhang} 
\def\Authors{Charles L. H. Hull\,$^{\dagger\,1,2,3}$, Qizhou Zhang\,$^{4}$}

\def\Address{\small $^{1}$National Astronomical Observatory of Japan, Alonso de C\'ordova 3788, Office 61B, Vitacura 763 0422, Santiago, Chile \smallskip \\
$^{2}$Joint ALMA Observatory, Alonso de C\'ordova 3107, Vitacura 763 0355, Santiago, Chile \smallskip \\
$^{3}$NAOJ Fellow \smallskip \\
$^{4}$Center for Astrophysics $|$ Harvard \& Smithsonian, 60 Garden Street, MS 42, Cambridge, MA 02138, USA \smallskip}

\def\corrAuthor{Charles L. H. Hull}
\def\corrEmail{chat.hull@nao.ac.jp \vspace{-1em}}

\newcommand{\skipthis}[1]{} 
\newcommand{\hi}{{\rm H}{\sc i}\xspace}
\newcommand{\hii}{{\rm H}{\sc ii}\xspace}
\newcommand{\uchii}{{\rm UCH}{\sc ii}\xspace} 
\def\NH3{NH$_3$} 
 
\def\msun{M$_\odot$}
 
\def\lsun{L$_\odot$} 
\def\kms-1{km~s$^{-1}$}
\def\h2o{$\rm{H_2O}$} 
\def\h2{$\rm{H_2}$} 
\def\CM2{$\rm{cm^{-2}}$}
\def\cm3{$\rm{cm^{-3}}$}

\def\n2h+{N$_2$H$^+$}
\def\ch3oh{CH$_3$OH}

\begin{document}
\onecolumn
\firstpage{1}
\title[Magnetic fields in forming stars]{Interferometric observations of magnetic fields in forming stars\footnote{The published, open-access version is available here: \href{https://www.frontiersin.org/articles/10.3389/fspas.2019.00003/full}{https://www.frontiersin.org/articles/10.3389/fspas.2019.00003/full} \vspace{-1.8em}}} 

\author[\firstAuthorLast ]{\Authors} 
\address{} 
\correspondance{} 

\extraAuth{}



\maketitle

\begin{abstract}
\section{}

\vspace{-2em}
\noindent
The magnetic field is a key ingredient in the recipe of star formation.  However, the importance of the magnetic field in the early stages of the formation of low- and high-mass stars is still far from certain.  Over the past two decades, the millimeter and submillimeter interferometers BIMA, OVRO, CARMA, SMA, and most recently ALMA have made major strides in unveiling the role of the magnetic field in star formation at progressively smaller spatial scales;  ALMA observations have recently achieved spatial resolutions of up to $\sim$\,100\,au and $\sim$\,1,000\,au in nearby low- and high-mass star-forming regions, respectively.  From the kiloparsec scale of molecular clouds down to the inner few hundred au immediately surrounding forming stars, the polarization at millimeter and submillimeter wavelengths is dominated by polarized thermal dust emission, where the dust grains are aligned relative to the magnetic field. Interferometric studies have focused on this dust polarization and occasionally on the polarization of spectral-line emission.  We review the current state of the field of magnetized star formation, from the first BIMA results through the latest ALMA observations, in the context of several questions that continue to motivate the studies of high- and low-mass star formation.  By aggregating and analyzing the results from individual studies, we come to several conclusions: (1) Magnetic fields and outflows from low-mass protostellar cores are randomly aligned, suggesting that the magnetic field at $\sim$\,1000\,au scales is not the dominant factor in setting the angular momentum of embedded disks and outflows.  (2) Recent measurements of the thermal and dynamic properties in high-mass star-forming regions reveal small virial parameters, challenging the assumption of equilibrium star formation.  However, we estimate that a magnetic field strength of a fraction of a mG to several mG in these objects could bring the dense gas close to a state of equilibrium.  Finally, (3) We find that the small number of sources with hourglass-shaped magnetic field morphologies at 0.01--0.1\,pc scales cannot be explained purely by projection effects, suggesting that while it does occur occasionally, magnetically dominated core collapse is not the predominant mode of low- or high-mass star formation.

\keyFont{\small \section{Keywords:} astronomy, low-mass star formation, high-mass star formation, polarization, magnetic fields, dust, interferometry, millimeter-wave observations} 
\end{abstract}

\section{Introduction}

Magnetic fields are known to play a critical role in many aspects of both low- and high-mass star formation.  Even weakly ionized star-forming material is coupled to the ambient magnetic field, and thus the field can regulate (or prevent) the collapse and fragmentation of star-forming clouds \citep[][in this volume]{HennebelleInutsuka2019, KrumholzFederrath2019, TeyssierCommercon2019}, can influence the formation of protoplanetary disks \citep[][in this volume]{WursterLi2018}, and can launch bipolar outflows and jets from young protostars \citep[][in this volume]{PudritzRay2019}.  Mapping the morphology of magnetic fields in low- and high-mass star-forming regions is therefore critical to better understand how magnetic fields affect the star-formation process at early times, and how the role of the field changes relative to other dynamical effects (e.g., turbulence, rotation, thermal and radiation pressure, and gravitational collapse) as a function of spatial scale, source environment, and source mass.  Over more than fifty years, studies of magnetic fields in low- and high-mass star-forming regions have been performed across a wide range of spatial scales, from the $>$\,100\,pc scale of molecular clouds, to the 1\,pc scale of clumps, to the 0.1\,pc scale of dense cores, and finally to the 1000--100\,au scale of protostellar envelopes surrounding forming protostellar systems.\footnote{\label{footnote:1} In this review we follow the nomenclature used in \citet{zhang2009} and \citet{Motte2018}, who refer to a \textit{cloud} as a structure 10\,--\,100\,pc in size; a molecular \textit{clump} as a structure 1\,pc in size that forms massive stars along with a population of lower mass stars; and a \textit{dense core} as a structure 0.01\,--\,0.1\,pc in size that forms one or a group of stars.  Following the nomenclature of, e.g., \citet{Hull2014}, a protostellar \textit{envelope} is a $\sim$\,1000\,au ($\sim$\,0.005\,pc) structure comprising the densest part of the dense core, inside of which one or a few protostars form.}

In this review we introduce the typical tracers of the magnetic field in star-forming regions, as well as the methods used to analyze the observations.  We then discuss the state of the field of magnetized star formation, focusing primarily on the last three decades of high-resolution polarization studies by millimeter and submillimeter (hereafter combined into ``(sub)millimeter'') interferometers including the Berkeley-Illinois-Maryland Association (BIMA) millimeter array, the Combined Array for Research in Millimeter-wave Astronomy (CARMA), the Submillimeter Array (SMA), and the Atacama Large Millimeter/submillimeter Array (ALMA).\footnote{Note that while we aim to provide an exhaustive review of the literature surrounding millimeter-wave interferometric observations of low- and high-mass stars, we mention only a few theoretical and single-dish polarization studies in order to support our narrative.  We do not discuss other types of millimeter-wave polarization observations, i.e., toward the galactic center, quasars, or evolved stars.}

\subsection{Magnetic field tracers (plane-of-sky component)}
\label{sec:intro_tracers_POS}

\subsubsection{Polarized dust emission}

Polarized thermal emission from dust grains is the main tracer of the magnetic field in star-forming regions observed at high resolution and at (sub)millimeter wavelengths.  Under most circumstances, oblong interstellar medium (ISM) dust grains with sizes $\ll$\,100\,$\mu$m are aligned with their long axes perpendicular to magnetic field lines \citep[e.g.,][]{Hildebrand1988}.  The currently accepted way to achieve this alignment is via the ``radiative torque'' (RAT) mechanism, where an anisotropic radiation field (e.g., the external UV field in the ISM, or the radiation from a deeply embedded protostar) causes grains to become aligned relative to the magnetic field \citep{Lazarian2007, Hoang2009, Andersson2015}.\footnote{\label{footnote:RATs} RATs that align grains with respect to the magnetic field $B$ are sometimes known as ``$B$-RATs,'' in contrast to other radiative torque mechanisms such as the ``$k$-RATs'' mentioned in Section \ref{sec:scattering}, where RATs align grains with respect to the radiation direction.}  Thus, at the physical scales of star-forming clouds, cores, and envelopes (i.e., scales $\gtrsim$\,100\,au), magnetically aligned dust grains emit thermal radiation that is polarized perpendicular to the magnetic field.  Observations of dust polarization, which comprise the vast majority of the single-dish and interferometric polarization observations to date, are discussed at length throughout this article.

\subsubsection{Polarized molecular-line emission}
\label{sec:intro_tracers_POS:GK}

Polarization of molecular-line emission is another tracer of the magnetic field in star-forming regions.  Molecular and atomic lines are sensitive to magnetic fields, which cause their spectral levels to split into magnetic sub-levels.   For some molecules, linear polarization can arise when an anisotropy in the radiation and/or velocity field yields a population of magnetic sub-levels that are not in local thermodynamic equilibrium (LTE); this is known as the Goldreich-Kylafis (G-K) effect.  Polarization from the G-K effect is  most easily detected where the spectral line emission has an optical depth $\tau \approx 1$, when the ratio of the collision rate to the radiative transition rate (i.e., the spontaneous emission rate) is $\sim$\,1, and where the gradient in the radiation and/or velocity field is large.  The effect is strongest in the lowest rotational transitions of simple molecules such as CO, CS, HCN, SiO, or HCO$^+$.  Spectral-line polarization from the G-K effect can be parallel or perpendicular to the magnetic field.  Due to the different optical depths of the parallel and perpendicularly polarized components in different locations on the sky, polarization from the G-K effect ultimately traces the plane-of-sky magnetic field orientation with an ambiguity of 90$^{\circ}$ \citep{Goldreich1981, Goldreich1982, Kylafis1983, Deguchi1984, Lis1988}.

The first detections of the G-K effect were by \citet{Glenn1997}, who presented National Radio Astronomy Observatory (NRAO) 12\,m telescope observations of the evolved star IRC +10216, and by \citet{Greaves1999}, who observed the galactic center and the high-mass star-forming clouds S140 and DR 21 using the James Clerk Maxwell Telescope (JCMT).  In addition to these first observations, others have detected the G-K effect in molecular outflows from low-mass protostellar sources \citep[e.g.,][]{Girart1999, Lai2002, Cortes2006b, CFLee2014, Ching2016, CFLee2018b} as well as in high-mass protostellar sources \citep[e.g.,][]{Lai2003, Cortes2005, Cortes2006a, Cortes2008}.  The G-K effect is a powerful way to trace the magnetic field in regions where the brightness of the thermal dust emission is too low to detect polarization at the few-percent level, either because the source is too distant or because the column density of dust is too low (e.g., in an outflow cavity where the gas and dust have been driven away).  

One way to resolve the 90$^{\circ}$ ambiguity in G-K observations is via radiative transfer modeling.  This was done in \citet{Cortes2005}, who expanded on the original G-K models (which assumed that gradients in the CO optical depth were responsible for the necessary anisotropy in the radiation field) by including bright, central sources as additional causes of anisotropy.  They used these models to successfully reproduce the 90$^{\circ}$ difference in polarization angles seen toward the high-mass star-forming region DR 21(OH) in their 3\,mm BIMA observations of CO\,($J$\,$=$\,1\,$\rightarrow$\,0) versus earlier 1.3\,mm observations of CO\,($J$\,$=$\,2\,$\rightarrow$\,1) by \citet{Lai2003}.  While the modeling in \citet{Cortes2005} was successful, in general it is difficult to know the structure of the radiation field, especially in high-mass star-forming regions.  Other methods exist to resolve the ambiguity, such as comparing polarization from both dust and spectral lines in the same region.  This was done in DR 21(OH) by \citet{Lai2003} and \citet{Ching2016}; however, it is not clear how strong the correlation should be between the polarization angles from the two types of emission, as the dust emission traces much denser material than the CO.

Looking to the future, in low-mass sources with well behaved outflows, or in protoplanetary disks, the radiation and velocity fields may be simple enough to allow us to more confidently break the ambiguity in polarization orientation. Given the high quality of ALMA data toward these types of sources and the fact that ALMA's sensitivity will allow us to probe polarization in multiple transitions of many different molecules, spectral-line polarization studies have a bright future.

Finally, one must exercise caution when interpreting polarized spectral-line observations, as linearly polarized spectral-line emission can be converted into circularly polarized emission via anisotropic resonant scattering.  Several studies have detected non-Zeeman circularly polarized emission, including \citet{Houde2013} in Orion KL and \citet{Hezareh2013} in the supernova remnant SNR IC 443. These observations were from the polarimeters at the Caltech Submillimeter Observatory (CSO) and the Institut de Radioastronomie Millim\'etrique (IRAM) 30\,m telescope, respectively (see \citealt{Chamma2018} for more results from the SMA).  The model described in those publications shows that resonant scattering can cause a rotation in the polarization position angle in maps of linear spectral-line polarization.  \citet{Hezareh2013} used the detected Stokes $V$ signal toward SNR IC 443 to correct the map of linear spectral-line polarization.  After doing so, the polarization was well aligned with dust polarization observations using the PolKA polarimeter at the Atacama Pathfinder Experiment (APEX) telescope.  The authors note that this effect is proportional to the square of the magnitude of the plane-of-sky magnetic field, which opens up the possibility of detecting the magnetic field via circular polarization observations of Zeeman-insensitive spectral lines.

\subsubsection{The Velocity-gradient technique}

Another method for probing the plane-of-sky magnetic field in star-forming regions is the ``velocity-gradient technique'' (VGT), which estimates the magnetic field orientation using the velocity gradients present in spectral-line observations.  In turbulent regions that are not gravitationally bound, \citet{GonzalezCasanovaLazarian2017} and \citet{YuenLazarian2017a} showed that the magnetic field is expected to be oriented perpendicular to velocity gradients.  The authors also extended their study to shocked and gravitationally bound regions \citep{YuenLazarian2017b}, and argued that the VGT provides an even better view of the plane-of-sky magnetic field morphology in the interstellar medium (particularly in diffuse regions with a low column density of dust) than both the \textit{Planck} polarization data \citep[e.g.,][]{PlanckXIX} and observations of neutral \hi ``fibers,'' which also trace the interstellar magnetic field \citep{Clark2014, Clark2015}.

\subsection{Magnetic field tracers (line-of-sight component): the Zeeman effect}
\label{sec:intro_tracers_LOS}

The Zeeman effect is another important tracer of the magnetic field that has been observed primarily with single-dish radio telescopes to infer the line-of-sight magnetic field strength\footnote{Note that under certain conditions one can derive the total magnetic field strength from the Zeeman effect \citep{Heiles1993,Heiles2005b}. This has been seen several times toward galactic OH masers \citep{Hutawarakorn2002, Fish2006, Caswell2014}.  However, in typical Zeeman observations of star-forming molecular clouds, the Zeeman signal is only strong enough to yield the strength of the magnetic field along the line of sight.} in molecular clouds \citep{Troland1986, Crutcher1993, Crutcher1999}.  When threaded by a magnetic field, atomic hydrogen and molecules with a strong magnetic dipole moment will have the degeneracy in magnetic sub-levels lifted for states with non-zero angular momentum.  This will split the radio frequency transitions into a number of linearly and elliptically polarized components separated slightly in frequency.  Measuring this Zeeman splitting is the only way to directly measure a component of the magnetic field strength.  However, we will not focus more on the thermal Zeeman effect in this review, as there are no reported observations with a (sub)millimeter-wavelength interferometer.  For reviews of single-dish observations of magnetic fields in molecular clouds via the Zeeman effect, see \citet{Crutcher2012} and \citet[][in this volume]{CrutcherKemball2019}.

\subsection{Analysis methods}

\subsubsection{Indirect estimates of magnetic field strength}
\label{sec:analysis_field_strength}

The polarization arising from magnetically aligned dust grains and from the G-K effect yields the plane-of-sky magnetic field morphology.  However, these observations do not contain information about the magnetic field strength.  Since knowing the field strength is critical to our understanding of the importance of magnetic fields, a variety of indirect methods have thus been developed to estimate the field strength from these types of observations.

The most longstanding of these methods is the Davis-Chandrasekhar-Fermi (DCF) method \citep{Davis1951, Chandrasekhar1953}, which uses the fact that turbulent motions should cause an observable scatter in what would otherwise be a well ordered mean magnetic field.  The original, simplest form of the DCF formula yields an estimate of the plane-of-sky magnetic field strength 

\begin{equation}
B_\textrm{POS} = \frac{\sqrt{4\pi\rho}}{\delta\phi} \delta V \,\,,
\end{equation}

\noindent
where $\rho$ is the gas density, $\delta$V is the one-dimensional velocity dispersion, and $\delta\phi$ is the dispersion in polarization position angles.  $\delta$V and $\delta\phi$ are both assumed to be caused by turbulent motions in the region being studied.  The derivation of this expression also assumes that there is a mean magnetic field in the region, that the turbulence is incompressible and isotropic, and that the turbulent components of the kinetic and magnetic energies are in equipartition.  Note that the DCF method also assumes that the dispersion is ``small,'' i.e., that the turbulent magnetic energy is small compared with the mean-field magnetic energy in the system.

The DCF method was originally developed to estimate magnetic field strengths in the diffuse ISM, where perturbations in the magnetic field can be assumed to be caused by turbulent motions in the magnetized medium.  Comparisons with MHD simulations of giant molecular clouds (GMCs) by \citet{Ostriker2001} found that at these scales, the original DCF method typically overestimates the magnetic field strength by a factor of $\sim$\,2 because line-of-sight field tangling is not taken into account.  Further simulations by \citet{Heitsch2001} also found that the true field strength is overestimated unless finite telescope resolution and self-gravity within the GMC are accounted for.

At the smaller scales of the individual star-forming clumps and cores that are probed by interferometers, gravity is almost always the dominant dynamical factor, and thus the structure of the background field must be removed before calculating the dispersion in polarization position angles.  In two ideal cases with very clean examples of hourglass-shaped magnetic fields, \citet{Girart2006} and \citet{Qiu2014} removed the background hourglass field by subtracting a set of parabolic field lines from the data, after which they calculated the magnetic field dispersion relative to the background structure.  In complicated cases, however, a more general method is necessary to remove arbitrary background field structure.  This has been achieved by employing a second-order structure function of the polarization angle that allows the separation of the turbulent and mean magnetic-field components, with the underlying assumption that the mean-field component has an intrinsically larger spatial scale than the turbulent component \citep{FalcetaGoncalves2008, Hildebrand2009, Houde2009, Houde2011, Chitsazzadeh2012}.  Further refinements of these studies have also taken into account interferometric filtering effects \citep{Houde2016}.

\citet{FalcetaGoncalves2008} used the structure-function approach to test the validity of the DCF technique using their MHD simulations.  In addition to analyzing the effect of different telescope resolutions (their conclusion: lower-resolution observations tend to overestimate the field strength), they also derived a generalized form of the DCF equation, which allows for the separation of the turbulent and mean magnetic field components, and yielded magnetic-field estimates that were accurate to within $\sim$\,20\%.  More recently, \citet{Juarez2017} performed SMA observations of a magnetized high-mass star-forming region and used the structure-function method to compare the data with synthetic observations of gravity-dominated MHD simulations.  They found the magnetic field strength estimates from both the observations and simulations to be in good agreement.

\citet{Koch2012} proposed another method of indirectly measuring the magnetic field strength that is also based on an ideal MHD framework.  They argue that a position-dependent magnetic field strength in a polarization map can be calculated using the angle between the magnetic field and the gradient of the total-intensity emission, and a 
second angle between the local gravity direction and the gradient of the total-intensity emission.  This is based on the assumption that the intensity gradient is a proxy for the direction of motion in the MHD force equation.  For example, in a centrally concentrated, gravitationally bound core, the infalling motion will be along the intensity gradient (across the iso-intensity contours).  This method requires observations of the magnetic-field morphology (i.e., via observations of thermal dust polarization or the G-K effect) in order to produce a spatial distribution of the magnetic field strength.  However, unlike the DCF method, this method does not require spectral-line observations in the analysis.  Estimations of magnetic field strengths using this method have been found to be consistent with previous estimations from the DCF method (see, e.g., Section \ref{sec:high_mass_cores}).

\subsubsection{The mass-to-flux ratio}
\label{sec:m2f}

Merely measuring the magnetic field strength does not allow us to determine immediately the importance of the magnetic field in a given star-forming region.  Therefore, in magnetic field studies it is common to compare the magnetic energy density with that of other dynamical quantities such as gravity, turbulence, and outflow feedback (see, e.g., \citealt{Hull2017b} for a comparison of magnetic energy with gravitational and outflow energy densities in the intermediate-mass Class 0 protostellar core Serpens SMM1).  

Historically, there has been a strong focus on the comparison of gravity (which causes inward motion) and the magnetic field (whose tension provides resistance against infall across the field lines).  The common quantity quoted as a metric for the importance of the magnetic field with respect to gravity is the ``mass-to-flux ratio'' $M/\Phi$, where $M$ is the mass of the object of interest and $\Phi$ is the flux of the magnetic field threading the object.  As discussed in \citet[][and references therein]{Crutcher2004}, the maximum mass that can be supported by a given magnetic flux is given by $M_\textrm{crit} = \Phi / 2 \pi \sqrt{G}$.  However, it is more useful to discuss the dynamical status of an object by measuring the ratio $\lambda$ of the observed mass-to-flux ratio to the critical mass-to-flux ratio:

\begin{equation}
\lambda = \frac{\left(M/\Phi \right)_\textrm{obs}}{\left(M/\Phi \right)_\textrm{crit}}\,\,.
\end{equation}

\noindent
Clouds that are supported by the magnetic field and are not collapsing are deemed ``subcritical'' ($\lambda$\,$<$\,1), whereas those where gravity has overcome the resistance of the magnetic field are referred to as ``supercritical'' ($\lambda$\,$>$\,1).  \citet{Crutcher2012} analyzed data across a wide range of spatial scales that trace more than five orders of magnitude in densities, and found that when the hydrogen column density $N_H > 10^{21}$\,cm$^{-2}$, all star-forming objects are supercritical (i.e., are collapsing).  This value is less than the typical column densities of low-mass ($N_H \sim 10^{22} - 10^{23}$\,cm$^{-2}$; e.g., \citealt{Girart2006, Hull2017b}) and high-mass ($N_H \sim 10^{23} - 10^{24}$\,cm$^{-2}$; e.g., \citealt{Girart2009}) protostellar cores, and thus the types of objects we review in this article are all supercritical.  This is reasonable, as most of them have already formed stars, as revealed by the presence of bipolar outflows.  Furthermore, due to the sensitivity limits of CARMA and the SMA, most of the sources in previous interferometric surveys of polarization \citep{Hull2014, Zhang2014} were chosen based on their strong millimeter flux, which correlates with the presence of embedded star formation.

\subsection{Core-mass estimates from dust emission}
\label{sec:core_mass}

In order to convert the observed millimeter-wave flux density $S_{\nu}$ contained within a given spatial area into a corresponding gas mass $M_{\mathrm{gas}}$, we can use the following relation:

\begin{equation}
M_{\mathrm{gas}} = \frac{S_{\nu}d^{2}}{\kappa_{\nu}B_{\nu}\left(T_d\right)}\,\,,
\end{equation}

\noindent 
where $d$ is the distance to the source, $\kappa_{\nu}$ is the opacity of the dust \citep{Ossenkopf1994}, and $B_{\nu}\left(T_d \right)$ is the Planck function at the frequency of the observations.  $T_d$ is the temperature of the dust, which is usually $\sim$\,20--50\,K in a low-mass protostellar core \citep{Girart2006}, and as high as (or greater than) 100\,K in a high-mass core \citep{Girart2009}.  Once the dust mass is calculated, a gas-to-dust mass ratio of 100 is usually assumed in order to calculate the total (gas\,$+$\,dust) mass of the protostellar core.  Note that such an estimate does not include the mass of the central star(s), which must be obtained by other means, e.g., via direct detection of a Keplerian disk around the source \citep[e.g.,][]{Tobin2012, Ohashi2014} or via determination of the source's bolometric luminosity.

\subsection{Motivating questions in low- and high-mass star formation}

The primary goal of observing the magnetic field at any spatial scale is to determine the importance (or lack thereof) of the magnetic field in the star-formation process.  The steady progress toward this goal over the last two decades began with single-dish submillimeter polarization surveys probing $\gtrsim$\,20\arcsec scales using the Viper 2\,m telescope at the South Pole (SPARO polarimeter, e.g., \citealt{Dotson1998, Renbarger2004, HBLi2006}), the JCMT (850\,$\mu$m SCUBA polarimeter, e.g., \citealt{Matthews2009}), and the CSO (350\,$\mu$m SHARP [\citealt{Li2008}] and Hertz [\citealt{Dotson2010}] polarimeters).  A resurgence of single-dish studies has been brought about by results from the PolKa polarimeter at the APEX telescope \citep{Siringo2004, Siringo2012, Hezareh2013, Wiesemeyer2014, Alves2014}, the BISTRO survey with the upgraded POL-2 polarimeter at the JCMT \citep[e.g.,][]{WardThompson2017, Pattle2017, JKwon2018, Pattle2018, Soam2018}, results from the polarimeter on the Balloon-borne Large Aperture Submillimeter Telescope \citep[BLAST;][]{Roy2011, Gandilo2016, Fissel2016}, observations from the HAWC+ polarimeter \citep{Vaillancourt2007} on the Stratospheric Observatory for Infrared Astronomy \citep[SOFIA; e.g.,][]{Chuss2018, Gordon2018, LopezRodriguez2018}, and the galactic polarization maps from the \textit{Planck} satellite \citep[e.g.,][]{PlanckXIX,PlanckXX,PlanckXXI,PlanckXXXII,PlanckXXXIII,PlanckXXXV}.  These new studies will pave the way for future work with even more sensitive instruments such at the next-generation BLAST instrument (BLAST-TNG; \citealt{Galitzki2014}), the NIKA2 polarimeter at the IRAM 30\,m telescope \citep{Ritacco2017}, and the TolTEC polarimeter at the Large Millimeter Telescope (LMT).  For reviews on multi-scale/multi-wavelength studies and single-dish observations of magnetic fields, respectively, see \citet{HBLiLaw2019} and \citet{PattleFissel2019}, both in this volume.

There are a number of questions applicable to both low- and high-mass star formation that have been investigated using (sub)millimeter polarimetric observations.  These include the (direct or indirect) measurements of the absolute magnetic field strength in star-forming material at different spatial scales, as well as the estimation of the dynamical importance of the magnetic field with respect to gravity (i.e., the mass-to-flux ratio; see Section \ref{sec:m2f}).  Observations of magnetic fields across multiple spatial scales toward both low-mass \citep{HBLi2009, Hull2014} and high-mass \citep{Zhang2014, HBLi2015} protostellar sources have also been used to constrain the dynamical importance of the magnetic field based on the morphological consistency (or lack thereof) of the field as a function of scale.  Generally, however, these observations have only compared two or three of the relevant spatial scales (i.e., 100\,pc cloud scales, 1\,pc clump scales, 0.1\,pc dense-core scales, 1000\,au protostellar envelope scales, and 100\,au disk scales).  A full characterization of the magnetic field from galactic (\textit{Planck}) scales down to scales approaching the 100\,au size of protoplanetary disks has yet to be accomplished, but will be possible in the near future when upcoming polarization surveys of the full populations of protostars in entire molecular clouds are completed.

In the low-mass regime, single-dish observations probe the magnetic field in star-forming clouds at large scales, revealing the magnetic field from the scale of entire molecular clouds (\textit{Planck}, BLAST) to the canonical, $\sim$\,0.1\,pc dense core, where one or a few protostars will form (JCMT, CSO).  One of the main benefits of single-dish studies is their ability to recover a larger range of spatial scales than interferometers, thus enabling an accurate characterization of the magnetic field in ambient cloud material.  However, higher resolution is needed in order to probe the environments of individual stars; this is where results from the BIMA, CARMA, SMA, and ALMA interferometers dominate the discussion, allowing us to characterize the magnetic field from scales of several $\times$ 1000\,au down to the scales of a few $\times$ 10\,au accessible to ALMA.\footnote{At the high resolutions achievable by ALMA, several studies have revealed polarization in well resolved maps of protoplanetary disks; however, it appears that in many cases the polarized emission is from dust scattering and not from magnetically aligned dust grains (see Section \ref{sec:scattering}).}  The main questions that have been tackled over the last two decades using data from these interferometers include:
\textbf{(1)} What is the importance of the magnetic field in regulating the collapse of star-forming cores? (Section \ref{sec:collapse});
\textbf{(2)} What is the relationship of bipolar outflows with the magnetic field? (Section \ref{sec:Bfield_outflow}); and
\textbf{(3)} What is the role of the magnetic field in the launching and collimation of bipolar outflows in low-mass protostars? (Section \ref{sec:outflows})

Moving to the high-mass regime: high-mass stars ($M_* >$ 8\,\msun) form predominantly in clustered environments where a population of stars are born with a range of stellar masses \citep{LadaLada2003}. These high-mass stellar populations form in dense cores that are embedded in parsec-scale, massive molecular clumps.   
This review will focus on several questions raised in recent studies of these high-mass sources, including 
\textbf{(4)} What is the dynamical role of magnetic fields in dense cores? (Section \ref{sec:high_mass_cores});
\textbf{(5)} What is the role of the magnetic field in the formation of disks and the launching of protostellar outflows in high-mass protostars? (Section \ref{sec:Bfield_outflow_high});
\textbf{(6)} Do magnetic fields play a significant role in the fragmentation of molecular clumps and the formation of dense cores? (Section \ref{sec:fragmentation_high}); and
\textbf{(7)} Does high-mass star formation proceed in virial equilibrium? (Section \ref{sec:virial_high})

The dense clustering nature of high-mass star formation implies considerable fragmentation within these massive molecular clumps, which distinguishes high-mass star formation from the more isolated process of low-mass star formation.  High-mass regions also tend to have much more intense radiation environments, hosting \hii regions whose radiative feedback and ionization can impact the ambient magnetic field.  Another clear difference arises from the fact that the best-studied low-mass stars are forming at distances $\sim$\,10\,$\times$ closer than typical high-mass star forming regions, allowing us to study the formation of individual low-mass protostellar systems and their associated outflows, jets, and disks in much greater detail than is possible in high-mass systems. Ultimately, however, studies of both low- and high-mass star formation use the same observing techniques and confront many of the same questions.  In this review we focus on those questions that have been of the most interest to both communities in recent years.

\section{Magnetic fields in low-mass star formation}
\label{sec:low_mass}

The revolution of high-resolution, interferometric observations of polarization began with BIMA and the Owens Valley Radio Observatory (OVRO).  These two sets of antennas were later combined into CARMA \citep{Bock2006}.  Early observations with BIMA and 
OVRO\footnote{The two polarization results from OVRO are observations toward NGC~1333-IRAS~4A and IRAS~16293 \citep{Akeson1996, Akeson1997}; however, OVRO was known to have issues with polarization calibration, which is the most likely explanation for the inconsistency of those results with later observations of the same sources \citep{Girart1999, Girart2006, Girart2008, Rao2009}.} 
covered a wide range of topics, including polarization observations of dust, SiO, CO, and SiO 
masers toward iconic regions in Orion \citep{Rao1998, Plambeck2003, Girart2004, Matthews2005} as well as observations of individual protostars \citep{Girart1999, Kwon2006, Cortes2006b}.  These first observations, combined with the extensive follow-up from CARMA, the SMA, ALMA form the body of work motivating this review.  Below we put this work in the context of a narrative addressing several of the major open questions in the field of magnetized low-mass\footnote{We do not treat the topic of intermediate-mass star formation separately in this review.  As many of the characteristics of the early stages of intermediate- and low-mass star formation are thought to be similar \citep{Beltran2015}, we include references to several intermediate-mass objects and regions in this section.  Many of these intermediate-mass sources are in Orion \citep[e.g.,][]{Takahashi2006, Takahashi2018, Hull2014}, but there are also objects in other regions, such as Serpens SMM1 in the Serpens Main molecular cloud \citep{vanKempen2016, Hull2017b}.} star formation.

\subsection{The role of the magnetic field in protostellar collapse}
\label{sec:collapse}

\begin{wrapfigure}[19]{r}{3.0in}
\vspace{-2em}
\includegraphics[scale=0.85, clip, trim=0cm 0cm 0cm 0cm]{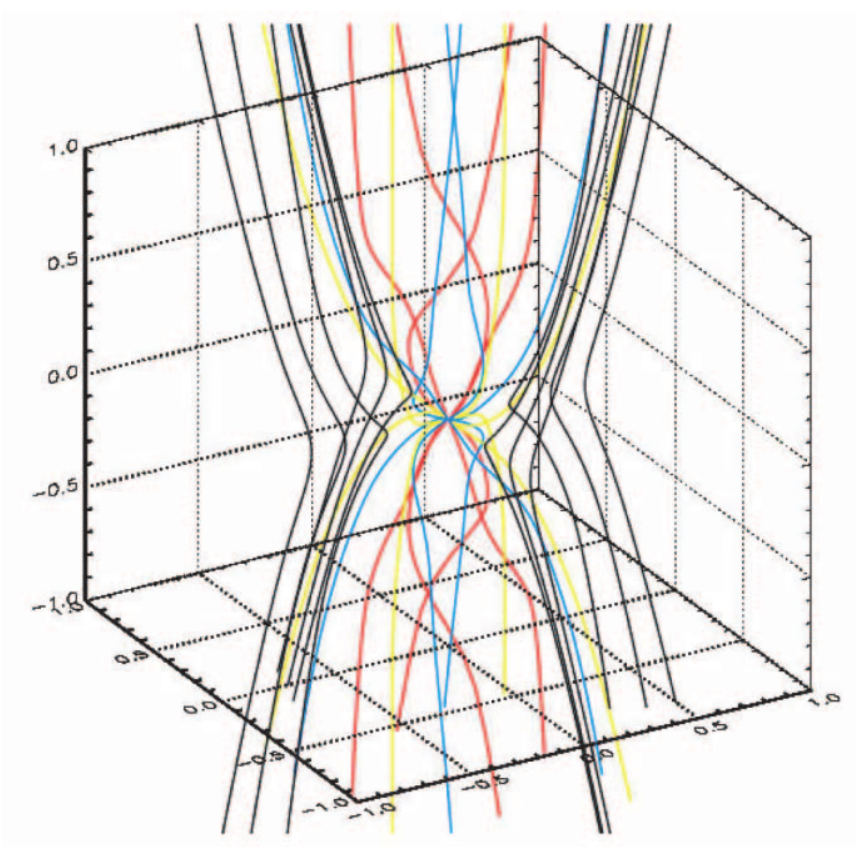}
\caption{ 
\small
A model from \citet{Allen2003} showing an hourglass-shaped magnetic field configuration in a collapsing, magnetized star-forming core.  Reproduced with permission from the American Astronomical Society (AAS).
}
\label{fig:hourglass_model}
\end{wrapfigure}

In models of magnetically regulated protostellar collapse \citep[e.g.,][]{Mouschovias1976a, Mouschovias1976b, Mouschovias1991, Mouschovias1999}, a strong, well ordered magnetic field provides outward pressure support of the infalling material.   This is because the field is coupled (or ``frozen'') to the small fraction of charged particles in the weakly ionized gas.  However, in non-turbulent models, the non-ideal MHD effect of ambipolar diffusion \citep{Mestel1956} enables the neutral material (which comprises the bulk of the star-forming core) to slip slowly past the magnetic field lines, thus removing magnetic flux and eventually allowing collapse to proceed once the mass-to-flux ratio exceeds the critical value.  

\begin{figure}
\includegraphics[width=1.0\textwidth, clip, trim=0cm 8.5cm 0cm -1cm]{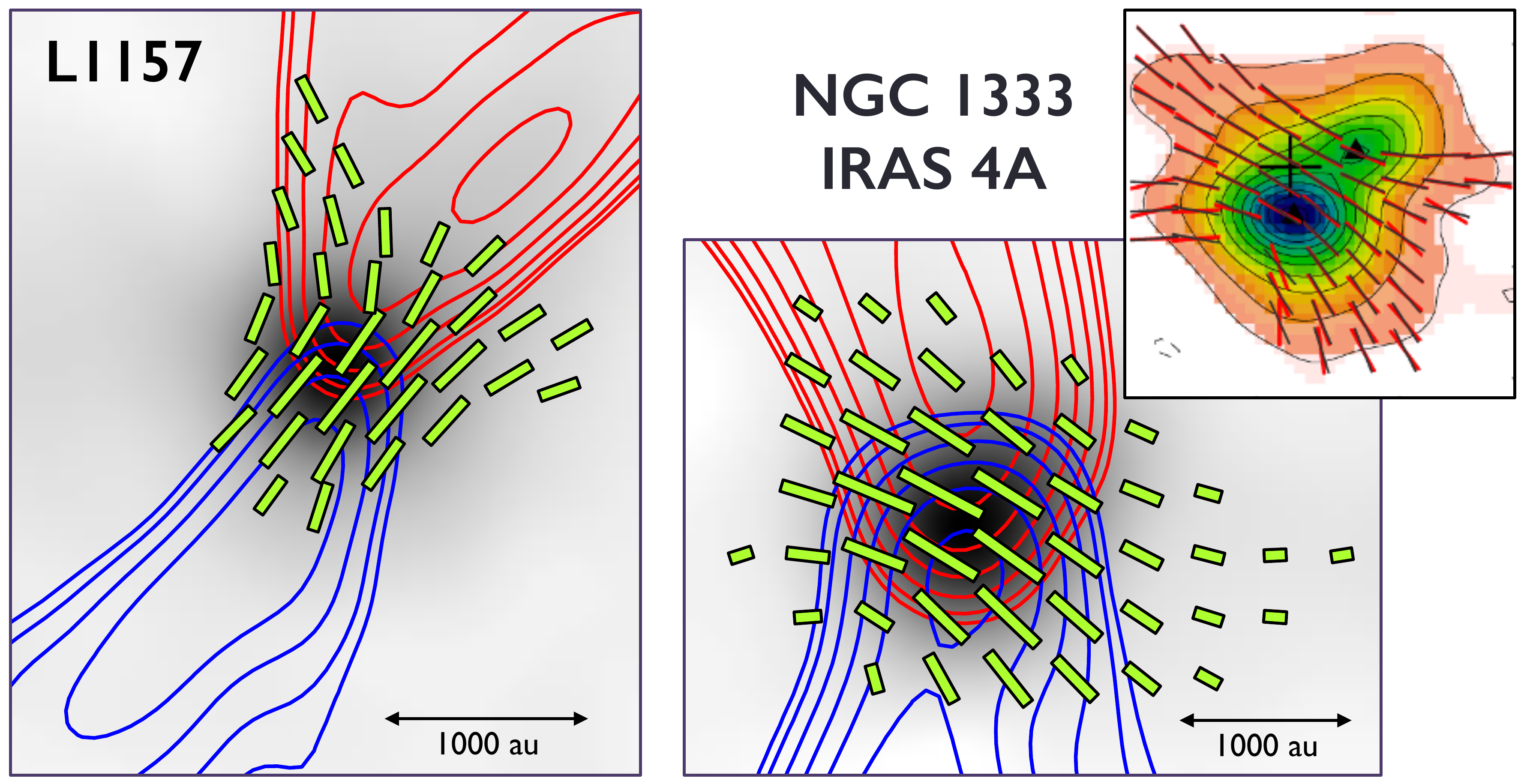}
\caption{
\small
Observations of hourglass-shaped magnetic field configurations (line segments).  The rightmost, overlapping panels are NGC 1333-IRAS 4A, observed by the SMA in \citet[][top]{Girart2006}, and again by CARMA in \citet[][bottom/background]{Hull2014}.  The SMA and CARMA observations are zoomed to the same spatial scale.  The left-hand panel is the isolated Class 0 protostar L1157 (adapted from \citealt{Hull2014}; see also \citealt{Stephens2013}).  Figures reproduced with permission from \textit{Science} magazine and the AAS.
}
\label{fig:hourglass_obs}
\end{figure}

One of the predicted signposts of highly magnetized star formation is that at high enough densities ($\gtrsim 10^4$\,cm$^{-3}$), the collapse of strongly magnetized gas should pinch the magnetic field into an ``hourglass'' shape with a symmetry axis perpendicular to the major axis of a flattened, $\sim$\,1000\,au ``pseudodisk'' \citep{Galli1993a, Galli1993b}.  The hourglass is expected to persist down to scales $<$\,1000\,au \citep{Fiedler1993, Galli1993b, Allen2003, Goncalves2008, Frau2011, Kataoka2012, Mocz2017}; see Figure \ref{fig:hourglass_model}.  And indeed, the predicted hourglass has now been seen in a number of interferometric observations of low-mass protostellar cores \citep[][see Figure \ref{fig:hourglass_obs}]{Girart1999, Girart2006, Girart2008, Rao2009, Stephens2013, Hull2014, Kwon2018, Sadavoy2018a, Maury2018}, suggesting that some protostellar cores do form in strongly magnetized regions.  For a discussion of constraining strong-field star formation via observations of hourglass-shaped magnetic fields, see Section \ref{sec:stat}.  

A second signpost of strong-field star formation is the consistency of the magnetic field orientation across multiple spatial scales. If the field is strong relative to other dynamical effects, observations at small scales should reveal a magnetic field whose original orientation is preserved from the parsec scale of the cloud in which the source is embedded.  To date, multi-scale studies of the magnetic field in low- and high-mass star-forming regions have compared two or three scales: i.e., $\sim$\,kpc galactic scales to $\sim$\,0.1\,pc dense-core scales in \citet{Stephens2011}; 100\,pc cloud scales to 0.1\,pc dense-core scales in \citet{HBLi2009, HBLi2015}; 1\,pc clump scales to 0.1\,pc dense-core scales in \citet{Zhang2014}, continuing down to $\sim$\,0.01\,pc scales in \citet{Girart2013} and \citet{Ching2017}; 0.1\,pc dense core scales to 1000\,au protostellar envelope scales in \citet{Hull2014} and \citet{Davidson2014}; and 0.1\,pc to 1000\,au to 100\,au scales in \citet{Hull2017a, Hull2017b}.  

In the low-mass regime, \citet{HBLi2009} found striking consistency between the magnetic field orientation in the Orion molecular cloud derived from background starlight polarization at $\sim$\,100\,pc scales versus polarized thermal dust emission at $\sim$\,0.1\,pc scales.  \citet{Hull2014} took this one step further, finding consistency in the field orientation in just a few of the low-mass protostars in their sample from scales of $\sim$\,0.1\,pc to $\sim$\,1000\,au.  The sample of sources from \citet{Hull2014} that maintained consistency in the magnetic field orientation down to scales of $\sim$\,1000\,au tended to be those sources with a higher polarization fraction, which implies that the magnetic fields in those sources are more ordered, and thus may be more dynamically important.  

Among those sources with consistent large-to-small scales magnetic fields are several with known hourglass morphologies, including OMC3 MMS6 \citep{Hull2014}, NGC~1333-IRAS~4A \citep{Girart2006}, L1157 \citep{Stephens2013}, and L1448 IRS 2 \citep{Kwon2018}.  The magnetic field strengths have been estimated toward the latter three objects, an are all relatively high, on the order of $\gtrsim$\,1\,mG, which is similar to values obtained in high-mass regions (see Section \ref{sec:high_mass_cores}).  However, while the values are high (5\,mG in IRAS~4A [\citealt{Girart2006}]; 1.3--3.5\,mG in L1157 [\citealt{Stephens2013}]; and 750\,$\mu$G in L1448 IRS 2 [\citealt{Kwon2018}]), the mass-to-flux ratios calculated for the two most magnetized sources (IRAS~4A and L1157) are both slightly greater than the critical value (1.7 and 1.1 for IRAS~4A and L1157, respectively), which is reasonable, considering that the objects have already collapsed to form protostars. 

\begin{figure}[hbt!]
\includegraphics[width=1.0\textwidth, clip, trim=5.5cm 0cm 5.0cm -1cm]{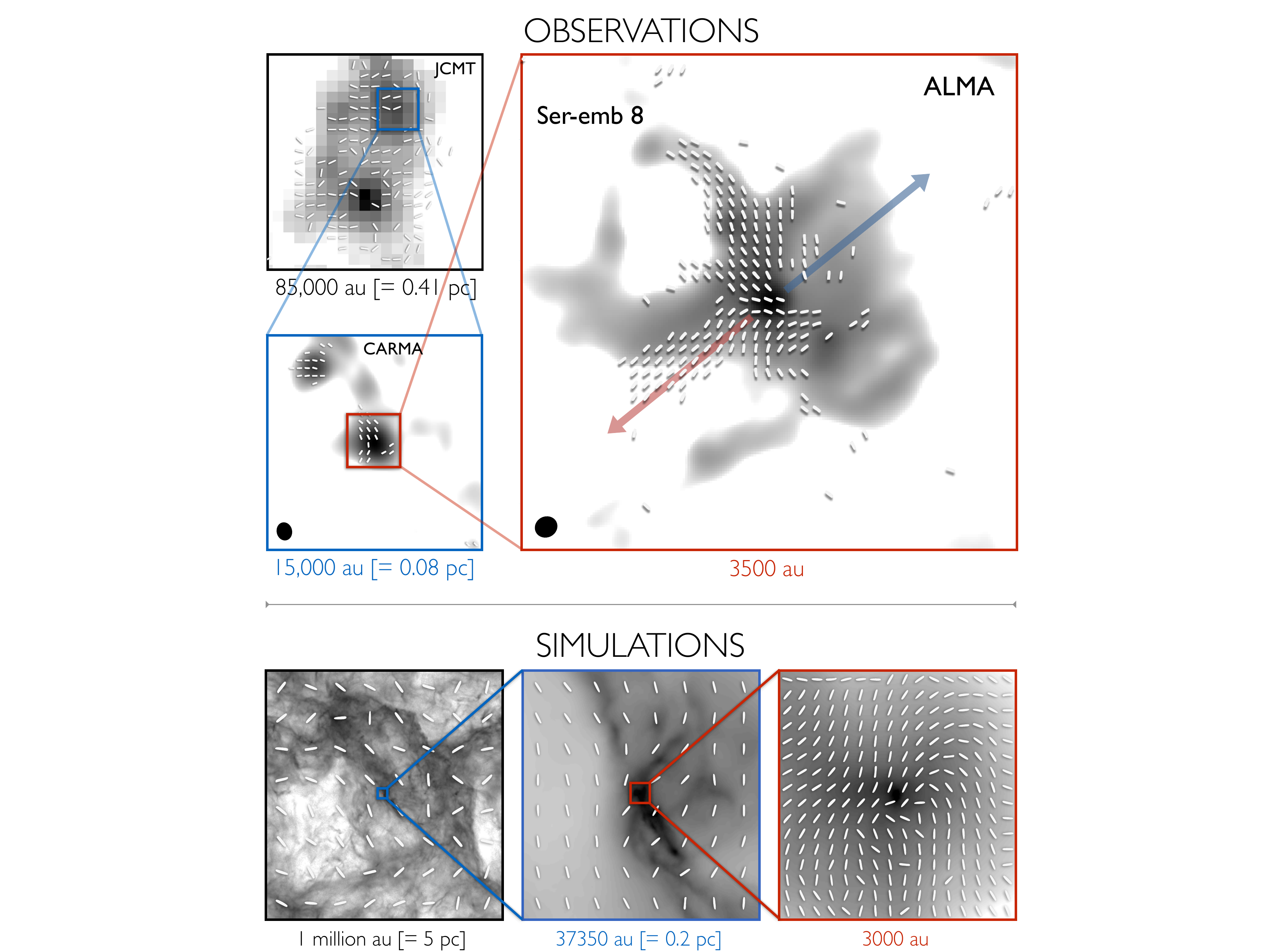}
\caption{
\small
\textit{Top:} Multi-scale observations of the magnetic field (line segments) toward the low-mass Class 0 protostar Ser-emb~8.  Grayscale is total-intensity (Stokes $I$) thermal dust emission.  Observations are from the JCMT at 850\,$\mu$m (top-left; \citealt{Matthews2009}), CARMA at 1.3\,mm (bottom-left; \citealt{Hull2014}), and ALMA at 870\,$\mu$m (right).  The red and blue arrows indicate the red- and blueshifted lobes of the bipolar outflow from Ser-emb~8 traced in CO\,($J$\,$=$\,2\,$\rightarrow$\,1).  The data show the inconsistency of the magnetic field across spatial scales, and are consistent with moving-mesh AREPO MHD simulations (\textit{bottom}) whose initial conditions include a magnetic field that is weak compared with turbulence at large (5\,pc) scales.  Adapted from \citet{Hull2017a}.  Reproduced with permission from the AAS.
}
\label{fig:ser8_multiscale}
\end{figure}

Follow-up studies with ALMA of individual sources from previous surveys (e.g., \citealt{Hull2014}) have suggested that consistency in the magnetic field orientation across spatial scales may be the exception rather than the rule at scales smaller than $\sim$\,0.1\,pc.  Indeed, detailed multi-scale follow-up studies by \citet{Hull2017a, Hull2017b} of Ser-emb 8 and Serpens SMM1, two Class 0 protostellar sources in the Serpens Main molecular cloud, found significant deviations in the magnetic field morphology across spatial scales ranging from $\sim$\,0.1\,pc down to $\sim$\,\,100\,au.

In the case of Serpens SMM1, it appears that the deviations at the $\sim$\,100\,au scales probed by ALMA are due to shaping of the magnetic field by the bipolar outflow. This is in spite of the fact that the magnetic field strength is estimated to be quite high, $\sim$\,5.7\,mG \citep{Hull2017b}.  The fact that dynamics are shaping the magnetic field morphology in this source is not surprising, however, as the magnetic is comparable to the kinetic energy in the outflow.  This is in contrast to sources like NGC 1333-IRAS 4A, where estimates by \citet{Girart2006} show that the magnetic energy is clearly dominant over turbulent motions in the source.

In the case of Ser-emb 8, the outflow does not have a clear effect on the magnetic field, and yet the field morphology is still not consistent across scales.  A comparison of the Ser-emb 8 data with moving-mesh AREPO gravoturbulent MHD simulations \citep{Mocz2017} suggests that Ser-emb 8 may have formed in an environment where dynamical effects such as turbulence and infall dominate the magnetic field, in contrast to the strong-field examples described above; see Figure \ref{fig:ser8_multiscale}.  These results are qualitatively consistent with other simulations studying star formation in weakly magnetized turbulent environments \citep[e.g.,][]{Seifried2015, Offner2017}, suggesting that the importance of the magnetic field at the smallest scales of star formation lies along a continuum, from the bright, highly polarized examples of strong-field star formation to the more complex examples of star formation in regions dominated by dynamical processes.  Large, high-resolution polarization surveys by ALMA will reveal the distribution of low-mass protostars across this continuum of magnetic importance, and will allow us to compare with results from high-mass regions such as those by \citet{Zhang2014}, who found that the magnetic field is dynamically important at the larger spatial scales probed by their observations (see Section \ref{sec:fragmentation_high}).

\subsection{Misalignment of outflows and magnetic fields}
\label{sec:Bfield_outflow}

For more than a decade, one of the primary ways that astronomers have tested the importance of the magnetic field in star-forming regions has been via comparisons of the orientations of bipolar outflows/jets and the ambient magnetic field.  This is because if a protostellar core is very strongly magnetized, the magnetic field has the ability to align all of the relevant axes: the axis of the (well ordered and poloidal) magnetic field, the symmetry axis of the pseudodisk, the rotational axis of the protostellar disk, and the axis of the outflow and/or jet emanating from the central source.  This happens because of the ``magnetic braking'' phenomenon \citep{Allen2003}, where a strong magnetic field removes angular momentum from the central source and causes the angular momentum (and thus disk/outflow) and magnetic axes to align \citep[e.g.,][]{Machida2006}.  

This same magnetic braking phenomenon can potentially lead to what has come to be known as the ``magnetic braking catastrophe,'' where a strong magnetic field aligned with the core rotation axis can suppress the formation of a Keplerian disk in MHD simulations \citep[e.g.,][]{Galli2006, Mellon2008, Li2011}.  This may lead to the formation of sources like L1157 \citep{Stephens2013} and B335 \citep{Maury2018}, which have outflows aligned with the magnetic field, and which have as-of-yet undetectably small disks ($\lesssim$\,10\,au).  However, since it is known that large Keplerian disks form around many protostellar sources, a variety of methods have been proposed to overcome this problem, including the introduction of an initial misalignment between the rotation axis and the magnetic field, which enhances disk formation \citep[e.g.,][]{Hennebelle2009, Joos2012, Krumholz2013, Li2013}), and the consideration of non-ideal MHD effects such as ambipolar diffusion \citep{Dapp2012, Masson2016, Tsukamoto2018}, Ohmic dissipation \citep{Dapp2012, Tomida2015, Tsukamoto2018}, the Hall effect \citep{Tsukamoto2015a, Tsukamoto2015b, Tsukamoto2017, Wurster2018}, and magnetic reconnection \citep{SantosLima2012, Li2014}.

\begin{figure}[hbt!]
\begin{center}
\includegraphics[width=0.85\textwidth, clip, trim=0cm 0cm 0cm 0cm]{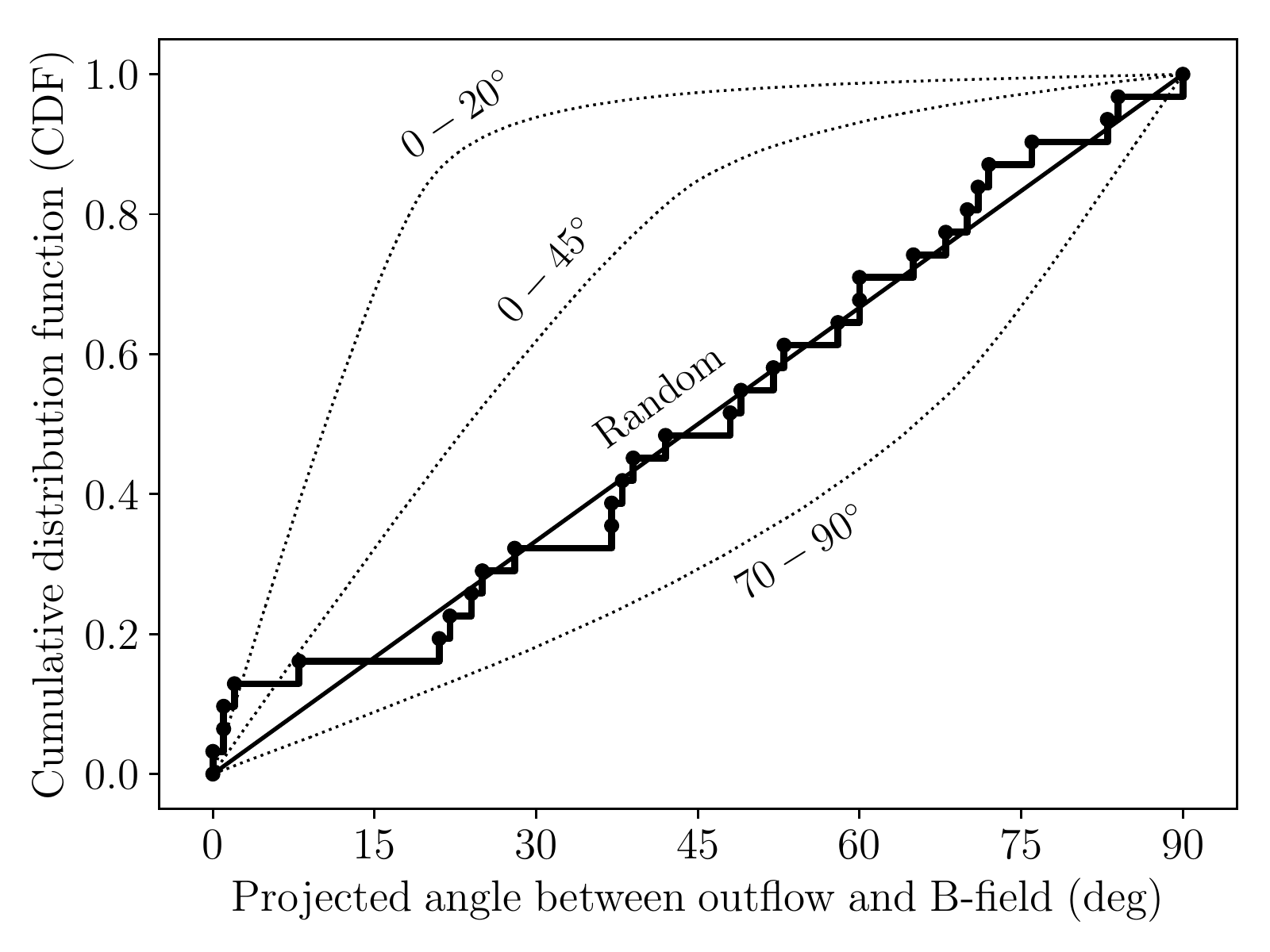}
\caption{
\small
The thick, stepped curve shows the cumulative distribution function (CDF) of the (projected) angles between the bipolar outflows and the mean magnetic-field orientations in the full sample of low-mass protostellar cores observed to date in full polarization with BIMA, the SMA, CARMA, and ALMA.  The dotted curves are the CDFs from Monte Carlo simulations where the magnetic fields and outflows are oriented within 20$^\circ$, 45$^\circ$, and 70--90$^\circ$ of one another, respectively. The straight line is the CDF for random orientation. The plot shows that outflows appear to be randomly aligned with magnetic fields in the sample of low-mass sources whose magnetic fields have been observed with $\sim$\,500--1000\,au resolution.
}
\label{fig:misalignment}
\end{center}
\end{figure}

Observationally, if very strong magnetic fields were the norm, then the rotational axes of protostellar disks, and the jets and outflows that emanate from them, should all be parallel with the ambient magnetic field.  A study of seven low-mass protostellar cores by \citet{Chapman2013} found a correlation between outflows and magnetic fields at $\sim$\,0.1\,pc scales.  However, the majority of the studies of this type have come to the opposite conclusion.  For example, \citet{Menard2004} found that the optical jets from classical T Tauri stars in the Taurus-Auriga molecular cloud are randomly oriented with respect to the parsec-scale magnetic field observed via background-starlight polarization observations.  \citet{Targon2011} obtained a similar result for 28 regions spread over the Galaxy, finding no strong correlations between protostellar jets and the ambient magnetic field.  On the $\sim$\,0.1\,pc scales of high-mass star-forming cores, \citet{Curran2007} and \citet{Zhang2014} used thermal dust polarization observations by the JCMT and the SMA, respectively, to determine that outflows and inferred magnetic fields are randomly aligned.  Finally, \citet{Hull2013, Hull2014} used the 1.3\,mm polarization system at CARMA \citep{Hull2015b} to observe dust polarization toward a sample of low- and high-mass sources, and found that the outflows and $\sim$\,1000\,au-scale magnetic fields in the low-mass sources were randomly aligned.  In Figure \ref{fig:misalignment}, we compile all of the outflow-versus-magnetic-field angles derivable to date from interferometric observations of low-mass protostellar cores \citep{Girart1999,Girart2006,Girart2008,Rao2009,Hull2013,Stephens2013,Hull2014,Hull2017a,Hull2017b,Cox2018,Galametz2018,Kwon2018,Maury2018,Sadavoy2018a,Harris2018}, and come to the same conclusion: while a few sources have well aligned outflows and magnetic fields (e.g., those on the very bottom-left of the plot in Figure \ref{fig:misalignment} that are climbing the 0--20$^\circ$ curve, several of which have hourglass-shaped field morphologies; see Section \ref{sec:stat}), overall protostellar outflows and magnetic fields measured at 1000\,au-scales are randomly aligned.

The finding from several studies that magnetic fields and outflows are randomly oriented suggests that most protostars form out of material with a magnetic field that is too weak to maintain a consistent orientation all the way down to the 0.1--10\,au scales where outflows are launched.  Rather, it seems likely that dynamical effects such as turbulence and infall are more important than the magnetic field when it comes to dictating the ultimate angular-momentum direction at the small ($<$\,1000\,au) scales relevant for the formation of protostellar disks and outflows/jets.  This claim is corroborated by the analysis by \citet{JLee2017} of the synthetic observations (produced using the \texttt{Dustpol} radiative transfer software \citealt{Padovani2012}) of two MHD simulations with different initial mass-to-flux ratios \citep{Offner2017}.  They concluded that while the protostar that formed in the strong-field case exhibited a correlation between the outflow and the magnetic field, the weaker-field case showed a random orientation.  This is most likely because asymmetric accretion from the turbulent envelope stochastically reoriented the disk/outflow during the earliest (Class 0) formation stage, ultimately decoupling the source from the natal magnetic field (see, e.g., simulations by \citealt{CYChen2018}).

Two non-magnetic studies focused on the alignment of protostellar outflows with respect to the natal filamentary structure in which the sources formed, and on the relative alignment of the outflows from wide ($\sim$\,1000\,au) and tight ($<$\,250\,au) binary/multiple systems.  Both studies used data from the MASSES survey at the SMA \citep{Lee2015, Stephens2018}.  Regarding outflows versus filaments, \citet{Stephens2017a} studied the relative orientation of 57 protostellar outflows in the Perseus molecular cloud (derived from the SMA MASSES data) with the local filamentary structure (derived from \textit{Herschel} observations), and found that the orientations are consistent with a random distribution.  Their results held regardless of protostellar age, multiplicity, or the opacity of the dense core, suggesting that the angular momentum of the protostellar cores and outflow-launching disks are independent of the large-scale structure out of which the protostars are forming.  Regarding the orientation of outflows from multiple systems, \citet{Lee2016} used SMA MASSES observations to determine that the outflows from proto-multiple systems in the Perseus molecular cloud are randomly aligned.  \citet{Offner2016} followed up with MHD simulations that are consistent with the SMA observations, arguing that multiple systems with randomly oriented angular momenta are likely to have formed from turbulent fragmentation.  

The turbulent fragmentation scenario is consistent with recent ALMA observations of multiple systems with misaligned protoplanetary disks \citep{JensenAkeson2014, JELee2017}. However, observations by \citet{Tobin2016b} have found evidence for an alternative theory of multiple formation: disk fragmentation \citep{Kratter2010}.  The VLA Nascent Disk and Multiplicity Survey of Perseus Protostars (VANDAM) by \citet{Tobin2016a} found a bimodality in the multiplicity of systems; they argued that the wide multiples are the result of turbulent fragmentation, whereas the tight multiples are the result of disk fragmentation.  This latter conclusion was strengthened in a follow-up study by \citet{Tobin2018}, who observed a sample of tight binaries from the VANDAM survey with ALMA in both continuum and spectral lines.

Recent work by \citet{Galametz2018} focused on the role of the magnetic field in the formation of multiple systems using an SMA survey of magnetic fields in protostellar cores.  They found tantalizing evidence that a large misalignment between the outflow and the magnetic field is found preferentially in protostars with higher rotational energies.  This $\sim$\,90$^\circ$ misalignment observed in some objects could thus be due to the winding of the magnetic field lines in the equatorial plane by strong rotation. Furthermore, they found hints that many of those same sources (i.e., those with approximately perpendicular outflows and magnetic fields) are wide multiple sources and/or have large disks, whereas the sources with well aligned magnetic fields and outflows tend to be single objects with small (or undetected) disks at the $\sim$\,500\,--\,1500\,au resolution of their observations.  The trend of large disks being associated with sources that have misaligned magnetic fields and outflows was also tentatively seen in an analysis of Karl G. Jansky Very Large Array (VLA) observations of Class 0 and I protostars by \citet{SeguraCox2018}.  These results suggest that the morphology and dynamical importance of magnetic fields at the scale of the protostellar envelope may significantly impact the outcome of protostellar collapse as well as the formation of disks and multiple systems.  Large ALMA surveys in polarization toward binary/multiple systems will shed light on the impact of the magnetic field on multiple formation via both turbulent and disk fragmentation.

\subsection{The importance of the magnetic field in launching jets and outflows}
\label{sec:outflows}

Magnetic fields play a critical role in launching and collimating both bipolar outflows and jets from young forming stars \citep{Frank2014}.  Several theories exist to explain how outflows and jets are generated, including the ``disk wind'' theory where an outflow is launched from the magnetized, rotating surface of a disk \citep{Konigl2000}, and the ``X-wind'' theory where jets are launched close to the central protostar itself \citep{Shu2000}.  Both of these theories require a magnetic field to function, and that magnetic field is expected to have both poloidal (i.e., along the outflow) and toroidal (perpendicular to the outflow) components due to the combination of infall, outflow, and rotational motions present near a forming star.  Characterizing the magnetic field in outflows and jets can thus allow us to investigate the origin of outflows in the context of these different theories.

Historically, observations of dust polarization have been used mainly to study the magnetic field morphology in the optically thin dense cores of dust and gas surrounding embedded protostars.  With the sensitivity and resolution of ALMA, it is now possible to detect polarized dust emission along the edges of outflow cavities \citep{Hull2017b, Maury2018}. However, there is not enough dust in the cavity itself (where the outflow has evacuated most of the material) to allow for a detection of the polarized emission.  Therefore, in order to probe the magnetic field in the outflowing material, one must turn to observations of spectral-line polarization (see Section \ref{sec:intro_tracers_POS:GK}).

\begin{figure*}[hbt!]
\begin{center}
\includegraphics[width=1.0\textwidth, clip, trim=5cm 0cm 3cm -1cm]{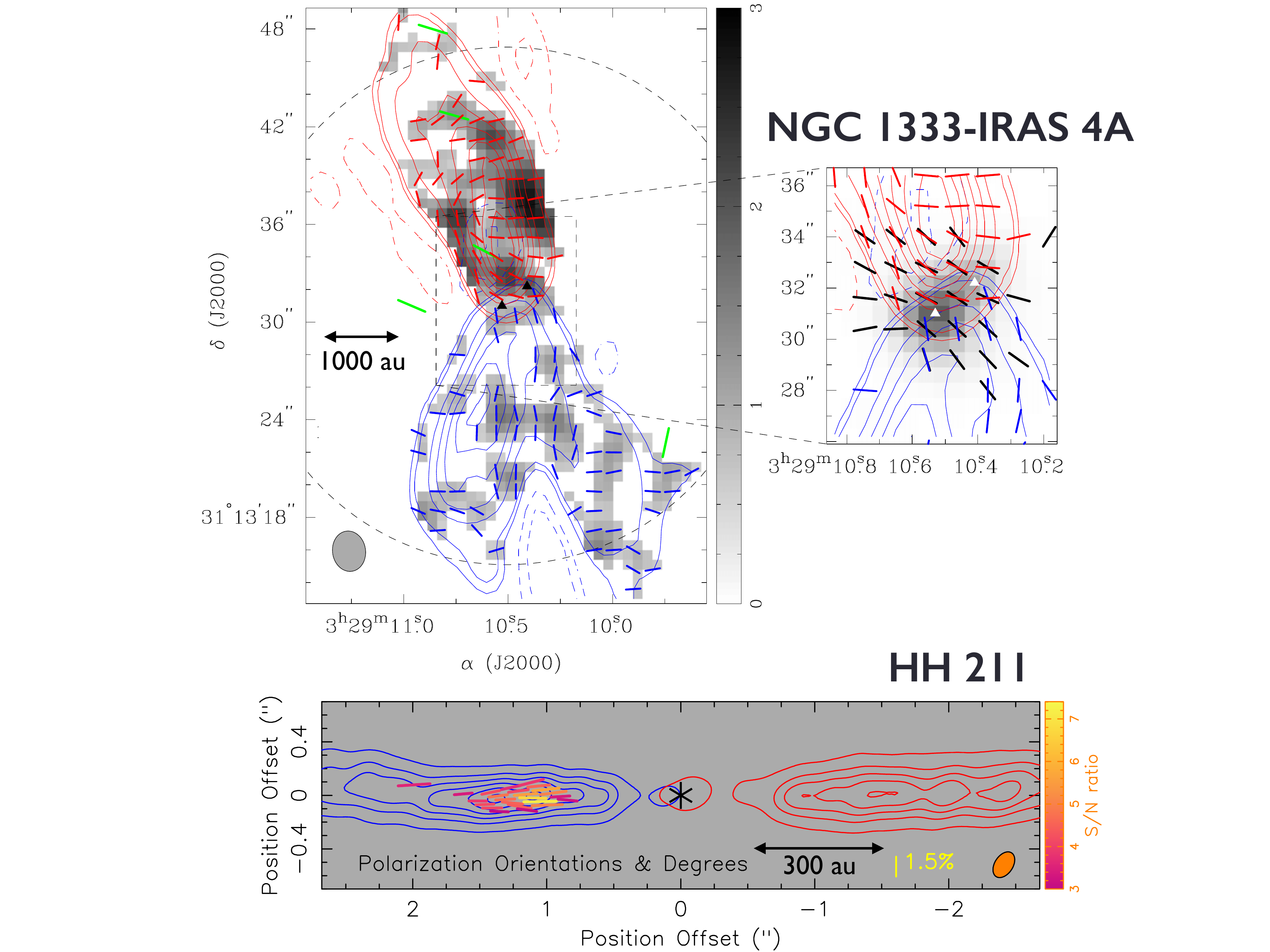}
\caption{
\small
Observations of spectral-line polarization from the G-K effect toward bipolar outflows from low-mass protostellar sources.  \textit{Top left:} SMA observations of polarization of the CO\,($J$\,$=$\,3\,$\rightarrow$\,2) line (red and blue line segments) toward NGC 1333-IRAS 4A, adapted from \citet{Ching2016}.  The grayscale shows the CO\,($J$\,$=$\,3\,$\rightarrow$\,2) polarized intensity in units of Jy\,beam$^{-1}$\,km\,s$^{-1}$.  The authors also compare with CO\,($J$\,$=$\,2\,$\rightarrow$\,1) BIMA polarization results from \citet{Girart1999}, plotted in green.  \textit{Top right:} comparison of the SMA CO polarization data with the 870\,$\mu$m dust polarization map (the black line segments have been rotated by 90$^\circ$ to reflect the inferred magnetic field direction).  
\textit{Bottom:} SMA observations of polarization in the SiO\,($J$\,$=$\,8\,$\rightarrow$\,7) line, adapted from \citet{CFLee2018b}.  The colors of the line segments indicate different levels of significance in the polarized intensity; the lengths of the segments are scaled to the polarization fraction.  The top panel was reproduced with permission from the AAS.  The bottom panel was reproduced, with modifications, in accordance with the Creative Commons Attribution 4.0 International License (\href{https://creativecommons.org/licenses/by/4.0/}{\texttt{creativecommons.org/licenses/by/4.0}}).
}
\label{fig:low_mass_SL_pol}
\end{center}
\end{figure*}

While several studies have focused on SiO maser polarization \citep[e.g.,][]{Plambeck2003}, the majority of spectral-line polarization studies toward low-mass forming stars have targeted thermal CO and SiO emission polarized due to the G-K effect.  Pioneering CO\,($J$\,$=$\,2\,$\rightarrow$\,1) polarization observations were performed with BIMA by \citet{Girart1999, Girart2004, Kwon2006}; and \citet{Cortes2006b}.  So far there has been one detection of SiO\,($J$\,$=$\,8\,$\rightarrow$\,7) polarization toward the low-mass protostar HH 211, tentatively detected by \citet{CFLee2014} using the SMA, and recently confirmed by ALMA observations reported in \citet{CFLee2018b}: see the bottom panel of Figure \ref{fig:low_mass_SL_pol}.

The BIMA observations of CO\,($J$\,$=$\,2\,$\rightarrow$\,1) polarization toward the iconic Class 0 protostar NGC~1333-IRAS~4A by \citet{Girart1999} were the first interferometric detection of the G-K effect.  Toward the central core, these observations are consistent with the magnetic field inferred from polarized dust emission.  The observations by \citeauthor{Girart1999} were followed up with the SMA by \citet{Ching2016} in the higher-energy transition CO\,($J$\,$=$\,3\,$\rightarrow$\,2). \citeauthor{Ching2016} found good consistency between their observations and those by \citeauthor{Girart1999} in the inner regions of the outflow where the polarization detections overlap.  These SMA observations, shown in Figure \ref{fig:low_mass_SL_pol} (top panel), allowed the authors to come to a number of conclusions.  First was the fact that the data are consistent with a magnetic field in IRAS~4A that is poloidal at the base of the outflows (there are two outflows, each launched by a member of the embedded binary) and toroidally wrapped up further out in the outflow cavity.  This observation, combined with the coexistence of a low-velocity outflow and a high-velocity jet in the source, led the authors to conclude that the outflows in IRAS~4A are most likely driven by MHD winds from the surface of a rotating disk.

\section{Magnetic fields in high-mass star formation}
\label{sec:high_mass}

\subsection{Magnetic field measurements at core scales}
\label{sec:high_mass_cores}

The first pioneering high-resolution observations of linearly polarized continuum and spectral-line emission toward high-mass star-forming regions were made with the BIMA interferometer. \citet{Rao1998} reported the first interferometric polarization observations of a high-mass star-forming region, toward Orion KL.  They detected linear polarization at both 3.3\,mm and 1.3\,mm in the BIMA data at a resolution of 1000\,--\,3000\,au, revealing abrupt changes in the magnetic field orientations among the continuum emission peaks. This chaotic distribution is in contrast to the uniform magnetic field topology in the lower resolution polarization maps revealed by single dish telescopes \citep[e.g.,][]{Schleuning1998, Houde2004, Pattle2017, WardThompson2017}. 


Shortly after the Orion KL study, \citet{Lai2001} reported polarization observations of W51 e1 and e2 in the 1.3\,mm continuum emission using BIMA. Later, \citet{Tang2009b} and \citet{Koch2018} published results toward the same source using the SMA and ALMA, respectively.  W51 is a cloud complex that harbors massive star formation at various evolutionary stages \citep{Ginsburg2015, Saral2017}. W51 east hosts an active star-forming molecular clump with as many as 10 compact radio continuum sources over the 0.2\,pc projected area of sources e1 and e2 \citep{Ginsburg2016}. \citet{Zhang1997} reported inverse P-Cygni profile in the e2 core in the \NH3\,($J$,$K$)\,=\,(1,1), (2,2) and (3,3) spectral lines, consistent with infall motions of the dense gas. \citet{Lai2001} detected linear polarization in the continuum emission in the e2 and e8 cores at a resolution of 14,000\,au using BIMA and found that the inferred plane-of-sky components of the magnetic fields are mostly uniform, with an average position angle of 113$^\circ$ in the e2 and 105$^\circ$ in the e8 core. Using the DCF method, the authors estimated a magnetic field strength of 0.8\,mG and 1.3\,mG in the e2 and e8 cores, respectively.

More sensitive observations of continuum emission at 870\,$\mu$m toward W51 using the SMA revealed a non-uniform magnetic field morphology at a higher resolution of 3300\,au \citep{Tang2009b}. \citeauthor{Tang2009b} explored the possible reasons for the different distributions between the BIMA and the SMA images, finding that interferometric spatial filtering is the most likely cause. This spatial filtering by interferometers can be an advantage, as it allows us to probe magnetic fields at different spatial scales, thus revealing the dynamical role of the magnetic field across the many spatial scales relevant to the star-formation process.

\begin{figure}[hbt!]
\includegraphics[width=1.0\textwidth, clip, trim=0cm 0cm 0cm 0.5cm]{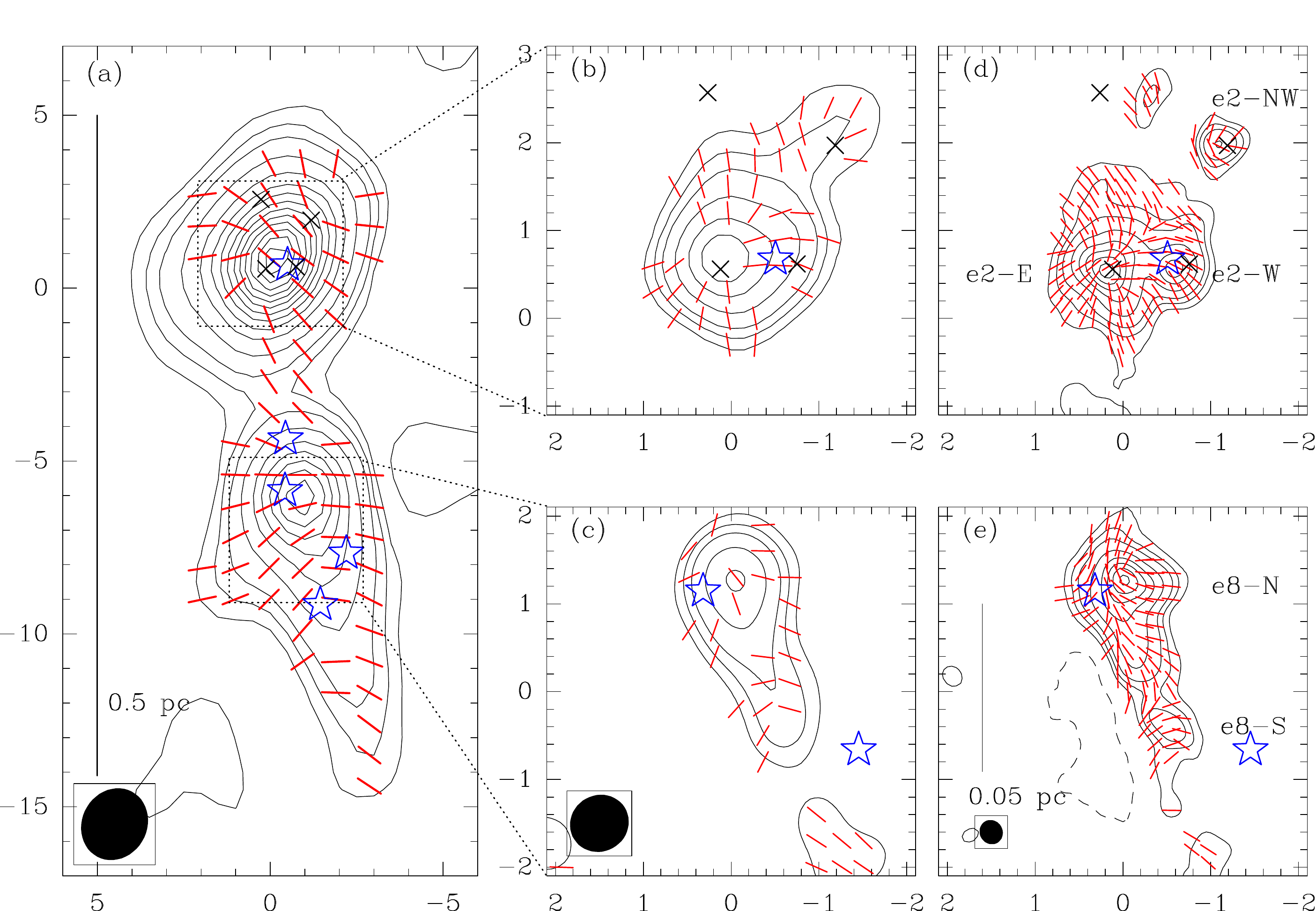}
\caption{
\small
Magnetic fields (red line segments) in W51 e2 and e8 obtained from the SMA and ALMA that exhibit more substructure at higher angular resolution. \textit{Panel (a):} SMA observations at 870\,$\mu$m probing large scales at a resolution of $2''$ ($\sim$\,10,000\,au). \textit{Panels (b) and (c):} SMA observations with a resolution of $0.\!\!^{\prime\prime}7$ ($\sim$\,3500\,au) at 870\,$\mu$m. \textit{Panels (d) and (e):} ALMA observations at 1.3\,mm with a resolution of  $0.\!\!^{\prime\prime}26$ ($\sim$\,1330\,au).  Contours are Stokes $I$ dust continuum intensity at levels of 3,\,6,\,10,\,20,\,35,\,50,\,65,\,80,\,95... times 75\,mJy\,beam$^{-1}$ in panel (a), 60\,mJy\,beam$^{-1}$ in panels (b) and (c), and 6\,mJy\,beam$^{-1}$ in panels (d) and (e).  $\times$ symbols in panels (a), (b), and (d) mark the known continuum sources W51 e2-E, e2-W, e2-NW and e2-N.  Star symbols indicate known ultra-compact \hii (\uchii) regions. Synthesized beams are shown as ellipses at the lower-left corner of panels (a), (c), and (e). Adapted from \citet{Koch2018}.  Reproduced with permission from the AAS.}
\label{fig:w51}
\end{figure}

Figure \ref{fig:w51} shows the magnetic field maps obtained by the SMA and later by ALMA at various spatial resolutions toward W51 e2 and e8. The figure shows a pinched, hourglass morphology in the e2 core at a resolution of 3300\,au (the top-middle panel), and significantly more substructures at a resolution of 1500\,au (top-right panel). \citet{Koch2018} speculate that the additional substructure in the magnetic field is the result of gravitational collapse at high densities that pulls and/or bends the field lines.

One of the key questions in the fields of magnetized low- and high-mass star formation is whether the magnetic field is dynamically important relative to turbulence and gravity. \citet{Koch2010} applied the two-point correlation function technique \citep{Hildebrand2009} and found a ratio of turbulent to magnetic energy ranging from 0.7 to 1.27 at scales ranging from $\sim$\,15,000\,au down to $\sim$\,500\,au.  \citet{Koch2012} proposed a polarization-intensity-gradient technique that can be used to derive a distribution of magnetic field strength in a given source (see Section \ref{sec:analysis_field_strength}). Applying the technique to the polarization map of W51 e2, \citet{Koch2012} found a field strength of 7.7\,mG. This value is in agreement with the estimate by \citet{Tang2009b}, who derived the field strength from the SMA polarization observations. However, it is higher than that found by \citet{Lai2001}, who used lower-resolution BIMA polarization observations probing larger spatial scales. 
The different values reported by \citet{Lai2001} and \citet{Koch2012} can be explained by the fact that the magnetic field strength increases toward smaller spatial scales where the gas density is higher.

Besides W51, another region of massive star formation whose magnetic field has been well studied is DR 21(OH) and its neighboring regions along the DR 21 filament \citep{Lai2003, Cortes2005, Girart2013, Hull2014, Houde2016, Ching2017, Ching2018}. DR 21(OH) was first imaged in linear polarization using BIMA in the CO\,($J$\,$=$\,2\,$\rightarrow$\,1) transition and 1.3\,mm continuum emission \citep{Lai2003}.  These results were followed up in the CO\,($J$\,$=$\,1\,$\rightarrow$\,0) transition and 3.4\,mm continuum emission using BIMA \citep{Cortes2005}; in the CO ($J$\,$=$\,3\,$\rightarrow$\,2) transition and 870\,$\mu$m continuum emission using the SMA \citep{Girart2013, Zhang2014}; and in the CO ($J$\,$=$\,2\,$\rightarrow$\,1) transition and 1.3\,mm continuum emission with CARMA \citep{Hull2014, Houde2016}. Strong polarization was detected in both 1.3\,mm and 870\,$\mu$m continuum emission. The field strength derived using the two-point correlation function is 2.1\,mG, yielding a mass-to-flux ratio of 6\,$\times$ the critical value \citep{Girart2013}. The field strength is consistent with the value reported by \citet{Lai2003} using the DCF method. \citet{Ching2017} imaged five additional regions along the DR 21 filament using the SMA.  The magnetic fields in all six cores display large dispersions, in contrast to the ordered magnetic field obtained with the JCMT at lower angular resolution \citep{Vallee2006, Matthews2009}. The field strength derived from the DCF method ranges from 0.4 to 1.7\,mG among the five cores, with mass-to-flux ratios from 1 to 4.3\,$\times$ the critical value.


As was the case in many low-mass star formation studies (see Section \ref{sec:collapse}), significant effort has been devoted to the search for pinched, hourglass-shaped magnetic field morphologies in massive dense cores. So far, the best examples of hourglass-shaped magnetic fields in massive dense cores are G31.41+0.31 and G240.31+0.07 \citep{Girart2009, Qiu2014}. 
G31.41+0.31 is a 500\,\msun\ hot molecular core with a luminosity of $3 \times 10^5$\,\lsun. Observed at resolutions as high as 2400\,au, the dense core does not appear to fragment in the dust continuum emission. However, sensitive observations with the VLA at 1.3 and 0.7\,cm wavelength revealed two compact continuum objects with a projected separation of 1300\,au \citep{Cesaroni2010}. The dense core exhibits infall motions as well as rotation over a scale of 14,000\,au. Magnetic fields inferred from the 870\,$\mu$m continuum emission reveal a distribution that is pinched along the major axis of the flattened core \citep{Girart2009}. No molecular outflows have been definitively identified in this region. Observations in the CO\,($J$\,$=$\,2\,$\rightarrow$\,1) transition found a velocity gradient along the major axis of the flattened core \citep{Cesaroni2011}; however, it is not clear if the gradient represents an outflow, or if it is due to core rotation. The strength of the plane-of-sky component of the magnetic field is 9.7\,mG, implying a turbulence-to-magnetic-energy ratio of 0.35. The rotational velocity within the core inferred from spectral-line observations of high-density tracers indicates significant magnetic braking. G31.41+0.31 is a case where the magnetic field dominates the turbulence and the dynamics in the system.

\begin{figure}[hbt!]
\includegraphics[width=1.0\textwidth, clip, trim=0cm 0cm 0cm 0cm]{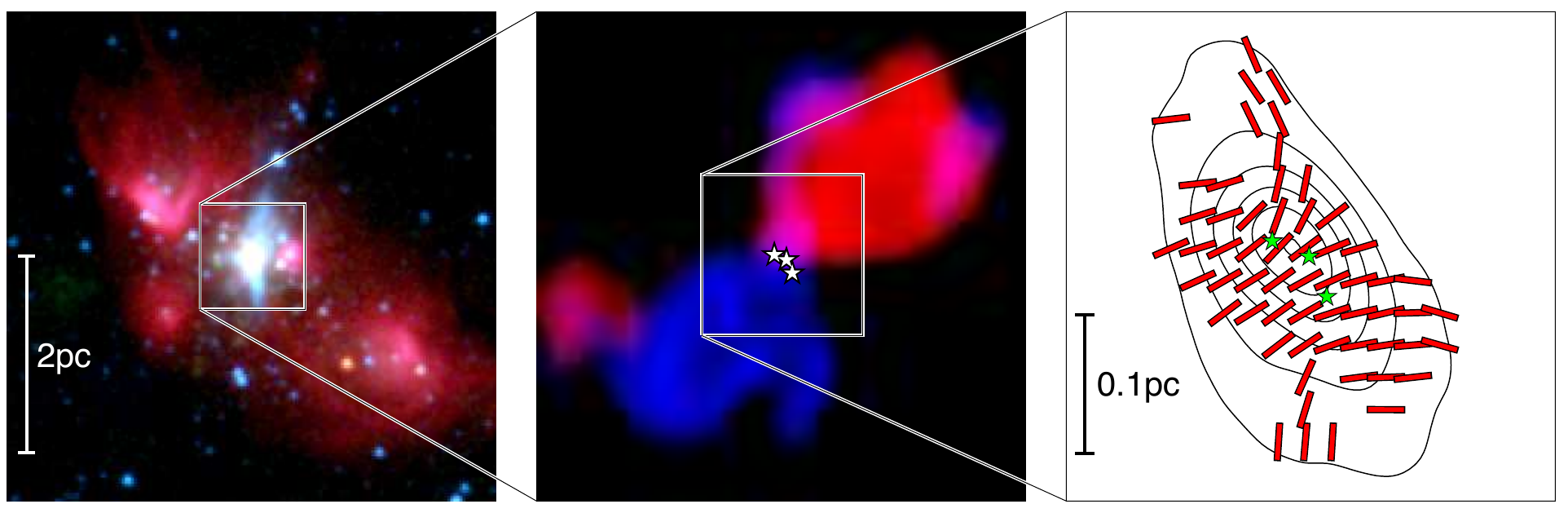}
\caption{
\small
Magnetic field (red line segments in the right panel) in the high-mass star formation region G240.31+0.07 inferred from dust continuum polarization at 870\,$\mu$m with the SMA \citep{Qiu2014, Zhang2014}. The left panel presents an overview of the region in a three-color composite of \textit{Spitzer} IRAC images in the 3.6\,$\mu$m (green), 4.5\,$\mu$m (blue) and 8.0\,$\mu$m (red) bands; the middle panel shows a wide-angle bipolar molecular outflow in the CO\,($J$\,$=$\,2\,$\rightarrow$\,1) line \citep{Qiu2009}. The star symbols indicate the known continuum peaks.  Reproduced with permission from the AAS.}
\label{fig:G240}
\end{figure}

Similar to G31.41+0.31, G240.31+0.07 is a massive star-forming region with an hourglass magnetic field morphology \citep{Qiu2014}. As shown in Figure \ref{fig:G240}, the dust continuum emission reveals a flattened structure extended along the northeast--southwest direction that has fragmented into three cores, each harboring at least one massive young star. A wide-angle, bipolar outflow is seen in the CO emission \citep{Qiu2009}, with the outflow axis parallel to the minor axis of the flattened dense core. Polarization observations at 870\,$\mu$m reveal a magnetic field topology pinched along the major axis of the core. The magnetic field strength estimated from the DCF method is 1.2\,mG, with a mass-to-flux ratio of 1.2\,$\times$ the critical value, and a turbulent-to-magnetic-energy ratio of 0.4. G240.31+0.07 is another clear example of a massive star-forming core in which the magnetic field dominates the turbulence and the dynamics in the system.

So far, the overwhelming majority of publications on interferometric observations of magnetic fields are studies of individual objects. The improvement in sensitivity with the advent of the CARMA and SMA polarimeters led to surveys of small---but significantly larger---samples of objects.  See Section \ref{sec:collapse} for a discussion of the TADPOL survey of predominantly low-mass sources by \citet{Hull2014}.  On the high-mass end, \citet{Zhang2014} reported polarization detections in 14 massive star-forming clumps from a survey of 21 star forming regions. This effort significantly increased the number of clumps with magnetic fields detections and consequently enabled statistical analyses of the ensemble behavior of magnetic fields in massive star-forming regions. To date, there are approximately 24 unique high-mass star forming clumps that have been observed in polarization with interferometers. They are Orion KL, NGC 2071, W3, W3(OH), DR 21(OH), DR 21 filament, G192, G30.79, NGC 6334 I/In/IV/V, W51 e and N, IRAS 18306, IRAS 18089, W43, NGC 7538, G5.89, NGC 2264C1, G34.4, G35.2N, G31.41+0.31, and G240.31+0.07
\citep{Rao1998,Lai2001,Lai2003,Cortes2006a,Cortes2006b,Cortes2008,Girart2009,Tang2009a,Tang2009b,Beuther2010,Tang2010,HRChen2012,Girart2013,Liu2013,Qiu2013,Frau2014,Hull2014,Qiu2014,Sridharan2014,Wright2014,Zhang2014,HBLi2015,Cortes2016,Houde2016,Ching2017,Juarez2017,Koch2018}
Magnetic fields toward these sources display diverse topologies.  In Sections \ref{sec:Bfield_outflow_high}, \ref{sec:fragmentation_high}, \ref{sec:virial_high}, and \ref{sec:stat} we discuss findings and interesting trends from the statistical analysis of this sample.  While the analysis represents a significant advance in the study of the role of the magnetic field in high-mass star formation, the limitations and biases in the sample used in the analysis cannot be overlooked. One of the most significant limitations is that non-detections are not included in the statistics. The advent of ALMA will increase the size of the sample drastically, enabling significantly more robust analysis within the next decade.

\subsection{Outflow--core magnetic field connection}
\label{sec:Bfield_outflow_high}

Linear polarization from CO rotational transitions (i.e., from the G-K effect) probes magnetic fields in molecular gas with densities from 10$^2$\,--\,10$^3$\,cm$^{-3}$. This can be an effective tool for tracing magnetic fields in protostellar outflows. In the context of high-mass star formation, \citet{Lai2003} reported the first detection of linear polarization in the CO\,($J$\,$=$\,2\,$\rightarrow$\,1) transition in DR 21(OH). The inferred magnetic field orientation is in the east--west direction, aligned with the major axis of the CO outflows. Subsequent polarimetric observations in the CO\,($J$\,$=$\,1\,$\rightarrow$\,0) transition by \citet{Cortes2005} found linear polarization perpendicular to that of the CO ($J$\,$=$\,2\,$\rightarrow$\,1) transition. While this disagreement may be reconciled by anisotropy in the CO optical depth and external radiation field (see Section \ref{sec:intro_tracers_POS:GK}), it highlights the difficulty of interpreting spectral-line polarization from the G-K effect. Finally, \citet{Beuther2010} reported detections of spectral-line polarization in the CO\,($J$\,$=$\,3\,$\rightarrow$\,2) transition in IRAS 18089-1732. Using the DCF method, they derived a magnetic field strength of 28\,$\mu$G. A similar field strength of 10\,$\mu$G is reported in DR 21(OH) by \citet{Cortes2005}.

Despite the early success in detecting spectral-line polarization primarily in DR 21(OH), very few protostellar sources have shown detectable spectral-line linear polarization since those early results. In the survey of 14 high-mass protostellar clumps by \citet{Zhang2014}, only DR 21(OH) had detectable polarization in the CO\,($J$\,$=$\,3\,$\rightarrow$\,2) transition \citep{Girart2013}. The limited sensitivity of the SMA may be a contributing factor to these non-detections. Furthermore, the CO emission is typically spatially extended, which presents an additional challenge when imaging linear polarization, since Stokes $Q$ and $U$ can be either positive or negative, and a lack of short-spacing information in sparsely sampled interferometric data can produce negative emission that may be confused with the polarization signal in Stokes $Q$ and $U$. Both factors are mitigated by ALMA, which provides significant improvements in sensitivity and imaging fidelity over BIMA, CARMA, and the SMA.

Probing magnetic fields in accretion disks around high-mass protostars is challenging in two regards. First, extremely high angular resolution observations are required to achieve the $\sim$\,100\,au linear scales of a disk at a typical source distance of several kpc. Second, at the typical high densities and optical depths in these disks, the polarized emission may be contaminated (and sometimes dominated) by self-scattering of large dust grains (see Section \ref{sec:scattering}). For example, \citet{Girart2018} observed polarized emission from the disk associated with the radio jet HH 80-81. These deep ALMA observations in the 1.14\,mm continuum emission spatially resolved a flattened disk with a radius of 291\,au. The disk is partially optically thick and its polarized emission appears to be dominated by self-scattering of large dust grains.

In the absence of magnetic field information in protostellar outflows and disks, the comparison of magnetic fields in cores with outflow axes offers an alternative to assess the importance of magnetic fields in the formation of disks and outflows in high-mass systems. As discussed in Section \ref{sec:collapse}, when collapsing to form stars, strongly magnetized cores are expected to form a disk and outflow system with the major axis of the outflow parallel to the magnetic field. \citet{Zhang2014} compared outflow axes with the plane-of-the-sky component of magnetic fields in cores from which the outflows originate. They found a slight preference around $0^\circ$ and $90^\circ$ in angles between the magnetic field and the outflow axis. However, due to the small sample size, the data are consistent with a random orientation of magnetic fields and outflows. This lack of correlation, if confirmed by more statistically significant samples, suggests that angular momentum in accretion disks is not dictated by the orientation of the magnetic field in the cores. The dynamical importance of the magnetic field thus appears to weaken relative to gravity and angular momentum from the core to the disk scales.

\subsection{Fragmentation and cluster formation}
\label{sec:fragmentation_high}

The clustering nature of massive stars implies that there must be considerable fragmentation in parsec-scale molecular clumps in order to produce a cluster of stars.  Gravity, turbulence, magnetic fields, and stellar feedback all influence fragmentation and the outcome of cluster formation.  The dynamical role of the magnetic field during the fragmentation of molecular clumps can be assessed if the magnetic field strength is measured directly.  As mentioned in Section \ref{sec:intro_tracers_LOS}, this can be achieved by observing circular polarization from the Zeeman effect.  While observations of the Zeeman effect have been carried in \hi and OH line emission using the VLA \citep{Crutcher2012, CrutcherKemball2019}, these observations probe a low-density medium that may not be directly involved in gravitational collapse. There are no reported interferometric observations of the Zeeman effect in dense molecular gas, although ALMA is likely to reshape this field once precise measurements of circular polarization become available to the user community.  Future interferometric observations of the Zeeman effect (measuring the line-of-sight magnetic field), when combined with linear dust and spectral-line polarization observations (measuring the plane-of-sky field), will allow much more robust estimates of the total magnetic field strengths in star-forming regions.

A powerful indirect method that can be used to assess the dynamical role of magnetic fields in protocluster formation is the analysis of the distribution of magnetic fields within cluster-forming molecular clumps.  Numerical simulations of turbulent, magnetized molecular clouds offer clues about magnetic field topologies in strong and weak field regimes. When the magnetic field is strong relative to turbulence, the field is less disturbed and appears to be ordered \citep[e.g.][]{Ostriker2001}.  

\begin{figure}[hbt!]
\begin{center}
\includegraphics[width=0.8\textwidth, clip, trim=3.5cm 2.5cm 2.75cm 3cm]{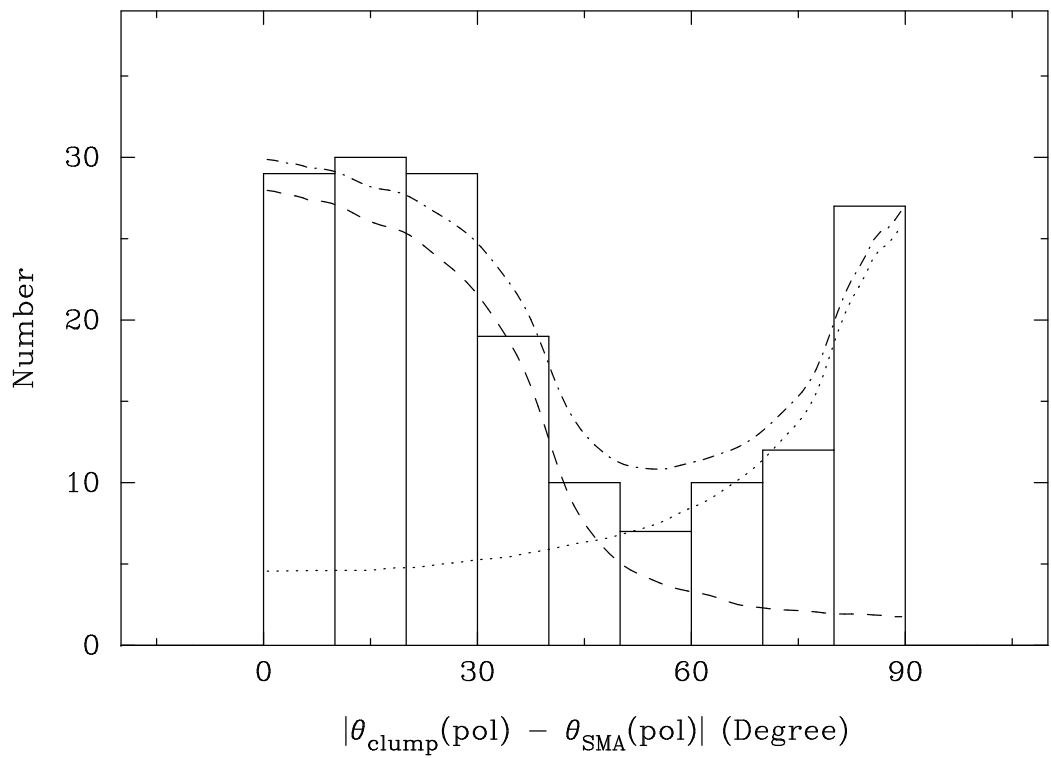}
\caption{
\small
The distribution of polarization angles in dense cores with respect to polarization angles of their parental clumps.  The dashed line represents the probability distribution of the plane-of-the-sky-projected orientations of pairs of vectors with $0^\circ - 40^\circ$ opening angles randomly oriented in space. The dotted line represents the probability distribution where the vectors are preferentially perpendicular, with $80^\circ - 90^\circ$ opening angles. The dashed-dotted line represents the contributions from the two distributions. Adapted from \citet{Zhang2014}.  Figures reproduced with permission from the AAS.}
\label{fig:Bclump_Bcore}
\end{center}
\end{figure}

Observationally, such a study becomes meaningful only when a statistically significant sample is involved.  \citet{Zhang2014} compared dust polarization in dense cores probed by the SMA with the polarization in the parental molecular clumps observed by single-dish telescopes in a sample of 14 high-mass star forming regions; the results show a bimodal distribution in polarization angles. As shown in Figure \ref{fig:Bclump_Bcore}, magnetic fields on dense-core scales are not randomly distributed, but are either parallel or perpendicular to the field orientations in their parental clumps.  A later study of a larger sample of 50 primarily high-mass sources by \citet{Koch2014} compared the magnetic field orientation with the gradient of the total dust emission  and came to the same conclusion.  These findings indicate that the magnetic fields are dynamically important in cluster-forming clumps, and that the field is strong enough on the clump scale to channel the material along the field lines into dense cores during the gravitational collapse.   While \citet{Hull2014} found hints of consistency in the magnetic field orientation from $\sim$\,0.1\,pc to $\sim$\,1000\,au scales in a few low-mass sources (see Section \ref{sec:collapse}), the results at the $\sim$\,1\,--\,0.1\,pc scales from \citet{Koch2014} and \citet{Zhang2014} are much more significant, suggesting that the magnetic field may be more dynamically important at parsec scales.

\subsection{Virial equilibrium in massive cores and cluster formation}
\label{sec:virial_high}

An essential component of the classical view of star formation is that dense cores should be in a state of an approximate virial equilibrium \citep{Larson1981, Shu1987, McKee2003}. However, there is increasing evidence from recent observations that cores forming high-mass protostars may not be in virial equilibrium. \citet{Pillai2011} carried out a stability analysis in two infrared dark clouds (IRDCs) and found that the virial parameter $\alpha_{\textrm{vir}}$, defined as the ratio between the virial mass $M_\textrm{Virial}$ and the gas mass $M_\textrm{gas}$, is typically between 0.1 to 0.3. These results are corroborated by \citet{Kauffmann2013}, who compiled a large sample of massive clumps from surveys and found that a significant fraction of them have virial parameters $\ll$\,2, where 2 is the value expected from a pressure confined, self-gravitating Bonnor-Ebert sphere.
More recently, \cite{Traficante2018} pointed out that the virial mass can be underestimated because the spectral line emission of a tracer molecule preferentially probes sections of a cloud with gas densities above the critical density of the line transition. This effect may result in an underestimate of the observed line width, and hence lead to a small virial parameter. However, since the effective excitation density of a line transition is typically lower than the critical density due to radiative trapping of photons when the optical depth is non-negligible, this effect may not be as significant. For example, many of the virial analyses in the literature use the NH$_3$ and N$_2$H$^+$ lines, which often exhibit consistent line widths despite having critical densities that differ by more than a factor of 5.

These small virial parameters appear to challenge the picture of star formation in which gas evolves in a state of equilibrium. However, the virial analyses discussed above do not include contributions from magnetic fields, which may provide significant support in dense cores.  \citet{Frau2014} carried out continuum polarization measurements of NGC 7538 with the SMA and found a magnetic field strength of 2.3--2.6\,mG in the region. \citeauthor{Frau2014} also performed a detailed energetics comparison of the gravitational potential energy, turbulent support, thermal pressure, and magnetic support.  Among the 13 dense cores analyzed, the magnetic support amounts to 0.2 to 2.4\,$\times$ the combined effect from turbulent and thermal pressure. More than half (eight out of 14) of the cores have magnetic support comparable to the turbulent and thermal support. In addition, the virial parameters including the magnetic support vary from 0.1 in the central region where star formation takes place to 8 in cores that are influenced significantly by molecular outflows.

\begin{figure}[hbt!]
\begin{center}
\includegraphics[width=0.8\textwidth, clip, trim=4.5cm 2.5cm 5cm 3.5cm]{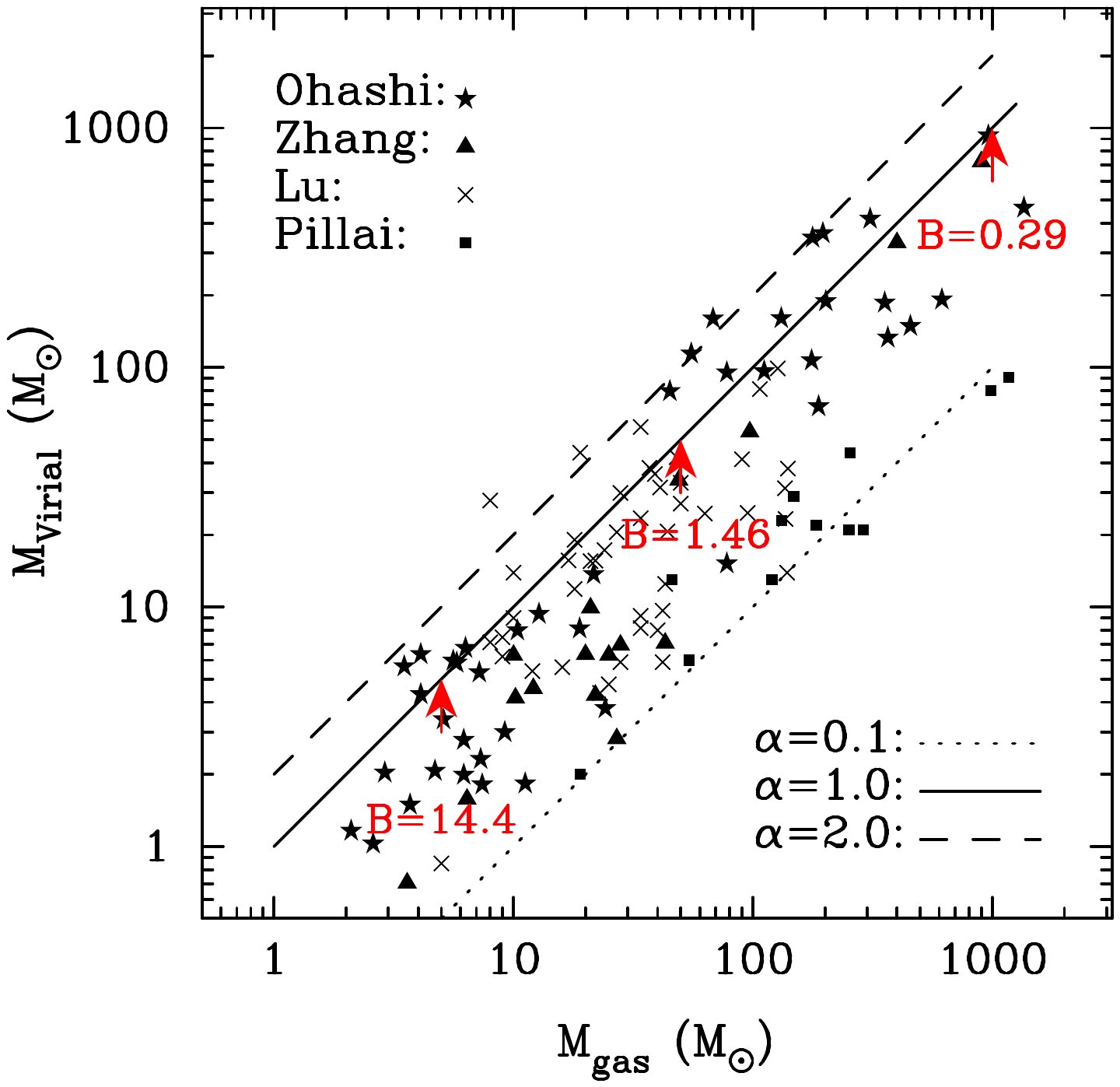}
\caption{
\small
Virial mass $M_\textrm{Virial}$ versus gas mass $M_\textrm{gas}$ for clumps and cores in 12 high-mass star forming regions. The data were taken from \citet{Pillai2011, Zhang2015,Ohashi2016}; and \citet{Lu2018}. The virial mass includes contributions from thermal and non-thermal line widths. The gas mass was derived from dust continuum emission. The dashed, solid, and dotted lines denote virial parameters $\alpha_{\textrm{vir}} = M_\textrm{Virial}/M_\textrm{gas}$ of 2, 1 and 0.1, respectively. The majority of the data points have $\alpha_{\textrm{vir}} < 1$, indicating that clumps and cores are far from virial equilibrium when contributions from magnetic fields are not included. The red arrows indicate the strength of the magnetic field (in mG) that are required to increase $\alpha_{\textrm{vir}}$ from 0.5 to 1.0 (see Section \ref{sec:virial_high}).}
\label{fig:virial}
\end{center}
\end{figure}

The studies of NGC 7538 by \citet{Frau2014} and \citet{Wright2014} demonstrate that magnetic fields can indeed be a significant contributor to the support of dense cores. Such an analysis should be extended to high-mass star-forming regions at earlier evolutionary stages when stellar feedback has not significantly altered the initial physical conditions.  To that end, \citet{Pillai2011, Zhang2015, Ohashi2016, Sanhueza2017}; and \citet{Lu2018} performed virial analyses in high-mass star forming regions prior to the development of strong \hii regions. The interferometric observations allowed the identification of structures $<$\,0.1\,pc in size within massive clumps. Figure \ref{fig:virial} presents a comparison of the virial mass and the gas mass for structures identified in 12 high-mass star forming regions \citep{Pillai2011, Zhang2015, Ohashi2016, Lu2018}. The virial mass is computed as

\begin{equation}
M_\textrm{Virial} = \frac{3 k R \sigma_v^2 }{G}\,\,,
\end{equation}

\noindent
where $\sigma_v$ is the line-of-sight velocity dispersion due to both non-thermal and thermal broadening, $R$ is the radius, and $G$ the gravitational constant. $k = \frac{5 - 2a}{3 - a}$ is a correction factor related to the density profile $\rho \propto r^{-a}$. We assume a constant density in the structure, i.e., $a = 0$, which leads to the maximum value in the correction factor ($k = 5/3$) and thus the maximum virial mass. The gas mass is derived from the dust continuum emission (see Section \ref{sec:core_mass}). Figure \ref{fig:virial} reveals that the dense gas structures in these high-mass star forming regions have virial parameters $<$\,2. Furthermore, a large fraction of the dense gas structures have virial parameters $\ll$\,2.  Since there are no direct measurements of magnetic fields for this sample, we compute the magnetic field required to increase the virial parameter from 0.5 to 1.0. The magnetic virial mass \citep{Hennebelle2008c} is computed as

\begin{equation}
M_\textrm{Mag} = \frac{5 R \sigma_A^2 }{6 G}\,\,,
\end{equation}

\noindent
where $\sigma_A = \frac{B }{\sqrt{4 \pi \rho}}$ is the Alfv\'enic velocity corresponding to a magnetic field strength $B$ and density $\rho$. We compute the magnetic field strengths when $M_\textrm{Mag} = M_\textrm{Virial}  = 0.5M_\textrm{gas}.$  Assuming a representative gas mass $M_\textrm{gas}$ and radius $R$ for the clumps ($M_\textrm{gas} = 1000$\,\msun, $R = 0.5$\,pc), cores ($M_\textrm{gas} = 50$\,\msun, $R = 0.05$\,pc), and condensations ($M_\textrm{gas} = 5$\,\msun, $R = 0.005$\,pc), we find required magnetic field strengths of 0.29\,mG, 1.46\,mG and 14.4\,mG, respectively. Under these conditions, the virial parameters would be 0.5 without the contribution of magnetic fields, and would increase to 1 after the inclusion of magnetic fields. The required field strengths are in broad agreement with typical literature values of  magnetic fields that were derived using the DCF method based on polarization observations of massive star-forming regions.

Recent measurements of the thermodynamic properties in high-mass star-forming regions reveal small virial parameters that appear to challenge the assumption of equilibrium star formation.  However, a lack of magnetic field measurements in these same regions leaves open the possibility of virialized star formation, since the field strength of a fraction of a mG to several mG could bring the dense gas close to a state of equilibrium. With the advent of ALMA, we expect significant progress to be made on this vital question in high-mass star formation as more observations of both spectral lines and polarization are carried out over the coming years.

\section{Constraining strong-field star formation with statistics of hourglass-shaped magnetic fields}
\label{sec:stat}

There are approximately 32 unique interferometric observations of low-mass (Class 0, 0/I, or I) star-forming cores exhibiting significant polarization detections that are not obviously dominated by dust scattering.  They are
L1448 IRS 2, 
L1448N(B), 
L1448C, 
NGC 1333-IRAS 2A, 
SVS 13A, 
SVS 13B, 
NGC 1333-IRAS 4A, 
Per-emb-21, 
NGC 1333-IRAS 7, 
NGC 1333-IRAS 4B, 
IRAS 03282+3035, 
IRAS 03292+3039, 
B1-c, 
B1-b, 
HH 211 mm, 
HH797, 
L1551 NE, 
NGC 2024 FIR 5, 
OMC3-MMS5, 
OMC3-MMS6, 
OMC2-FIR3, 
OMC2-FIR4, 
VLA 1623, 
IRAS 16293A, 
IRAS 16293B, 
Ser-emb 8, 
Ser-emb 8 (N), 
Serpens SMM1-a, 
Serpens SMM1-b, 
B335, 
L1157, 
and CB 230 
\citep{Girart1999, Lai2002, Girart2006, Rao2009, Alves2011, Stephens2013, Hull2013, Hull2014, Hull2017a, Hull2017b, SeguraCox2015, Cox2018, Galametz2018, Harris2018, Kwon2018, Maury2018, Sadavoy2018a, Sadavoy2018c}.

Among them, 
NGC~1333-IRAS~4A, 
IRAS~16293A, 
L1157, 
NGC 2024 FIR 5,
OMC3 MMS 6, 
L1448 IRS 2, 
B335, 
VLA 1623, 
and B1-c 
(9/32 $\approx$ 28\%)
exhibit hourglass morphologies, consistent with magnetically dominant star formation
\citep{Girart1999, Lai2002, Girart2006, Rao2009, Alves2011, Stephens2013, Hull2014, Cox2018, Galametz2018, Kwon2018, Maury2018, Sadavoy2018a}.  
 
On the high-mass side, there are 24 unique high-mass star forming clumps observed in polarization with interferometers. They are 
Orion KL, 
NGC 2071, 
W3, 
W3(OH), 
DR 21(OH), 
DR 21 filament, 
G192, 
G30.79, 
NGC 6334 I/In/IV/V, 
W51 e and N, 
IRAS 18306, 
IRAS 18089, 
W43, 
NGC 7538, 
G5.89, 
NGC 2264C1, 
G34.43+0.24, 
G35.2N, 
G31.41+0.31, 
and G240.31+0.07 
\citep{Rao1998,Lai2001,Lai2003,Cortes2006a,Cortes2006b,Cortes2008,Girart2009,Tang2009a,Tang2009b,Beuther2010,Tang2010,HRChen2012,Girart2013,Liu2013,Qiu2013,Frau2014,Hull2014,Qiu2014,Sridharan2014,Wright2014,Zhang2014,HBLi2015,Cortes2016,Houde2016,Ching2017,Juarez2017,Koch2018}.

Of these, 
G31.41+0.31, 
G240.31+0.07, 
NGC 6334 I and In, 
and W51 e2 
(5/24 $\approx$ 21\%)
exhibit hourglass magnetic field configurations
\citep{Girart2009,Tang2009a,Qiu2014,HBLi2015}.

A pinched/hourglass-shaped magnetic field configuration can yield a range of magnetic-field morphologies when projected in the plane of the sky \citep{Frau2011}. When the main axis of the system is aligned with the line of sight, the plane-of-sky component of the field is oriented radially outward. This scenario was proposed by \citet{Girart2013} to explain the magnetic field topology in DR 21(OH). Among the low- and high-mass objects with robust detections of polarization (and thus inferred magnetic fields), 28\% of the low-mass sources and 21\% of the high-mass sources exhibit hourglass configurations.  Assuming that the sample is randomly oriented in three dimensions, 
the probability than an hourglass axis will be oriented within 30$^\circ$ of the line of sight is $\sim$\,13\% (\citealt{Frau2011} show that an angle closer to the line of sight than $\sim$\,30$^\circ$ yields a mostly radial pattern).  Therefore, the majority of the objects ($\sim$\,87\%) should display an hourglass shape if the variations in magnetic field morphologies were solely due to projection effects.  


The fact that only a small fraction of the objects exhibit hourglass morphologies suggests that the scenario of magnetically dominant core collapse is not the predominant mode of low- or high-mass star formation.  However, we should note that when taking into account the projection effect, the fraction of the detected hourglass reaches 28\%/87\% $\approx$\,1/3 for low-mass and 21\%/87\% $\approx$\,1/4 for high-mass sources. Such fractions are not negligible, and are even more significant considering the various environmental and dynamical differences between sources, as well as the observational limitations that may hinder the detection of an hourglass---in particular the sensitivity and dynamic range limitations of the pre-ALMA interferometers.

Of the sources observed by ALMA that clearly lack hourglass magnetic field morphologies, only one so far (Ser-emb 8 in \citealt{Hull2017a}; see Figure \ref{fig:ser8_multiscale}) shows a truly chaotic morphology, presumably having been influenced by turbulence and infall.  However, the few ALMA polarization observations published to date have shown more examples of magnetic fields that seem to have been shaped by bipolar outflows.  These observations are challenging to interpret, as magnetic fields aligned with the walls of an outflow cavity can look quite similar to an hourglass when observed with low resolution.  Thus, future studies will need to carefully consider how to determine which ``hourglasses'' are preserved from the natal collapse phase, such as in NGC 1333-IRAS 4A, where at $\sim$\,1000\,au scales the hourglass appears to be unperturbed by the two outflows emanating at different angles from the embedded binary \citep{Girart2006}; versus which are caused by outflow/jet feedback, such as in Serpens SMM1 \citep{Hull2017b} and B335 \citep{Maury2018}.

We further caution that the above values are lower limits that are limited by telescope resolution: i.e., in the high-mass case, we consider each source as unique, despite the fact that each clump is likely to have more than one protostellar object.  This is even true in the low-mass case: for example, SMA observations were able to distinguish the magnetic fields in SVS 13A and B, whereas CARMA observations were not; and ALMA observations were able to map the magnetic fields in Serpens SMM1-a and b separately, whereas CARMA and SMA observations were not.  Finally, and perhaps most important, the number of objects observed with interferometers remains small: observations by ALMA are likely to increase the sample size significantly in the near future, enabling studies that will establish far better statistical constraints.

\section{Polarization from dust scattering}
\label{sec:scattering}

While polarized emission from dust scattering is independent of the magnetic field, we briefly mention it here because this young, quickly growing subfield emerged from studies that were attempting to make resolved maps of the magnetic field in protoplanetary disks, which is one of the longstanding goals of disk- and star-formation studies.  \citet{Rao2014} reported a tentative detection of a toroidally wrapped magnetic field in the Class 0 protostar IRAS~16293B at $\sim$\,75\,au resolution using the SMA. However, on the whole, pre-ALMA full polarization observations of disks did not have the sensitivity or the resolution necessary to make well resolved maps of low-level ($\lesssim$\,1\%) polarized dust emission \citep{Hughes2009b,Hughes2013}.  

Around the same time that ALMA started producing polarization results, several theoretical studies suggested that polarized (sub)millimeter-wave emission from protoplanetary disks could be caused by the self-scattering of emission from (sub)millimeter-sized dust grains \citep{Kataoka2015, Kataoka2016, Pohl2016, Yang2016a, Yang2016b}, consistent with previous work by \citet{Cho2007}.  \citet{Kataoka2016} and \citet{Yang2016b} explained the 1.3\,mm CARMA polarization observations by \citet{Stephens2014} of the Class I/II source HL~Tau in the context of this self-scattering mechanism.  Self-scattering can also explain the polarization pattern observed in the Class II protoplanetary disk IM Lup in 870\,$\mu$m ALMA observations by \citet{Hull2018a}, and similar observations of HD 163296 by \citet{Dent2019}.  

The situation became even more complex when \citet{Tazaki2017} proposed yet another mechanism that can cause polarization in disks: dust grains whose minor axes are aligned with the gradient in the dust emission (this work is rooted in the radiative torque model by \citealt{Lazarian2007}).  This mechanism, which is sometimes referred to as ``$k$-RATs'' (i.e., where $k$ is the orientation of the gradient in the dust emission; see also Footnote \ref{footnote:RATs}), is different from self-scattering by dust grains. However, similar to self-scattering, it is independent of any magnetic field that might be in the disk.  Polarization from $k$-RATs, which has a morphology that is azimuthal, is broadly consistent with 3\,mm ALMA observations of polarization in the HL~Tau disk \citep{Kataoka2017}, although \citet{Yang2019} found that their $k$-RAT model did not reproduce the data when examined in detail.  The transition from possible $k$-RAT alignment at long (3\,mm) wavelengths to scattering at short (870\,$\mu$m) wavelengths was shown clearly by \citet{Stephens2017b}, who reported high-resolution ALMA polarization results at 870\,$\mu$m and 1.3\,mm, complementing the 3\,mm observations reported by \citet{Kataoka2017}.  The intermediate-wavelength 1.3\,mm data exhibit a morphology consistent with roughly equal contributions of self-scattering and $k$-RATs.

Other recent results have interpretations that are not as clear.  These include observations of the edge-on HH 111 and HH 212 disks \citep{CFLee2018a}, the high-mass protostellar disk Cepheus A HW2 \citep{FernandezLopez2016}, and the transition disk HD 142527 \citep{Kataoka2016b, SOhashi2018}; the polarized emission from these objects may be caused by either scattering or magnetically aligned grains ($B$-RATs).  Observations of the low-mass disks CW Tau and DG Tau \citep{Bacciotti2018} and the massive GGD27 MM1 disk \citep{Girart2018} associated with the iconic HH 80-81 radio jet show polarization that may be caused by dust self-scattering and/or $k$-RATs.  Polarization observations of a number of embedded Class 0 protostars by \citet{SeguraCox2015, Cox2018, Harris2018, Sadavoy2018a, Sadavoy2018c} and \citet{Takahashi2018} sometimes show signs of scattering in the inner regions of sources and magnetic alignment in the outer regions.  Finally, \citet{Alves2018} reported $\sim$\,30\,au-resolution observations of the Class I binary source BHB07-11, which exhibits consistent polarization patterns at 3\,mm, 1.3\,mm, and 850\,$\mu$m. They argue that the polarized emission most likely arises from magnetically aligned dust grains.  

We emphasize that studies of (sub)millimeter polarization from disks are in their infancy. More observations and more mature, detailed models are necessary before we will be able to determine whether there is, in fact, any contribution from magnetic fields in the polarized emission from protostellar disks.

\section{Summary}

The steady improvement in telescope sensitivity at the (sub)millimeter-wave bands where dust continuum emission is strong has led to a revolution in interferometric observations of linear polarization over the past two decades. In recent years, both CARMA and the SMA conducted surveys of magnetic fields in samples of $>$\,10 star-forming cores/clumps. These studies have proven insightful in revealing the statistical behavior of magnetic fields in star formation. However, while individual case studies continue to be fruitful, large, less biased surveys that target statistically significant samples of sources are the only way to further constrain the dynamical role of magnetic fields in the star formation process. 

Studies of low-mass star formation have found that while there are a few cases where the magnetic field seems to regulate the collapse of star-forming material across the scales from clouds/clumps to cores to envelopes, there are also many cases where the multi-scale magnetic field morphology shows no consistency. Furthermore, the resolution and sensitivity of ALMA have yielded new observations that show evidence of shaping of the magnetic field by bipolar outflows at the $\sim$\,100\,au scale in some (but not all) sources.  A few recent observations of spectral-line polarization by the SMA and ALMA have shed light on the importance of magnetic fields in the launching of outflows.  More observations of this type with ALMA will enhance this subfield significantly in the coming years.

Studies of high-mass star formation found that magnetic fields are not the dominant force within the parsec scale of molecular clumps. However, fields appear to be dynamically important in the fragmentation of clumps into 0.1\,pc-scale cores. Furthermore, molecular gas with densities $>$ $10^4$\,cm$^{-3}$ in high-mass star formation appears to be far from virial equilibrium if magnetic fields do not contribute significantly to the dynamical process. This significant departure from the state of virialization appears to challenge the basic assumption of equilibrium star formation. Direct measurements of magnetic fields will shed more insight on this important issue. 

In both the low- and high-mass regimes, comparisons of outflow versus magnetic field orientations have yielded random alignment.  The outflow is a probe of the angular momentum at the smallest scales in the source, and thus the limited data currently available point to a scenario where angular momentum is far more important than the magnetic field at the small ($<$\,1000\,au) scales of individual forming protostars.  

The role of magnetic fields in the formation of high- and low-mass disks is less clear due to a small number of observations, and the issue of scattering by dust. Near-future studies targeting the Zeeman and G-K effects may finally be able to access information about the magnetic field in disks \citep[e.g.,][]{Brauer2017}.  Furthermore, high resolution studies at longer wavelengths in regions that are optically thick (and thus dominated by scattering) in the ALMA data will be made possible by future telescopes such as the Next Generation Very Large Array (ngVLA; \citealt{Carilli2015}).  The ngVLA will be a powerful tool for studying magnetized star formation \citep{Isella2015, Hull2018b}, as it will offer dramatic improvements in the sensitivity, resolution, and image fidelity over the current VLA, which has enabled just a few studies of polarization in the very brightest low- and high-mass star-forming sources \citep{CarrascoGonzalez2010, Cox2015, Liu2016, Liu2018}.

Despite major advances in recent years, the studies presented in this review---in particular the survey studies---are biased toward objects with detectable polarization. Nearly all low- and high-mass star forming objects have embedded protostars (and sometimes \hii regions, in the case of high-mass sources). A crucial, under-explored area of star formation involves studying the initial conditions of the magnetic field before feedback (e.g., infall, outflows, and radiation) has altered its morphology. However, this has rarely been achieved due to a lack of sensitivity. In the coming years, large, multi-scale, multi-telescope surveys of magnetic fields in star-forming clouds covering the full range of evolutionary stages will allow us to break new ground in our studies of magnetized star formation.

\section*{Author Contributions}

CH led the writing of the sections about low-mass star formation and scattering.  QZ led the section about high-mass star formation.  Both authors wrote the abstract, introduction, hourglass-statistics section, and summary.

\section*{Funding}

CH acknowledges the support of both the NAOJ Fellowship as well as JSPS KAKENHI grant 18K13586.  QZ acknowledges the support of the the Smithsonian Institute SSA grant, ``Are Magnetic Fields Dynamically Important in Massive Star Formation?''

\section*{Acknowledgments}
CH and QZ acknowledge the two anonymous referees for their careful reading of the manuscript and for the insightful comments, which led to a more clear and thorough presentation of this review.
CH acknowledges Paulo Cort\'es, Martin Houde, and Josep Miquel Girart for the useful discussion.
The BIMA array was operated by the Berkeley-Illinois-Maryland Association with support from the National Science Foundation.
Support for CARMA construction was derived from the states of California, Illinois, and Maryland, the James S. McDonnell Foundation, the Gordon and Betty Moore Foundation, the Kenneth T. and Eileen L. Norris Foundation, the University of Chicago, the Associates of the California Institute of Technology, and the National Science Foundation.
The Submillimeter Array is a joint project between the Smithsonian Astrophysical Observatory and the Academia Sinica Institute of Astronomy and Astrophysics, and is funded by the Smithsonian Institution and the Academia Sinica.
ALMA is a partnership of ESO (representing
   its member states), NSF (USA) and NINS (Japan), together with NRC
   (Canada), NSC and ASIAA (Taiwan), and KASI (Republic of Korea),
   in cooperation with the Republic of Chile. The Joint ALMA
   Observatory is operated by ESO, AUI/NRAO and NAOJ.
The National Radio Astronomy Observatory is a facility of the National Science Foundation operated under cooperative agreement by Associated Universities, Inc. 

\textit{Facilities:}  ALMA, APEX, BIMA, BLAST, BLAST-TNG, CARMA, CSO, IRAM 30\,m, JCMT, LMT, NRAO 12\,m, SMA, SOFIA, Viper, VLA.  

\textit{Software:} APLpy, an open-source plotting package for Python hosted at \href{http://aplpy.github.com}{\texttt{aplpy.github.com}} \citep{Robitaille2012}.  Common Astronomy Software Applications (CASA, \citet{McMullin2007}).  Astropy \citep{Astropy2018}.

\section*{Conflict of Interest Statement}

The authors declare that the research was conducted in the absence of any commercial or financial relationships that could be construed as a potential conflict of interest.


\begin{thebibliography}{260}
\providecommand{\natexlab}[1]{#1}
\expandafter\ifx\csname urlstyle\endcsname\relax
  \providecommand{\doi}[1]{doi:\discretionary{}{}{}#1}\else
  \providecommand{\doi}{doi:\discretionary{}{}{}\begingroup
  \urlstyle{rm}\Url}\fi
\providecommand{\selectlanguage}[1]{\relax}
\providecommand{\bibAnnoteFile}[1]{%
  \IfFileExists{#1}{\begin{quotation}\noindent\textsc{Key:} #1\\
  \textsc{Annotation:}\ \input{#1}\end{quotation}}{}}
\providecommand{\bibAnnote}[2]{%
  \begin{quotation}\noindent\textsc{Key:} #1\\
  \textsc{Annotation:}\ #2\end{quotation}}

\bibitem[{{Akeson} and {Carlstrom}(1997)}]{Akeson1997}
{Akeson}, R.~L. and {Carlstrom}, J.~E. (1997).
\newblock {Magnetic Field Structure in Protostellar Envelopes}.
\newblock \emph{\apj} 491, 254--266.
\newblock \doi{10.1086/304949}
\input{Akeson1997}

\bibitem[{{Akeson} et~al.(1996){Akeson}, {Carlstrom}, {Phillips}, and
  {Woody}}]{Akeson1996}
{Akeson}, R.~L., {Carlstrom}, J.~E., {Phillips}, J.~A., and {Woody}, D.~P.
  (1996).
\newblock {Millimeter Interferometric Polarization Imaging of the Young Stellar
  Object NGC 1333/IRAS 4A}.
\newblock \emph{\apjl} 456, L45.
\newblock \doi{10.1086/309856}
\input{Akeson1996}

\bibitem[{{Allen} et~al.(2003){Allen}, {Li}, and {Shu}}]{Allen2003}
{Allen}, A., {Li}, Z.-Y., and {Shu}, F.~H. (2003).
\newblock {Collapse of Magnetized Singular Isothermal Toroids. II. Rotation and
  Magnetic Braking}.
\newblock \emph{\apj} 599, 363--379.
\newblock \doi{10.1086/379243}
\input{Allen2003}

\bibitem[{{Alves} et~al.(2014){Alves}, {Frau}, {Girart}, {Franco}, {Santos},
  and {Wiesemeyer}}]{Alves2014}
{Alves}, F.~O., {Frau}, P., {Girart}, J.~M., {Franco}, G.~A.~P., {Santos},
  F.~P., and {Wiesemeyer}, H. (2014).
\newblock {On the radiation driven alignment of dust grains: Detection of the
  polarization hole in a starless core}.
\newblock \emph{\aap} 569, L1.
\newblock \doi{10.1051/0004-6361/201424678}
\input{Alves2014}

\bibitem[{{Alves} et~al.(2011){Alves}, {Girart}, {Lai}, {Rao}, and
  {Zhang}}]{Alves2011}
{Alves}, F.~O., {Girart}, J.~M., {Lai}, S.-P., {Rao}, R., and {Zhang}, Q.
  (2011).
\newblock {The Magnetic Field in the NGC 2024 FIR 5 Dense Core}.
\newblock \emph{\apj} 726, 63.
\newblock \doi{10.1088/0004-637X/726/2/63}
\input{Alves2011}

\bibitem[{{Alves} et~al.(2018){Alves}, {Girart}, {Padovani}, {Galli}, {Franco},
  {Caselli} et~al.}]{Alves2018}
{Alves}, F.~O., {Girart}, J.~M., {Padovani}, M., {Galli}, D., {Franco},
  G.~A.~P., {Caselli}, P., et~al. (2018).
\newblock {Magnetic field in a young circumbinary disk}.
\newblock \emph{\aap} 616, A56.
\newblock \doi{10.1051/0004-6361/201832935}
\input{Alves2018}

\bibitem[{{Andersson} et~al.(2015){Andersson}, {Lazarian}, and
  {Vaillancourt}}]{Andersson2015}
{Andersson}, B.-G., {Lazarian}, A., and {Vaillancourt}, J.~E. (2015).
\newblock {Interstellar Dust Grain Alignment}.
\newblock \emph{\araa} 53, 501--539.
\newblock \doi{10.1146/annurev-astro-082214-122414}
\input{Andersson2015}

\bibitem[{{Astropy Collaboration} et~al.(2018){Astropy Collaboration},
  {Price-Whelan}, {Sip{\H o}cz}, {G{\"u}nther}, {Lim}, {Crawford}
  et~al.}]{Astropy2018}
{Astropy Collaboration}, {Price-Whelan}, A.~M., {Sip{\H o}cz}, B.~M.,
  {G{\"u}nther}, H.~M., {Lim}, P.~L., {Crawford}, S.~M., et~al. (2018).
\newblock {The Astropy Project: Building an Open-science Project and Status of
  the v2.0 Core Package}.
\newblock \emph{\aj} 156, 123.
\newblock \doi{10.3847/1538-3881/aabc4f}
\input{Astropy2018}

\bibitem[{{Bacciotti} et~al.(2018){Bacciotti}, {Girart}, {Padovani}, {Podio},
  {Paladino}, {Testi} et~al.}]{Bacciotti2018}
{Bacciotti}, F., {Girart}, J.~M., {Padovani}, M., {Podio}, L., {Paladino}, R.,
  {Testi}, L., et~al. (2018).
\newblock {ALMA Observations of Polarized Emission toward the CW Tau and DG Tau
  Protoplanetary Disks: Constraints on Dust Grain Growth and Settling}.
\newblock \emph{\apjl} 865, L12.
\newblock \doi{10.3847/2041-8213/aadf87}
\input{Bacciotti2018}

\bibitem[{{Beltr{\'a}n}(2015)}]{Beltran2015}
{Beltr{\'a}n}, M.~T. (2015).
\newblock {Observational perspective of the youngest phases of
  intermediate-mass stars}.
\newblock \emph{\apss} 355, 283--290.
\newblock \doi{10.1007/s10509-014-2151-0}
\input{Beltran2015}

\bibitem[{{Beuther} et~al.(2010){Beuther}, {Vlemmings}, {Rao}, and {van der
  Tak}}]{Beuther2010}
{Beuther}, H., {Vlemmings}, W.~H.~T., {Rao}, R., and {van der Tak}, F.~F.~S.
  (2010).
\newblock {Magnetic Field Structure in a High-mass Outflow/Disk System}.
\newblock \emph{\apjl} 724, L113--L117.
\newblock \doi{10.1088/2041-8205/724/1/L113}
\input{Beuther2010}

\bibitem[{{Bock} et~al.(2006){Bock}, {Bolatto}, {Hawkins}, {Kemball}, {Lamb},
  {Plambeck} et~al.}]{Bock2006}
{Bock}, D., {Bolatto}, A.~D., {Hawkins}, D.~W., {Kemball}, A.~J., {Lamb},
  J.~W., {Plambeck}, R.~L., et~al. (2006).
\newblock {First results from CARMA: the combined array for research in
  millimeter-wave astronomy}.
\newblock In \emph{Society of Photo-Optical Instrumentation Engineers (SPIE)
  Conference Series}. vol. 6267, 13.
\newblock \doi{10.1117/12.674051}
\input{Bock2006}

\bibitem[{{Brauer} et~al.(2017){Brauer}, {Wolf}, and {Flock}}]{Brauer2017}
{Brauer}, R., {Wolf}, S., and {Flock}, M. (2017).
\newblock {Magnetic fields in circumstellar disks. The potential of Zeeman
  observations}.
\newblock \emph{\aap} 607, A104.
\newblock \doi{10.1051/0004-6361/201731140}
\input{Brauer2017}

\bibitem[{{Carilli} et~al.(2015){Carilli}, {McKinnon}, {Ott}, {Beasley},
  {Isella}, {Murphy} et~al.}]{Carilli2015}
{Carilli}, C.~L., {McKinnon}, M., {Ott}, J., {Beasley}, A., {Isella}, A.,
  {Murphy}, E., et~al. (2015).
\newblock {Next Generation Very Large Array Memo No. 5: Science Working Groups
  -- Project Overview}.
\newblock \emph{ArXiv e-prints}
\input{Carilli2015}

\bibitem[{{Carrasco-Gonz{\'a}lez} et~al.(2010){Carrasco-Gonz{\'a}lez},
  {Rodr{\'{\i}}guez}, {Anglada}, {Mart{\'{\i}}}, {Torrelles}, and
  {Osorio}}]{CarrascoGonzalez2010}
{Carrasco-Gonz{\'a}lez}, C., {Rodr{\'{\i}}guez}, L.~F., {Anglada}, G.,
  {Mart{\'{\i}}}, J., {Torrelles}, J.~M., and {Osorio}, M. (2010).
\newblock {A Magnetized Jet from a Massive Protostar}.
\newblock \emph{Science} 330, 1209.
\newblock \doi{10.1126/science.1195589}
\input{CarrascoGonzalez2010}

\bibitem[{{Caswell} et~al.(2014){Caswell}, {Green}, and
  {Phillips}}]{Caswell2014}
{Caswell}, J.~L., {Green}, J.~A., and {Phillips}, C.~J. (2014).
\newblock {Parkes full polarization spectra of OH masers - II. Galactic
  longitudes 240{$\deg$} to 350{$\deg$}}.
\newblock \emph{\mnras} 439, 1680--1739.
\newblock \doi{10.1093/mnras/stu046}
\input{Caswell2014}

\bibitem[{{Cesaroni} et~al.(2011){Cesaroni}, {Beltr{\'a}n}, {Zhang}, {Beuther},
  and {Fallscheer}}]{Cesaroni2011}
{Cesaroni}, R., {Beltr{\'a}n}, M.~T., {Zhang}, Q., {Beuther}, H., and
  {Fallscheer}, C. (2011).
\newblock {Dissecting a hot molecular core: the case of G31.41+0.31}.
\newblock \emph{\aap} 533, A73.
\newblock \doi{10.1051/0004-6361/201117206}
\input{Cesaroni2011}

\bibitem[{{Cesaroni} et~al.(2010){Cesaroni}, {Hofner}, {Araya}, and
  {Kurtz}}]{Cesaroni2010}
{Cesaroni}, R., {Hofner}, P., {Araya}, E., and {Kurtz}, S. (2010).
\newblock {The structure of hot molecular cores over 1000 AU}.
\newblock \emph{\aap} 509, A50.
\newblock \doi{10.1051/0004-6361/200912877}
\input{Cesaroni2010}

\bibitem[{{Chamma} et~al.(2018){Chamma}, {Houde}, {Girart}, and
  {Rao}}]{Chamma2018}
{Chamma}, M.~A., {Houde}, M., {Girart}, J.~M., and {Rao}, R. (2018).
\newblock {Non-Zeeman circular polarization of molecular spectral lines in the
  ISM}.
\newblock \emph{\mnras} 480, 3123--3131.
\newblock \doi{10.1093/mnras/sty2068}
\input{Chamma2018}

\bibitem[{{Chandrasekhar} and {Fermi}(1953)}]{Chandrasekhar1953}
{Chandrasekhar}, S. and {Fermi}, E. (1953).
\newblock {Magnetic Fields in Spiral Arms.}
\newblock \emph{\apj} 118, 113.
\newblock \doi{10.1086/145731}
\input{Chandrasekhar1953}

\bibitem[{{Chapman} et~al.(2013){Chapman}, {Davidson}, {Goldsmith}, {Houde},
  {Kwon}, {Li} et~al.}]{Chapman2013}
{Chapman}, N.~L., {Davidson}, J.~A., {Goldsmith}, P.~F., {Houde}, M., {Kwon},
  W., {Li}, Z.-Y., et~al. (2013).
\newblock {Alignment between Flattened Protostellar Infall Envelopes and
  Ambient Magnetic Fields}.
\newblock \emph{\apj} 770, 151.
\newblock \doi{10.1088/0004-637X/770/2/151}
\input{Chapman2013}

\bibitem[{{Chen} and {Ostriker}(2018)}]{CYChen2018}
{Chen}, C.-Y. and {Ostriker}, E.~C. (2018).
\newblock {Geometry, Kinematics, and Magnetization of Simulated Prestellar
  Cores}.
\newblock \emph{\apj} 865, 34.
\newblock \doi{10.3847/1538-4357/aad905}
\input{CYChen2018}

\bibitem[{{Chen} et~al.(2012){Chen}, {Rao}, {Wilner}, and {Liu}}]{HRChen2012}
{Chen}, H.-R., {Rao}, R., {Wilner}, D.~J., and {Liu}, S.-Y. (2012).
\newblock {The Magnetized Environment of the W3(H$_{2}$O) Protostars}.
\newblock \emph{\apjl} 751, L13.
\newblock \doi{10.1088/2041-8205/751/1/L13}
\input{HRChen2012}

\bibitem[{{Ching} et~al.(2017){Ching}, {Lai}, {Zhang}, {Girart}, {Qiu}, and
  {Liu}}]{Ching2017}
{Ching}, T.-C., {Lai}, S.-P., {Zhang}, Q., {Girart}, J.~M., {Qiu}, K., and
  {Liu}, H.~B. (2017).
\newblock {Magnetic Fields in the Massive Dense Cores of the DR21 Filament:
  Weakly Magnetized Cores in a Strongly Magnetized Filament}.
\newblock \emph{\apj} 838, 121.
\newblock \doi{10.3847/1538-4357/aa65cc}
\input{Ching2017}

\bibitem[{{Ching} et~al.(2018){Ching}, {Lai}, {Zhang}, {Girart}, {Qiu}, and
  {Liu}}]{Ching2018}
{Ching}, T.-C., {Lai}, S.-P., {Zhang}, Q., {Girart}, J.~M., {Qiu}, K., and
  {Liu}, H.~B. (2018).
\newblock {Interactions Between Gas Dynamics and Magnetic Fields in the Massive
  Dense Cores of the DR21 Filament}.
\newblock \emph{\apj} 865, 110.
\newblock \doi{10.3847/1538-4357/aad9fc}
\input{Ching2018}

\bibitem[{{Ching} et~al.(2016){Ching}, {Lai}, {Zhang}, {Yang}, {Girart}, and
  {Rao}}]{Ching2016}
{Ching}, T.-C., {Lai}, S.-P., {Zhang}, Q., {Yang}, L., {Girart}, J.~M., and
  {Rao}, R. (2016).
\newblock {Helical Magnetic Fields in the NGC 1333 IRAS 4A Protostellar
  Outflows}.
\newblock \emph{\apj} 819, 159.
\newblock \doi{10.3847/0004-637X/819/2/159}
\input{Ching2016}

\bibitem[{{Chitsazzadeh} et~al.(2012){Chitsazzadeh}, {Houde}, {Hildebrand}, and
  {Vaillancourt}}]{Chitsazzadeh2012}
{Chitsazzadeh}, S., {Houde}, M., {Hildebrand}, R.~H., and {Vaillancourt}, J.
  (2012).
\newblock {Characterization of Turbulence from Submillimeter Dust Emission}.
\newblock \emph{\apj} 749, 45.
\newblock \doi{10.1088/0004-637X/749/1/45}
\input{Chitsazzadeh2012}

\bibitem[{{Cho} and {Lazarian}(2007)}]{Cho2007}
{Cho}, J. and {Lazarian}, A. (2007).
\newblock {Grain Alignment and Polarized Emission from Magnetized T Tauri
  Disks}.
\newblock \emph{\apj} 669, 1085--1097.
\newblock \doi{10.1086/521805}
\input{Cho2007}

\bibitem[{{Chuss} et~al.(2018){Chuss}, {Andersson}, {Bally}, {Dotson},
  {Dowell}, {Guerra} et~al.}]{Chuss2018}
{Chuss}, D.~T., {Andersson}, B., {Bally}, J., {Dotson}, J.~L., {Dowell}, C.~D.,
  {Guerra}, J.~A., et~al. (2018).
\newblock {HAWC+/SOFIA Multiwavelength Polarimetric Observations of OMC-1}.
\newblock \emph{ArXiv e-prints}
\input{Chuss2018}

\bibitem[{{Clark} et~al.(2015){Clark}, {Hill}, {Peek}, {Putman}, and
  {Babler}}]{Clark2015}
{Clark}, S.~E., {Hill}, J.~C., {Peek}, J.~E.~G., {Putman}, M.~E., and {Babler},
  B.~L. (2015).
\newblock {Neutral Hydrogen Structures Trace Dust Polarization Angle:
  Implications for Cosmic Microwave Background Foregrounds}.
\newblock \emph{Physical Review Letters} 115, 241302.
\newblock \doi{10.1103/PhysRevLett.115.241302}
\input{Clark2015}

\bibitem[{{Clark} et~al.(2014){Clark}, {Peek}, and {Putman}}]{Clark2014}
{Clark}, S.~E., {Peek}, J.~E.~G., and {Putman}, M.~E. (2014).
\newblock {Magnetically Aligned H I Fibers and the Rolling Hough Transform}.
\newblock \emph{\apj} 789, 82.
\newblock \doi{10.1088/0004-637X/789/1/82}
\input{Clark2014}

\bibitem[{{Cortes} and {Crutcher}(2006)}]{Cortes2006a}
{Cortes}, P. and {Crutcher}, R.~M. (2006).
\newblock {Interferometric Mapping of Magnetic Fields: G30.79 FIR 10}.
\newblock \emph{\apj} 639, 965--968.
\newblock \doi{10.1086/498971}
\input{Cortes2006a}

\bibitem[{{Cortes} et~al.(2006){Cortes}, {Crutcher}, and
  {Matthews}}]{Cortes2006b}
{Cortes}, P.~C., {Crutcher}, R.~M., and {Matthews}, B.~C. (2006).
\newblock {Interferometric Mapping of Magnetic Fields: NGC 2071IR}.
\newblock \emph{\apj} 650, 246--251.
\newblock \doi{10.1086/507460}
\input{Cortes2006b}

\bibitem[{{Cortes} et~al.(2008){Cortes}, {Crutcher}, {Shepherd}, and
  {Bronfman}}]{Cortes2008}
{Cortes}, P.~C., {Crutcher}, R.~M., {Shepherd}, D.~S., and {Bronfman}, L.
  (2008).
\newblock {Interferometric Mapping of Magnetic Fields: The Massive Star-forming
  Region G34.4+0.23 MM}.
\newblock \emph{\apj} 676, 464--471.
\newblock \doi{10.1086/524355}
\input{Cortes2008}

\bibitem[{{Cortes} et~al.(2005){Cortes}, {Crutcher}, and {Watson}}]{Cortes2005}
{Cortes}, P.~C., {Crutcher}, R.~M., and {Watson}, W.~D. (2005).
\newblock {Line Polarization of Molecular Lines at Radio Frequencies: The Case
  of DR 21(OH)}.
\newblock \emph{\apj} 628, 780--788.
\newblock \doi{10.1086/430815}
\input{Cortes2005}

\bibitem[{{Cortes} et~al.(2016){Cortes}, {Girart}, {Hull}, {Sridharan},
  {Louvet}, {Plambeck} et~al.}]{Cortes2016}
{Cortes}, P.~C., {Girart}, J.~M., {Hull}, C.~L.~H., {Sridharan}, T.~K.,
  {Louvet}, F., {Plambeck}, R., et~al. (2016).
\newblock {Interferometric Mapping of Magnetic Fields: The ALMA View of the
  Massive Star-forming Clump W43-MM1}.
\newblock \emph{\apjl} 825, L15.
\newblock \doi{10.3847/2041-8205/825/1/L15}
\input{Cortes2016}

\bibitem[{{Cox} et~al.(2018){Cox}, {Harris}, {Looney}, {Li}, {Yang}, {Tobin}
  et~al.}]{Cox2018}
{Cox}, E.~G., {Harris}, R.~J., {Looney}, L.~W., {Li}, Z.-Y., {Yang}, H.,
  {Tobin}, J.~J., et~al. (2018).
\newblock {ALMA's Polarized View of 10 Protostars in the Perseus Molecular
  Cloud}.
\newblock \emph{\apj} 855, 92.
\newblock \doi{10.3847/1538-4357/aaacd2}
\input{Cox2018}

\bibitem[{{Cox} et~al.(2015){Cox}, {Harris}, {Looney}, {Segura-Cox}, {Tobin},
  {Li} et~al.}]{Cox2015}
{Cox}, E.~G., {Harris}, R.~J., {Looney}, L.~W., {Segura-Cox}, D.~M., {Tobin},
  J., {Li}, Z.-Y., et~al. (2015).
\newblock {High-resolution 8 mm and 1 cm Polarization of IRAS 4A from the VLA
  Nascent Disk and Multiplicity (VANDAM) Survey}.
\newblock \emph{\apjl} 814, L28.
\newblock \doi{10.1088/2041-8205/814/2/L28}
\input{Cox2015}

\bibitem[{{Crutcher}(1999)}]{Crutcher1999}
{Crutcher}, R.~M. (1999).
\newblock {Magnetic Fields in Molecular Clouds: Observations Confront Theory}.
\newblock \emph{\apj} 520, 706--713.
\newblock \doi{10.1086/307483}
\input{Crutcher1999}

\bibitem[{{Crutcher}(2004)}]{Crutcher2004}
{Crutcher}, R.~M. (2004).
\newblock {What Drives Star Formation?}
\newblock \emph{\apss} 292, 225--237.
\newblock \doi{10.1023/B:ASTR.0000045021.42255.95}
\input{Crutcher2004}

\bibitem[{{Crutcher}(2012)}]{Crutcher2012}
{Crutcher}, R.~M. (2012).
\newblock {Magnetic Fields in Molecular Clouds}.
\newblock \emph{\araa} 50, 29--63.
\newblock \doi{10.1146/annurev-astro-081811-125514}
\input{Crutcher2012}

\bibitem[{{Crutcher} and {Kemball}(2019)}]{CrutcherKemball2019}
{Crutcher}, R.~M. and {Kemball}, A.~J. (2019).
\newblock Zeeman effect observations of regions of star formation.
\newblock \emph{{Frontiers in Astronomy and Space Sciences}} under review
\input{CrutcherKemball2019}

\bibitem[{{Crutcher} et~al.(1993){Crutcher}, {Troland}, {Goodman}, {Heiles},
  {Kazes}, and {Myers}}]{Crutcher1993}
{Crutcher}, R.~M., {Troland}, T.~H., {Goodman}, A.~A., {Heiles}, C., {Kazes},
  I., and {Myers}, P.~C. (1993).
\newblock {OH Zeeman observations of dark clouds}.
\newblock \emph{\apj} 407, 175--184.
\newblock \doi{10.1086/172503}
\input{Crutcher1993}

\bibitem[{{Curran} and {Chrysostomou}(2007)}]{Curran2007}
{Curran}, R.~L. and {Chrysostomou}, A. (2007).
\newblock {Magnetic fields in massive star-forming regions}.
\newblock \emph{\mnras} 382, 699--716.
\newblock \doi{10.1111/j.1365-2966.2007.12399.x}
\input{Curran2007}

\bibitem[{{Dapp} et~al.(2012){Dapp}, {Basu}, and {Kunz}}]{Dapp2012}
{Dapp}, W.~B., {Basu}, S., and {Kunz}, M.~W. (2012).
\newblock {Bridging the gap: disk formation in the Class 0 phase with ambipolar
  diffusion and Ohmic dissipation}.
\newblock \emph{\aap} 541, A35.
\newblock \doi{10.1051/0004-6361/201117876}
\input{Dapp2012}

\bibitem[{{Davidson} et~al.(2014){Davidson}, {Li}, {Hull}, {Plambeck}, {Kwon},
  {Crutcher} et~al.}]{Davidson2014}
{Davidson}, J.~A., {Li}, Z.-Y., {Hull}, C.~L.~H., {Plambeck}, R.~L., {Kwon},
  W., {Crutcher}, R.~M., et~al. (2014).
\newblock {Testing Magnetic Field Models for the Class 0 Protostar L1527}.
\newblock \emph{\apj} 797, 74.
\newblock \doi{10.1088/0004-637X/797/2/74}
\input{Davidson2014}

\bibitem[{{Davis}(1951)}]{Davis1951}
{Davis}, L. (1951).
\newblock {The Strength of Interstellar Magnetic Fields}.
\newblock \emph{Physical Review} 81, 890--891.
\newblock \doi{10.1103/PhysRev.81.890.2}
\input{Davis1951}

\bibitem[{{Deguchi} and {Watson}(1984)}]{Deguchi1984}
{Deguchi}, S. and {Watson}, W.~D. (1984).
\newblock {Linear polarization of molecular lines at radio frequencies}.
\newblock \emph{\apj} 285, 126--133.
\newblock \doi{10.1086/162483}
\input{Deguchi1984}

\bibitem[{{Dent} et~al.(2019){Dent}, {Pinte}, {Cortes}, {M{\'e}nard}, {Hales},
  {Fomalont} et~al.}]{Dent2019}
{Dent}, W.~R.~F., {Pinte}, C., {Cortes}, P.~C., {M{\'e}nard}, F., {Hales}, A.,
  {Fomalont}, E., et~al. (2019).
\newblock {Submillimetre dust polarization and opacity in the HD163296
  protoplanetary ring system}.
\newblock \emph{\mnras} 482, L29--L33.
\newblock \doi{10.1093/mnrasl/sly181}
\input{Dent2019}

\bibitem[{{Dotson} et~al.(1998){Dotson}, {Novak}, {Renbarger}, {Pernic}, and
  {Sundwall}}]{Dotson1998}
{Dotson}, J.~L., {Novak}, G., {Renbarger}, T., {Pernic}, D., and {Sundwall},
  J.~L. (1998).
\newblock {SPARO: the submillimeter polarimeter for Antarctic remote
  observing}.
\newblock In \emph{Advanced Technology MMW, Radio, and Terahertz Telescopes},
  ed. T.~G. {Phillips}. vol. 3357 of \emph{\procspie}, 543--547.
\newblock \doi{10.1117/12.317388}
\input{Dotson1998}

\bibitem[{{Dotson} et~al.(2010){Dotson}, {Vaillancourt}, {Kirby}, {Dowell},
  {Hildebrand}, and {Davidson}}]{Dotson2010}
{Dotson}, J.~L., {Vaillancourt}, J.~E., {Kirby}, L., {Dowell}, C.~D.,
  {Hildebrand}, R.~H., and {Davidson}, J.~A. (2010).
\newblock {350 {$\mu$}m Polarimetry from the Caltech Submillimeter
  Observatory}.
\newblock \emph{\apjs} 186, 406--426.
\newblock \doi{10.1088/0067-0049/186/2/406}
\input{Dotson2010}

\bibitem[{{Falceta-Gon{\c c}alves} et~al.(2008){Falceta-Gon{\c c}alves},
  {Lazarian}, and {Kowal}}]{FalcetaGoncalves2008}
{Falceta-Gon{\c c}alves}, D., {Lazarian}, A., and {Kowal}, G. (2008).
\newblock {Studies of Regular and Random Magnetic Fields in the ISM: Statistics
  of Polarization Vectors and the Chandrasekhar-Fermi Technique}.
\newblock \emph{\apj} 679, 537-551.
\newblock \doi{10.1086/587479}
\input{FalcetaGoncalves2008}

\bibitem[{{Fern{\'a}ndez-L{\'o}pez} et~al.(2016){Fern{\'a}ndez-L{\'o}pez},
  {Stephens}, {Girart}, {Looney}, {Curiel}, {Segura-Cox}
  et~al.}]{FernandezLopez2016}
{Fern{\'a}ndez-L{\'o}pez}, M., {Stephens}, I.~W., {Girart}, J.~M., {Looney},
  L., {Curiel}, S., {Segura-Cox}, D., et~al. (2016).
\newblock {1.3 mm Polarized Emission in the Circumstellar Disk of a Massive
  Protostar}.
\newblock \emph{\apj} 832, 200.
\newblock \doi{10.3847/0004-637X/832/2/200}
\input{FernandezLopez2016}

\bibitem[{{Fiedler} and {Mouschovias}(1993)}]{Fiedler1993}
{Fiedler}, R.~A. and {Mouschovias}, T.~C. (1993).
\newblock {Ambipolar Diffusion and Star Formation: Formation and Contraction of
  Axisymmetric Cloud Cores. II. Results}.
\newblock \emph{\apj} 415, 680.
\newblock \doi{10.1086/173193}
\input{Fiedler1993}

\bibitem[{{Fish} and {Reid}(2006)}]{Fish2006}
{Fish}, V.~L. and {Reid}, M.~J. (2006).
\newblock {Full-Polarization Observations of OH Masers in Massive Star-forming
  Regions. II. Maser Properties and the Interpretation of Polarization}.
\newblock \emph{\apjs} 164, 99--123.
\newblock \doi{10.1086/502650}
\input{Fish2006}

\bibitem[{{Fissel} et~al.(2016){Fissel}, {Ade}, {Angil{\`e}}, {Ashton},
  {Benton}, {Devlin} et~al.}]{Fissel2016}
{Fissel}, L.~M., {Ade}, P.~A.~R., {Angil{\`e}}, F.~E., {Ashton}, P., {Benton},
  S.~J., {Devlin}, M.~J., et~al. (2016).
\newblock {Balloon-Borne Submillimeter Polarimetry of the Vela C Molecular
  Cloud: Systematic Dependence of Polarization Fraction on Column Density and
  Local Polarization-Angle Dispersion}.
\newblock \emph{\apj} 824, 134.
\newblock \doi{10.3847/0004-637X/824/2/134}
\input{Fissel2016}

\bibitem[{{Frank} et~al.(2014){Frank}, {Ray}, {Cabrit}, {Hartigan}, {Arce},
  {Bacciotti} et~al.}]{Frank2014}
{Frank}, A., {Ray}, T.~P., {Cabrit}, S., {Hartigan}, P., {Arce}, H.~G.,
  {Bacciotti}, F., et~al. (2014).
\newblock {Jets and Outflows from Star to Cloud: Observations Confront Theory}.
\newblock In \emph{Protostars and Planets VI}, eds. H.~{Beuther}, R.~S.
  {Klessen}, C.~P. {Dullemond}, and T.~{Henning} (Tucson, Arizona: University
  of Arizona Press), 451--474.
\newblock \doi{10.2458/azu_uapress_9780816531240-ch020}
\input{Frank2014}

\bibitem[{{Frau} et~al.(2011){Frau}, {Galli}, and {Girart}}]{Frau2011}
{Frau}, P., {Galli}, D., and {Girart}, J.~M. (2011).
\newblock {Comparing star formation models with interferometric observations of
  the protostar NGC 1333 IRAS 4A. I. Magnetohydrodynamic collapse models}.
\newblock \emph{\aap} 535, A44.
\newblock \doi{10.1051/0004-6361/201117813}
\input{Frau2011}

\bibitem[{{Frau} et~al.(2014){Frau}, {Girart}, {Zhang}, and {Rao}}]{Frau2014}
{Frau}, P., {Girart}, J.~M., {Zhang}, Q., and {Rao}, R. (2014).
\newblock {Shaping a high-mass star-forming cluster through stellar feedback.
  The case of the NGC 7538 IRS 1-3 complex}.
\newblock \emph{\aap} 567, A116.
\newblock \doi{10.1051/0004-6361/201423917}
\input{Frau2014}

\bibitem[{{Galametz} et~al.(2018){Galametz}, {Maury}, {Girart}, {Rao}, {Zhang},
  {Gaudel} et~al.}]{Galametz2018}
{Galametz}, M., {Maury}, A., {Girart}, J.~M., {Rao}, R., {Zhang}, Q., {Gaudel},
  M., et~al. (2018).
\newblock {SMA observations of polarized dust emission in solar-type Class 0
  protostars: Magnetic field properties at envelope scales}.
\newblock \emph{\aap} 616, A139.
\newblock \doi{10.1051/0004-6361/201833004}
\input{Galametz2018}

\bibitem[{{Galitzki} et~al.(2014){Galitzki}, {Ade}, {Angil{\`e}}, {Ashton},
  {Beall}, {Becker} et~al.}]{Galitzki2014}
{Galitzki}, N., {Ade}, P.~A.~R., {Angil{\`e}}, F.~E., {Ashton}, P., {Beall},
  J.~A., {Becker}, D., et~al. (2014).
\newblock {The Next Generation BLAST Experiment}.
\newblock \emph{Journal of Astronomical Instrumentation} 3, 1440001.
\newblock \doi{10.1142/S2251171714400017}
\input{Galitzki2014}

\bibitem[{{Galli} et~al.(2006){Galli}, {Lizano}, {Shu}, and
  {Allen}}]{Galli2006}
{Galli}, D., {Lizano}, S., {Shu}, F.~H., and {Allen}, A. (2006).
\newblock {Gravitational Collapse of Magnetized Clouds. I. Ideal
  Magnetohydrodynamic Accretion Flow}.
\newblock \emph{\apj} 647, 374--381.
\newblock \doi{10.1086/505257}
\input{Galli2006}

\bibitem[{{Galli} and {Shu}(1993{\natexlab{a}})}]{Galli1993a}
{Galli}, D. and {Shu}, F.~H. (1993{\natexlab{a}}).
\newblock {Collapse of Magnetized Molecular Cloud Cores. I. Semianalytical
  Solution}.
\newblock \emph{\apj} 417, 220.
\newblock \doi{10.1086/173305}
\input{Galli1993a}

\bibitem[{{Galli} and {Shu}(1993{\natexlab{b}})}]{Galli1993b}
{Galli}, D. and {Shu}, F.~H. (1993{\natexlab{b}}).
\newblock {Collapse of Magnetized Molecular Cloud Cores. II. Numerical
  Results}.
\newblock \emph{\apj} 417, 243.
\newblock \doi{10.1086/173306}
\input{Galli1993b}

\bibitem[{{Gandilo} et~al.(2016){Gandilo}, {Ade}, {Angil{\`e}}, {Ashton},
  {Benton}, {Devlin} et~al.}]{Gandilo2016}
{Gandilo}, N.~N., {Ade}, P.~A.~R., {Angil{\`e}}, F.~E., {Ashton}, P., {Benton},
  S.~J., {Devlin}, M.~J., et~al. (2016).
\newblock {Submillimeter Polarization Spectrum in the Vela C Molecular Cloud}.
\newblock \emph{\apj} 824, 84.
\newblock \doi{10.3847/0004-637X/824/2/84}
\input{Gandilo2016}

\bibitem[{{Ginsburg} et~al.(2015){Ginsburg}, {Bally}, {Battersby},
  {Youngblood}, {Darling}, {Rosolowsky} et~al.}]{Ginsburg2015}
{Ginsburg}, A., {Bally}, J., {Battersby}, C., {Youngblood}, A., {Darling}, J.,
  {Rosolowsky}, E., et~al. (2015).
\newblock {The dense gas mass fraction in the W51 cloud and its protoclusters}.
\newblock \emph{\aap} 573, A106.
\newblock \doi{10.1051/0004-6361/201424979}
\input{Ginsburg2015}

\bibitem[{{Ginsburg} et~al.(2016){Ginsburg}, {Goss}, {Goddi},
  {Galv{\'a}n-Madrid}, {Dale}, {Bally} et~al.}]{Ginsburg2016}
{Ginsburg}, A., {Goss}, W.~M., {Goddi}, C., {Galv{\'a}n-Madrid}, R., {Dale},
  J.~E., {Bally}, J., et~al. (2016).
\newblock {Toward gas exhaustion in the W51 high-mass protoclusters}.
\newblock \emph{\aap} 595, A27.
\newblock \doi{10.1051/0004-6361/201628318}
\input{Ginsburg2016}

\bibitem[{{Girart} et~al.(2009){Girart}, {Beltr{\'a}n}, {Zhang}, {Rao}, and
  {Estalella}}]{Girart2009}
{Girart}, J.~M., {Beltr{\'a}n}, M.~T., {Zhang}, Q., {Rao}, R., and {Estalella},
  R. (2009).
\newblock {Magnetic Fields in the Formation of Massive Stars}.
\newblock \emph{Science} 324, 1408.
\newblock \doi{10.1126/science.1171807}
\input{Girart2009}

\bibitem[{{Girart} et~al.(1999){Girart}, {Crutcher}, and {Rao}}]{Girart1999}
{Girart}, J.~M., {Crutcher}, R.~M., and {Rao}, R. (1999).
\newblock {Detection of Polarized CO Emission from the Molecular Outflow in NGC
  1333 IRAS 4A}.
\newblock \emph{\apjl} 525, L109--L112.
\newblock \doi{10.1086/312345}
\input{Girart1999}

\bibitem[{{Girart} et~al.(2018){Girart}, {Fern{\'a}ndez-L{\'o}pez}, {Li},
  {Yang}, {Estalella}, {Anglada} et~al.}]{Girart2018}
{Girart}, J.~M., {Fern{\'a}ndez-L{\'o}pez}, M., {Li}, Z.-Y., {Yang}, H.,
  {Estalella}, R., {Anglada}, G., et~al. (2018).
\newblock {Resolving the Polarized Dust Emission of the Disk around the Massive
  Star Powering the HH 80-81 Radio Jet}.
\newblock \emph{\apjl} 856, L27.
\newblock \doi{10.3847/2041-8213/aab76b}
\input{Girart2018}

\bibitem[{{Girart} et~al.(2013){Girart}, {Frau}, {Zhang}, {Koch}, {Qiu}, {Tang}
  et~al.}]{Girart2013}
{Girart}, J.~M., {Frau}, P., {Zhang}, Q., {Koch}, P.~M., {Qiu}, K., {Tang},
  Y.-W., et~al. (2013).
\newblock {DR 21(OH): A Highly Fragmented, Magnetized, Turbulent Dense Core}.
\newblock \emph{\apj} 772, 69.
\newblock \doi{10.1088/0004-637X/772/1/69}
\input{Girart2013}

\bibitem[{{Girart} et~al.(2004){Girart}, {Greaves}, {Crutcher}, and
  {Lai}}]{Girart2004}
{Girart}, J.~M., {Greaves}, J.~S., {Crutcher}, R.~M., and {Lai}, S.-P. (2004).
\newblock {BIMA and JCMT Spectropolarimetric Observations of the CO J = 2 - 1
  Line Towards Orion KL/IRc2}.
\newblock \emph{\apss} 292, 119--125.
\newblock \doi{10.1023/B:ASTR.0000045007.35868.17}
\input{Girart2004}

\bibitem[{{Girart} et~al.(2006){Girart}, {Rao}, and {Marrone}}]{Girart2006}
{Girart}, J.~M., {Rao}, R., and {Marrone}, D.~P. (2006).
\newblock {Magnetic Fields in the Formation of Sun-Like Stars}.
\newblock \emph{Science} 313, 812--814.
\newblock \doi{10.1126/science.1129093}
\input{Girart2006}

\bibitem[{{Girart} et~al.(2008){Girart}, {Rao}, and {Marrone}}]{Girart2008}
{Girart}, J.~M., {Rao}, R., and {Marrone}, D.~P. (2008).
\newblock {SMA observations of the magnetic fields around a low-mass
  protostellar system}.
\newblock \emph{\apss} 313, 87--90.
\newblock \doi{10.1007/s10509-007-9592-7}
\input{Girart2008}

\bibitem[{{Glenn} et~al.(1997){Glenn}, {Walker}, {Bieging}, and
  {Jewell}}]{Glenn1997}
{Glenn}, J., {Walker}, C.~K., {Bieging}, J.~H., and {Jewell}, P.~R. (1997).
\newblock {Millimeter-Wave Spectropolarimetry of Evolved Stars: Evidence for
  Polarized Molecular Line Emission}.
\newblock \emph{\apjl} 487, L89--L92.
\newblock \doi{10.1086/310863}
\input{Glenn1997}

\bibitem[{{Goldreich} and {Kylafis}(1981)}]{Goldreich1981}
{Goldreich}, P. and {Kylafis}, N.~D. (1981).
\newblock {On mapping the magnetic field direction in molecular clouds by
  polarization measurements}.
\newblock \emph{\apjl} 243, L75--L78.
\newblock \doi{10.1086/183446}
\input{Goldreich1981}

\bibitem[{{Goldreich} and {Kylafis}(1982)}]{Goldreich1982}
{Goldreich}, P. and {Kylafis}, N.~D. (1982).
\newblock {Linear polarization of radio frequency lines in molecular clouds and
  circumstellar envelopes}.
\newblock \emph{\apj} 253, 606--621.
\newblock \doi{10.1086/159663}
\input{Goldreich1982}

\bibitem[{{Gon{\c c}alves} et~al.(2008){Gon{\c c}alves}, {Galli}, and
  {Girart}}]{Goncalves2008}
{Gon{\c c}alves}, J., {Galli}, D., and {Girart}, J.~M. (2008).
\newblock {Modeling the magnetic field in the protostellar source NGC 1333 IRAS
  4A}.
\newblock \emph{\aap} 490, L39--L42.
\newblock \doi{10.1051/0004-6361:200810861}
\input{Goncalves2008}

\bibitem[{{Gonz{\'a}lez-Casanova} and
  {Lazarian}(2017)}]{GonzalezCasanovaLazarian2017}
{Gonz{\'a}lez-Casanova}, D.~F. and {Lazarian}, A. (2017).
\newblock {Velocity Gradients as a Tracer for Magnetic Fields}.
\newblock \emph{\apj} 835, 41.
\newblock \doi{10.3847/1538-4357/835/1/41}
\input{GonzalezCasanovaLazarian2017}

\bibitem[{{Gordon} et~al.(2018){Gordon}, {Lopez-Rodriguez}, {Andersson},
  {Clarke}, {Coude}, {Moullet} et~al.}]{Gordon2018}
{Gordon}, M.~S., {Lopez-Rodriguez}, E., {Andersson}, B.-G., {Clarke}, M.,
  {Coude}, S., {Moullet}, A., et~al. (2018).
\newblock {SOFIA Community Science I: HAWC+ Polarimetry of 30 Doradus}.
\newblock \emph{ArXiv e-prints}
\input{Gordon2018}

\bibitem[{{Greaves} et~al.(1999){Greaves}, {Holland}, {Friberg}, and
  {Dent}}]{Greaves1999}
{Greaves}, J.~S., {Holland}, W.~S., {Friberg}, P., and {Dent}, W.~R.~F. (1999).
\newblock {Polarized CO Emission from Molecular Clouds}.
\newblock \emph{\apjl} 512, L139--L142.
\newblock \doi{10.1086/311888}
\input{Greaves1999}

\bibitem[{{Harris} et~al.(2018){Harris}, {Cox}, {Looney}, {Li}, {Yang},
  {Fern{\'a}ndez-L{\'o}pez} et~al.}]{Harris2018}
{Harris}, R.~J., {Cox}, E.~G., {Looney}, L.~W., {Li}, Z.-Y., {Yang}, H.,
  {Fern{\'a}ndez-L{\'o}pez}, M., et~al. (2018).
\newblock {ALMA Observations of Polarized 872 {$\mu$}m Dust Emission from the
  Protostellar Systems VLA 1623 and L1527}.
\newblock \emph{\apj} 861, 91.
\newblock \doi{10.3847/1538-4357/aac6ec}
\input{Harris2018}

\bibitem[{{Heiles} and {Crutcher}(2005)}]{Heiles2005b}
{Heiles}, C. and {Crutcher}, R. (2005).
\newblock {Magnetic Fields in Diffuse HI and Molecular Clouds}.
\newblock In \emph{Cosmic Magnetic Fields}, eds. R.~{Wielebinski} and
  R.~{Beck}. vol. 664 of \emph{Lecture Notes in Physics, Berlin Springer
  Verlag}, 137.
\newblock \doi{10.1007/11369875_7}
\input{Heiles2005b}

\bibitem[{{Heiles} et~al.(1993){Heiles}, {Goodman}, {McKee}, and
  {Zweibel}}]{Heiles1993}
{Heiles}, C., {Goodman}, A.~A., {McKee}, C.~F., and {Zweibel}, E.~G. (1993).
\newblock {Magnetic fields in star-forming regions - Observations}.
\newblock In \emph{Protostars and Planets III}, eds. E.~H. {Levy} and J.~I.
  {Lunine}. 279--326
\input{Heiles1993}

\bibitem[{{Heitsch} et~al.(2001){Heitsch}, {Zweibel}, {Mac Low}, {Li}, and
  {Norman}}]{Heitsch2001}
{Heitsch}, F., {Zweibel}, E.~G., {Mac Low}, M.-M., {Li}, P., and {Norman},
  M.~L. (2001).
\newblock {Magnetic Field Diagnostics Based on Far-Infrared Polarimetry: Tests
  Using Numerical Simulations}.
\newblock \emph{\apj} 561, 800--814.
\newblock \doi{10.1086/323489}
\input{Heitsch2001}

\bibitem[{{Hennebelle} and {Chabrier}(2008)}]{Hennebelle2008c}
{Hennebelle}, P. and {Chabrier}, G. (2008).
\newblock {Analytical Theory for the Initial Mass Function: CO Clumps and
  Prestellar Cores}.
\newblock \emph{\apj} 684, 395--410.
\newblock \doi{10.1086/589916}
\input{Hennebelle2008c}

\bibitem[{{Hennebelle} and {Ciardi}(2009)}]{Hennebelle2009}
{Hennebelle}, P. and {Ciardi}, A. (2009).
\newblock {Disk formation during collapse of magnetized protostellar cores}.
\newblock \emph{\aap} 506, L29--L32.
\newblock \doi{10.1051/0004-6361/200913008}
\input{Hennebelle2009}

\bibitem[{{Hennebelle} and {Inutsuka}(2019)}]{HennebelleInutsuka2019}
{Hennebelle}, P. and {Inutsuka}, S.-i. (2019).
\newblock {The role of magnetic field in molecular cloud formation and
  evolution}.
\newblock \emph{arXiv e-prints} , arXiv:1902.00798
\input{HennebelleInutsuka2019}

\bibitem[{{Hezareh} et~al.(2013){Hezareh}, {Wiesemeyer}, {Houde}, {Gusdorf},
  and {Siringo}}]{Hezareh2013}
{Hezareh}, T., {Wiesemeyer}, H., {Houde}, M., {Gusdorf}, A., and {Siringo}, G.
  (2013).
\newblock {Non-Zeeman circular polarization of CO rotational lines in SNR IC
  443}.
\newblock \emph{\aap} 558, A45.
\newblock \doi{10.1051/0004-6361/201321900}
\input{Hezareh2013}

\bibitem[{{Hildebrand}(1988)}]{Hildebrand1988}
{Hildebrand}, R.~H. (1988).
\newblock {Magnetic fields and stardust}.
\newblock \emph{\qjras} 29, 327--351
\input{Hildebrand1988}

\bibitem[{{Hildebrand} et~al.(2009){Hildebrand}, {Kirby}, {Dotson}, {Houde},
  and {Vaillancourt}}]{Hildebrand2009}
{Hildebrand}, R.~H., {Kirby}, L., {Dotson}, J.~L., {Houde}, M., and
  {Vaillancourt}, J.~E. (2009).
\newblock {Dispersion of Magnetic Fields in Molecular Clouds. I}.
\newblock \emph{\apj} 696, 567--573.
\newblock \doi{10.1088/0004-637X/696/1/567}
\input{Hildebrand2009}

\bibitem[{{Hoang} and {Lazarian}(2009)}]{Hoang2009}
{Hoang}, T. and {Lazarian}, A. (2009).
\newblock {Grain Alignment Induced by Radiative Torques: Effects of Internal
  Relaxation of Energy and Complex Radiation Field}.
\newblock \emph{\apj} 697, 1316--1333.
\newblock \doi{10.1088/0004-637X/697/2/1316}
\input{Hoang2009}

\bibitem[{{Houde} et~al.(2004){Houde}, {Dowell}, {Hildebrand}, {Dotson},
  {Vaillancourt}, {Phillips} et~al.}]{Houde2004}
{Houde}, M., {Dowell}, C.~D., {Hildebrand}, R.~H., {Dotson}, J.~L.,
  {Vaillancourt}, J.~E., {Phillips}, T.~G., et~al. (2004).
\newblock {Tracing the Magnetic Field in Orion A}.
\newblock \emph{\apj} 604, 717--740.
\newblock \doi{10.1086/382067}
\input{Houde2004}

\bibitem[{{Houde} et~al.(2013){Houde}, {Hezareh}, {Jones}, and
  {Rajabi}}]{Houde2013}
{Houde}, M., {Hezareh}, T., {Jones}, S., and {Rajabi}, F. (2013).
\newblock {Non-Zeeman Circular Polarization of Molecular Rotational Spectral
  Lines}.
\newblock \emph{\apj} 764, 24.
\newblock \doi{10.1088/0004-637X/764/1/24}
\input{Houde2013}

\bibitem[{{Houde} et~al.(2016){Houde}, {Hull}, {Plambeck}, {Vaillancourt}, and
  {Hildebrand}}]{Houde2016}
{Houde}, M., {Hull}, C.~L.~H., {Plambeck}, R.~L., {Vaillancourt}, J.~E., and
  {Hildebrand}, R.~H. (2016).
\newblock {Dispersion of Magnetic Fields in Molecular Clouds. IV. Analysis of
  Interferometry Data}.
\newblock \emph{\apj} 820, 38.
\newblock \doi{10.3847/0004-637X/820/1/38}
\input{Houde2016}

\bibitem[{{Houde} et~al.(2011){Houde}, {Rao}, {Vaillancourt}, and
  {Hildebrand}}]{Houde2011}
{Houde}, M., {Rao}, R., {Vaillancourt}, J.~E., and {Hildebrand}, R.~H. (2011).
\newblock {Dispersion of Magnetic Fields in Molecular Clouds. III.}
\newblock \emph{\apj} 733, 109.
\newblock \doi{10.1088/0004-637X/733/2/109}
\input{Houde2011}

\bibitem[{{Houde} et~al.(2009){Houde}, {Vaillancourt}, {Hildebrand},
  {Chitsazzadeh}, and {Kirby}}]{Houde2009}
{Houde}, M., {Vaillancourt}, J.~E., {Hildebrand}, R.~H., {Chitsazzadeh}, S.,
  and {Kirby}, L. (2009).
\newblock {Dispersion of Magnetic Fields in Molecular Clouds. II.}
\newblock \emph{\apj} 706, 1504--1516.
\newblock \doi{10.1088/0004-637X/706/2/1504}
\input{Houde2009}

\bibitem[{{Hughes} et~al.(2013){Hughes}, {Hull}, {Wilner}, and
  {Plambeck}}]{Hughes2013}
{Hughes}, A.~M., {Hull}, C.~L.~H., {Wilner}, D.~J., and {Plambeck}, R.~L.
  (2013).
\newblock {Interferometric Upper Limits on Millimeter Polarization of the Disks
  around DG Tau, GM Aur, and MWC 480}.
\newblock \emph{\aj} 145, 115.
\newblock \doi{10.1088/0004-6256/145/4/115}
\input{Hughes2013}

\bibitem[{{Hughes} et~al.(2009){Hughes}, {Wilner}, {Cho}, {Marrone},
  {Lazarian}, {Andrews} et~al.}]{Hughes2009b}
{Hughes}, A.~M., {Wilner}, D.~J., {Cho}, J., {Marrone}, D.~P., {Lazarian}, A.,
  {Andrews}, S.~M., et~al. (2009).
\newblock {Stringent Limits on the Polarized Submillimeter Emission from
  Protoplanetary Disks}.
\newblock \emph{\apj} 704, 1204--1217.
\newblock \doi{10.1088/0004-637X/704/2/1204}
\input{Hughes2009b}

\bibitem[{{Hull} et~al.(2018{\natexlab{a}}){Hull}, {Carrasco-Gonz{\'a}lez},
  {Williams}, {Girart}, {Robishaw}, {Galv{\'a}n-Madrid} et~al.}]{Hull2018b}
{Hull}, C.~L.~H., {Carrasco-Gonz{\'a}lez}, C., {Williams}, P.~K.~G., {Girart},
  J.~M., {Robishaw}, T., {Galv{\'a}n-Madrid}, R., et~al. (2018{\natexlab{a}}).
\newblock {Magnetic fields in forming stars with the ngVLA}.
\newblock \emph{ArXiv e-prints}
\input{Hull2018b}

\bibitem[{{Hull} et~al.(2017{\natexlab{b}}){Hull}, {Girart}, {Tychoniec},
  {Rao}, {Cort{\'e}s}, {Pokhrel} et~al.}]{Hull2017b}
{Hull}, C.~L.~H., {Girart}, J.~M., {Tychoniec}, {\L}., {Rao}, R., {Cort{\'e}s},
  P.~C., {Pokhrel}, R., et~al. (2017{\natexlab{b}}).
\newblock {ALMA Observations of Dust Polarization and Molecular Line Emission
  from the Class 0 Protostellar Source Serpens SMM1}.
\newblock \emph{\apj} 847, 92.
\newblock \doi{10.3847/1538-4357/aa7fe9}
\input{Hull2017b}

\bibitem[{{Hull} et~al.(2017{\natexlab{a}}){Hull}, {Mocz}, {Burkhart},
  {Goodman}, {Girart}, {Cort{\'e}s} et~al.}]{Hull2017a}
{Hull}, C.~L.~H., {Mocz}, P., {Burkhart}, B., {Goodman}, A.~A., {Girart},
  J.~M., {Cort{\'e}s}, P.~C., et~al. (2017{\natexlab{a}}).
\newblock {Unveiling the Role of the Magnetic Field at the Smallest Scales of
  Star Formation}.
\newblock \emph{\apjl} 842, L9.
\newblock \doi{10.3847/2041-8213/aa71b7}
\input{Hull2017a}

\bibitem[{{Hull} and {Plambeck}(2015)}]{Hull2015b}
{Hull}, C.~L.~H. and {Plambeck}, R.~L. (2015).
\newblock {The 1.3mm Full-Stokes Polarization System at CARMA}.
\newblock \emph{Journal of Astronomical Instrumentation} 4, 1550005.
\newblock \doi{10.1142/S2251171715500051}
\input{Hull2015b}

\bibitem[{{Hull} et~al.(2013){Hull}, {Plambeck}, {Bolatto}, {Bower},
  {Carpenter}, {Crutcher} et~al.}]{Hull2013}
{Hull}, C.~L.~H., {Plambeck}, R.~L., {Bolatto}, A.~D., {Bower}, G.~C.,
  {Carpenter}, J.~M., {Crutcher}, R.~M., et~al. (2013).
\newblock {Misalignment of Magnetic Fields and Outflows in Protostellar Cores}.
\newblock \emph{\apj} 768, 159.
\newblock \doi{10.1088/0004-637X/768/2/159}
\input{Hull2013}

\bibitem[{{Hull} et~al.(2014){Hull}, {Plambeck}, {Kwon}, {Bower}, {Carpenter},
  {Crutcher} et~al.}]{Hull2014}
{Hull}, C.~L.~H., {Plambeck}, R.~L., {Kwon}, W., {Bower}, G.~C., {Carpenter},
  J.~M., {Crutcher}, R.~M., et~al. (2014).
\newblock {TADPOL: A 1.3 mm Survey of Dust Polarization in Star-forming Cores
  and Regions}.
\newblock \emph{\apjs} 213, 13.
\newblock \doi{10.1088/0067-0049/213/1/13}
\input{Hull2014}

\bibitem[{{Hull} et~al.(2018{\natexlab{b}}){Hull}, {Yang}, {Li}, {Kataoka},
  {Stephens}, {Andrews} et~al.}]{Hull2018a}
{Hull}, C.~L.~H., {Yang}, H., {Li}, Z.-Y., {Kataoka}, A., {Stephens}, I.~W.,
  {Andrews}, S., et~al. (2018{\natexlab{b}}).
\newblock {ALMA Observations of Polarization from Dust Scattering in the IM Lup
  Protoplanetary Disk}.
\newblock \emph{\apj} 860, 82.
\newblock \doi{10.3847/1538-4357/aabfeb}
\input{Hull2018a}

\bibitem[{{Hutawarakorn} et~al.(2002){Hutawarakorn}, {Cohen}, and
  {Brebner}}]{Hutawarakorn2002}
{Hutawarakorn}, B., {Cohen}, R.~J., and {Brebner}, G.~C. (2002).
\newblock {OH masers and magnetic fields in the bipolar outflow source W75N}.
\newblock \emph{\mnras} 330, 349--364.
\newblock \doi{10.1046/j.1365-8711.2002.05068.x}
\input{Hutawarakorn2002}

\bibitem[{{Isella} et~al.(2015){Isella}, {Hull}, {Moullet},
  {Galv{\'a}n-Madrid}, {Johnstone}, {Ricci} et~al.}]{Isella2015}
{Isella}, A., {Hull}, C.~L.~H., {Moullet}, A., {Galv{\'a}n-Madrid}, R.,
  {Johnstone}, D., {Ricci}, L., et~al. (2015).
\newblock {Next Generation Very Large Array Memo No. 6, Science Working Group
  1: The Cradle of Life}.
\newblock \emph{ArXiv e-prints}
\input{Isella2015}

\bibitem[{{Jensen} and {Akeson}(2014)}]{JensenAkeson2014}
{Jensen}, E.~L.~N. and {Akeson}, R. (2014).
\newblock {Misaligned protoplanetary disks in a young binary star system}.
\newblock \emph{\nat} 511, 567--569.
\newblock \doi{10.1038/nature13521}
\input{JensenAkeson2014}

\bibitem[{{Joos} et~al.(2012){Joos}, {Hennebelle}, and {Ciardi}}]{Joos2012}
{Joos}, M., {Hennebelle}, P., and {Ciardi}, A. (2012).
\newblock {Protostellar disk formation and transport of angular momentum during
  magnetized core collapse}.
\newblock \emph{\aap} 543, A128.
\newblock \doi{10.1051/0004-6361/201118730}
\input{Joos2012}

\bibitem[{{Ju{\'a}rez} et~al.(2017){Ju{\'a}rez}, {Girart}, {Zamora-Avil{\'e}s},
  {Tang}, {Koch}, {Liu} et~al.}]{Juarez2017}
{Ju{\'a}rez}, C., {Girart}, J.~M., {Zamora-Avil{\'e}s}, M., {Tang}, Y.-W.,
  {Koch}, P.~M., {Liu}, H.~B., et~al. (2017).
\newblock {Magnetized Converging Flows toward the Hot Core in the
  Intermediate/High-mass Star-forming Region NGC 6334 V}.
\newblock \emph{\apj} 844, 44.
\newblock \doi{10.3847/1538-4357/aa78a6}
\input{Juarez2017}

\bibitem[{{Kataoka} et~al.(2012){Kataoka}, {Machida}, and
  {Tomisaka}}]{Kataoka2012}
{Kataoka}, A., {Machida}, M.~N., and {Tomisaka}, K. (2012).
\newblock {Exploring Magnetic Field Structure in Star-forming Cores with
  Polarization of Thermal Dust Emission}.
\newblock \emph{\apj} 761, 40.
\newblock \doi{10.1088/0004-637X/761/1/40}
\input{Kataoka2012}

\bibitem[{{Kataoka} et~al.(2016{\natexlab{a}}){Kataoka}, {Muto}, {Momose},
  {Tsukagoshi}, and {Dullemond}}]{Kataoka2016}
{Kataoka}, A., {Muto}, T., {Momose}, M., {Tsukagoshi}, T., and {Dullemond},
  C.~P. (2016{\natexlab{a}}).
\newblock {Grain Size Constraints on HL Tau with Polarization Signature}.
\newblock \emph{\apj} 820, 54.
\newblock \doi{10.3847/0004-637X/820/1/54}
\input{Kataoka2016}

\bibitem[{{Kataoka} et~al.(2015){Kataoka}, {Muto}, {Momose}, {Tsukagoshi},
  {Fukagawa}, {Shibai} et~al.}]{Kataoka2015}
{Kataoka}, A., {Muto}, T., {Momose}, M., {Tsukagoshi}, T., {Fukagawa}, M.,
  {Shibai}, H., et~al. (2015).
\newblock {Millimeter-wave Polarization of Protoplanetary Disks due to Dust
  Scattering}.
\newblock \emph{\apj} 809, 78.
\newblock \doi{10.1088/0004-637X/809/1/78}
\input{Kataoka2015}

\bibitem[{{Kataoka} et~al.(2016{\natexlab{b}}){Kataoka}, {Tsukagoshi},
  {Momose}, {Nagai}, {Muto}, {Dullemond} et~al.}]{Kataoka2016b}
{Kataoka}, A., {Tsukagoshi}, T., {Momose}, M., {Nagai}, H., {Muto}, T.,
  {Dullemond}, C.~P., et~al. (2016{\natexlab{b}}).
\newblock {Submillimeter Polarization Observation of the Protoplanetary Disk
  around HD 142527}.
\newblock \emph{\apjl} 831, L12.
\newblock \doi{10.3847/2041-8205/831/2/L12}
\input{Kataoka2016b}

\bibitem[{{Kataoka} et~al.(2017){Kataoka}, {Tsukagoshi}, {Pohl}, {Muto},
  {Nagai}, {Stephens} et~al.}]{Kataoka2017}
{Kataoka}, A., {Tsukagoshi}, T., {Pohl}, A., {Muto}, T., {Nagai}, H.,
  {Stephens}, I.~W., et~al. (2017).
\newblock {The Evidence of Radio Polarization Induced by the Radiative Grain
  Alignment and Self-scattering of Dust Grains in a Protoplanetary Disk}.
\newblock \emph{\apjl} 844, L5.
\newblock \doi{10.3847/2041-8213/aa7e33}
\input{Kataoka2017}

\bibitem[{{Kauffmann} et~al.(2013){Kauffmann}, {Pillai}, and
  {Goldsmith}}]{Kauffmann2013}
{Kauffmann}, J., {Pillai}, T., and {Goldsmith}, P.~F. (2013).
\newblock {Low Virial Parameters in Molecular Clouds: Implications for
  High-mass Star Formation and Magnetic Fields}.
\newblock \emph{\apj} 779, 185.
\newblock \doi{10.1088/0004-637X/779/2/185}
\input{Kauffmann2013}

\bibitem[{{Koch} et~al.(2010){Koch}, {Tang}, and {Ho}}]{Koch2010}
{Koch}, P.~M., {Tang}, Y., and {Ho}, P.~T.~P. (2010).
\newblock {Magnetic Field Properties in High-mass Star Formation from Large to
  Small Scales: A Statistical Analysis from Polarization Data}.
\newblock \emph{\apj} 721, 815--827.
\newblock \doi{10.1088/0004-637X/721/1/815}
\input{Koch2010}

\bibitem[{{Koch} et~al.(2012){Koch}, {Tang}, and {Ho}}]{Koch2012}
{Koch}, P.~M., {Tang}, Y.-W., and {Ho}, P.~T.~P. (2012).
\newblock {Magnetic Field Strength Maps for Molecular Clouds: A New Method
  Based on a Polarization-Intensity Gradient Relation}.
\newblock \emph{\apj} 747, 79.
\newblock \doi{10.1088/0004-637X/747/1/79}
\input{Koch2012}

\bibitem[{{Koch} et~al.(2018){Koch}, {Tang}, {Ho}, {Yen}, {Su}, and
  {Takakuwa}}]{Koch2018}
{Koch}, P.~M., {Tang}, Y.-W., {Ho}, P.~T.~P., {Yen}, H.-W., {Su}, Y.-N., and
  {Takakuwa}, S. (2018).
\newblock {Polarization Properties and Magnetic Field Structures in the
  High-mass Star-forming Region W51 Observed with ALMA}.
\newblock \emph{\apj} 855, 39.
\newblock \doi{10.3847/1538-4357/aaa4c1}
\input{Koch2018}

\bibitem[{{Koch} et~al.(2014){Koch}, {Tang}, {Ho}, {Zhang}, {Girart}, {Chen}
  et~al.}]{Koch2014}
{Koch}, P.~M., {Tang}, Y.-W., {Ho}, P.~T.~P., {Zhang}, Q., {Girart}, J.~M.,
  {Chen}, H.-R.~V., et~al. (2014).
\newblock {The Importance of the Magnetic Field from an SMA-CSO-combined Sample
  of Star-forming Regions}.
\newblock \emph{\apj} 797, 99.
\newblock \doi{10.1088/0004-637X/797/2/99}
\input{Koch2014}

\bibitem[{{Konigl} and {Pudritz}(2000)}]{Konigl2000}
{Konigl}, A. and {Pudritz}, R.~E. (2000).
\newblock {Disk Winds and the Accretion-Outflow Connection}.
\newblock \emph{Protostars and Planets IV} , 759
\input{Konigl2000}

\bibitem[{{Kratter} et~al.(2010){Kratter}, {Matzner}, {Krumholz}, and
  {Klein}}]{Kratter2010}
{Kratter}, K.~M., {Matzner}, C.~D., {Krumholz}, M.~R., and {Klein}, R.~I.
  (2010).
\newblock {On the Role of Disks in the Formation of Stellar Systems: A
  Numerical Parameter Study of Rapid Accretion}.
\newblock \emph{\apj} 708, 1585--1597.
\newblock \doi{10.1088/0004-637X/708/2/1585}
\input{Kratter2010}

\bibitem[{{Krumholz} et~al.(2013){Krumholz}, {Crutcher}, and
  {Hull}}]{Krumholz2013}
{Krumholz}, M.~R., {Crutcher}, R.~M., and {Hull}, C. L.~H. (2013).
\newblock {Protostellar Disk Formation Enabled by Weak, Misaligned Magnetic
  Fields}.
\newblock \emph{\apj} 767, L11.
\newblock \doi{10.1088/2041-8205/767/1/L11}
\input{Krumholz2013}

\bibitem[{{Krumholz} and {Federrath}(2019)}]{KrumholzFederrath2019}
{Krumholz}, M.~R. and {Federrath}, C. (2019).
\newblock The role of magnetic fields in setting the star formation rate and
  the initial mass function.
\newblock \emph{{Frontiers in Astronomy and Space Sciences}} in press
\input{KrumholzFederrath2019}

\bibitem[{{Kwon} et~al.(2018){Kwon}, {Doi}, {Tamura},
  {Matsumura}, {Pattle}, {Berry} et~al.}]{JKwon2018}
{Kwon}, J., {Doi}, Y., {Tamura}, M., {Matsumura}, M., {Pattle}, K., {Berry},
  D., et~al. (2018).
\newblock {A First Look at BISTRO Observations of the {$\rho$} Oph-A core}.
\newblock \emph{\apj} 859, 4.
\newblock \doi{10.3847/1538-4357/aabd82}
\input{JKwon2018}

\bibitem[{{Kwon} et~al.(2006){Kwon}, {Looney}, {Crutcher}, and
  {Kirk}}]{Kwon2006}
{Kwon}, W., {Looney}, L.~W., {Crutcher}, R.~M., and {Kirk}, J.~M. (2006).
\newblock {Two Bipolar Outflows and Magnetic Fields in the Multiple Protostar
  System L1448 IRS 3}.
\newblock \emph{\apj} 653, 1358--1368.
\newblock \doi{10.1086/508920}
\input{Kwon2006}

\bibitem[{{Kwon} et~al.(2018){Kwon}, {Stephens}, {Tobin},
  {Looney}, {Li}, {van der Tak} et~al.}]{Kwon2018}
{Kwon}, W., {Stephens}, I., {Tobin}, J., {Looney}, L., {Li}, Z.-Y., {van der
  Tak}, F., et~al. (2018).
\newblock {Saving Early Disk Formation of Young Stellar Objects from the
  Magnetic Braking Catastrophe}.
\newblock \emph{ArXiv e-prints}
\input{Kwon2018}

\bibitem[{{Kylafis}(1983)}]{Kylafis1983}
{Kylafis}, N.~D. (1983).
\newblock {Polarization of interstellar radio-frequency lines and magnetic
  field direction}.
\newblock \emph{\apj} 267, 137--150.
\newblock \doi{10.1086/160851}
\input{Kylafis1983}

\bibitem[{{Lada} and {Lada}(2003)}]{LadaLada2003}
{Lada}, C.~J. and {Lada}, E.~A. (2003).
\newblock {Embedded Clusters in Molecular Clouds}.
\newblock \emph{\araa} 41, 57--115.
\newblock \doi{10.1146/annurev.astro.41.011802.094844}
\input{LadaLada2003}

\bibitem[{{Lai} et~al.(2001){Lai}, {Crutcher}, {Girart}, and {Rao}}]{Lai2001}
{Lai}, S.-P., {Crutcher}, R.~M., {Girart}, J.~M., and {Rao}, R. (2001).
\newblock {Interferometric Mapping of Magnetic Fields in Star-forming Regions.
  I. W51 e1/e2 Molecular Cores}.
\newblock \emph{\apj} 561, 864--870.
\newblock \doi{10.1086/323372}
\input{Lai2001}

\bibitem[{{Lai} et~al.(2002){Lai}, {Crutcher}, {Girart}, and {Rao}}]{Lai2002}
{Lai}, S.-P., {Crutcher}, R.~M., {Girart}, J.~M., and {Rao}, R. (2002).
\newblock {Interferometric Mapping of Magnetic Fields in Star-forming Regions.
  II. NGC 2024 FIR 5}.
\newblock \emph{\apj} 566, 925--930.
\newblock \doi{10.1086/338336}
\input{Lai2002}

\bibitem[{{Lai} et~al.(2003){Lai}, {Girart}, and {Crutcher}}]{Lai2003}
{Lai}, S.-P., {Girart}, J.~M., and {Crutcher}, R.~M. (2003).
\newblock {Interferometric Mapping of Magnetic Fields in Star-forming Regions.
  III. Dust and CO Polarization in DR 21(OH)}.
\newblock \emph{\apj} 598, 392--399.
\newblock \doi{10.1086/378769}
\input{Lai2003}

\bibitem[{{Larson}(1981)}]{Larson1981}
{Larson}, R.~B. (1981).
\newblock {Turbulence and star formation in molecular clouds}.
\newblock \emph{\mnras} 194, 809--826.
\newblock \doi{10.1093/mnras/194.4.809}
\input{Larson1981}

\bibitem[{{Lazarian}(2007)}]{Lazarian2007}
{Lazarian}, A. (2007).
\newblock {Tracing magnetic fields with aligned grains}.
\newblock \emph{J.~Quant.~Spec.~Radiat.~Transf.} 106, 225--256.
\newblock \doi{10.1016/j.jqsrt.2007.01.038}
\input{Lazarian2007}

\bibitem[{Lee et~al.(2018)Lee, Hwang, Ching, Hirano, Lai, Rao
  et~al.}]{CFLee2018b}
Lee, C.-F., Hwang, H.-C., Ching, T.-C., Hirano, N., Lai, S.-P., Rao, R., et~al.
  (2018).
\newblock Unveiling a magnetized jet from a low-mass protostar.
\newblock \emph{Nature Communications} 9, 4636.
\newblock \doi{10.1038/s41467-018-07143-8}
\input{CFLee2018b}

\bibitem[{{Lee} et~al.(2018){Lee}, {Li}, {Ching}, {Lai}, and
  {Yang}}]{CFLee2018a}
{Lee}, C.-F., {Li}, Z.-Y., {Ching}, T.-C., {Lai}, S.-P., and {Yang}, H. (2018).
\newblock {ALMA Dust Polarization Observations of Two Young Edge-on
  Protostellar Disks}.
\newblock \emph{\apj} 854, 56.
\newblock \doi{10.3847/1538-4357/aaa769}
\input{CFLee2018a}

\bibitem[{{Lee} et~al.(2014){Lee}, {Rao}, {Ching}, {Lai}, {Hirano}, {Ho}
  et~al.}]{CFLee2014}
{Lee}, C.-F., {Rao}, R., {Ching}, T.-C., {Lai}, S.-P., {Hirano}, N., {Ho},
  P.~T.~P., et~al. (2014).
\newblock {Magnetic Field Structure in the Flattened Envelope and Jet in the
  Young Protostellar System HH 211}.
\newblock \emph{\apjl} 797, L9.
\newblock \doi{10.1088/2041-8205/797/1/L9}
\input{CFLee2014}

\bibitem[{{Lee} et~al.(2017){Lee}, {Lee}, {Dunham}, {Tatematsu},
  {Choi}, {Bergin} et~al.}]{JELee2017}
{Lee}, J.-E., {Lee}, S., {Dunham}, M.~M., {Tatematsu}, K., {Choi}, M.,
  {Bergin}, E.~A., et~al. (2017).
\newblock {Formation of wide binaries by turbulent fragmentation}.
\newblock \emph{Nature Astronomy} 1, 0172.
\newblock \doi{10.1038/s41550-017-0172}
\input{JELee2017}

\bibitem[{{Lee} et~al.(2017){Lee}, {Hull}, and
  {Offner}}]{JLee2017}
{Lee}, J.~W.~Y., {Hull}, C.~L.~H., and {Offner}, S.~S.~R. (2017).
\newblock {Synthetic Observations of Magnetic Fields in Protostellar Cores}.
\newblock \emph{\apj} 834, 201.
\newblock \doi{10.3847/1538-4357/834/2/201}
\input{JLee2017}

\bibitem[{{Lee} et~al.(2016){Lee}, {Dunham}, {Myers}, {Arce}, {Bourke},
  {Goodman} et~al.}]{Lee2016}
{Lee}, K.~I., {Dunham}, M.~M., {Myers}, P.~C., {Arce}, H.~G., {Bourke}, T.~L.,
  {Goodman}, A.~A., et~al. (2016).
\newblock {Misalignment of Outflow Axes in the Proto-multiple Systems in
  Perseus}.
\newblock \emph{\apjl} 820, L2.
\newblock \doi{10.3847/2041-8205/820/1/L2}
\input{Lee2016}

\bibitem[{{Lee} et~al.(2015){Lee}, {Dunham}, {Myers}, {Tobin}, {Kristensen},
  {Pineda} et~al.}]{Lee2015}
{Lee}, K.~I., {Dunham}, M.~M., {Myers}, P.~C., {Tobin}, J.~J., {Kristensen},
  L.~E., {Pineda}, J.~E., et~al. (2015).
\newblock {Mass Assembly of Stellar Systems and Their Evolution with the SMA
  (MASSES). Multiplicity and the Physical Environment in L1448N}.
\newblock \emph{\apj} 814, 114.
\newblock \doi{10.1088/0004-637X/814/2/114}
\input{Lee2015}

\bibitem[{{Li} et~al.(2008){Li}, {Dowell}, {Kirby}, {Novak}, and
  {Vaillancourt}}]{Li2008}
{Li}, H., {Dowell}, C.~D., {Kirby}, L., {Novak}, G., and {Vaillancourt}, J.~E.
  (2008).
\newblock {Design and initial performance of SHARP, a polarimeter for the
  SHARC-II camera at the Caltech Submillimeter Observatory}.
\newblock \emph{\ao} 47, 422--430.
\newblock \doi{10.1364/AO.47.000422}
\input{Li2008}

\bibitem[{{Li} et~al.(2006){Li}, {Griffin}, {Krejny}, {Novak}, {Loewenstein},
  {Newcomb} et~al.}]{HBLi2006}
{Li}, H., {Griffin}, G.~S., {Krejny}, M., {Novak}, G., {Loewenstein}, R.~F.,
  {Newcomb}, M.~G., et~al. (2006).
\newblock {Results of SPARO 2003: Mapping Magnetic Fields in Giant Molecular
  Clouds}.
\newblock \emph{\apj} 648, 340--354.
\newblock \doi{10.1086/505858}
\input{HBLi2006}

\bibitem[{{Li} et~al.(2009){Li}, {Dowell}, {Goodman}, {Hildebrand}, and
  {Novak}}]{HBLi2009}
{Li}, H.-b., {Dowell}, C.~D., {Goodman}, A., {Hildebrand}, R., and {Novak}, G.
  (2009).
\newblock {Anchoring Magnetic Field in Turbulent Molecular Clouds}.
\newblock \emph{\apj} 704, 891--897.
\newblock \doi{10.1088/0004-637X/704/2/891}
\input{HBLi2009}

\bibitem[{{Li} and {Law}(2019)}]{HBLiLaw2019}
{Li}, H.-B. and {Law}, C.~Y. (2019).
\newblock Observing the impacts of magnetic fields on molecular clouds.
\newblock \emph{{Frontiers in Astronomy and Space Sciences}} under review
\input{HBLiLaw2019}

\bibitem[{{Li} et~al.(2015){Li}, {Yuen}, {Otto}, {Leung}, {Sridharan}, {Zhang}
  et~al.}]{HBLi2015}
{Li}, H.-B., {Yuen}, K.~H., {Otto}, F., {Leung}, P.~K., {Sridharan}, T.~K.,
  {Zhang}, Q., et~al. (2015).
\newblock {Self-similar fragmentation regulated by magnetic fields in a region
  forming massive stars}.
\newblock \emph{\nat} 520, 518--521.
\newblock \doi{10.1038/nature14291}
\input{HBLi2015}

\bibitem[{{Li} et~al.(2011){Li}, {Krasnopolsky}, and {Shang}}]{Li2011}
{Li}, Z.-Y., {Krasnopolsky}, R., and {Shang}, H. (2011).
\newblock {Non-ideal MHD Effects and Magnetic Braking Catastrophe in
  Protostellar Disk Formation}.
\newblock \emph{\apj} 738, 180.
\newblock \doi{10.1088/0004-637X/738/2/180}
\input{Li2011}

\bibitem[{{Li} et~al.(2013){Li}, {Krasnopolsky}, and {Shang}}]{Li2013}
{Li}, Z.-Y., {Krasnopolsky}, R., and {Shang}, H. (2013).
\newblock {Does Magnetic-field-Rotation Misalignment Solve the Magnetic Braking
  Catastrophe in Protostellar Disk Formation?}
\newblock \emph{\apj} 774, 82.
\newblock \doi{10.1088/0004-637X/774/1/82}
\input{Li2013}

\bibitem[{{Li} et~al.(2014){Li}, {Krasnopolsky}, {Shang}, and {Zhao}}]{Li2014}
{Li}, Z.-Y., {Krasnopolsky}, R., {Shang}, H., and {Zhao}, B. (2014).
\newblock {On the Role of Pseudodisk Warping and Reconnection in Protostellar
  Disk Formation in Turbulent Magnetized Cores}.
\newblock \emph{\apj} 793, 130.
\newblock \doi{10.1088/0004-637X/793/2/130}
\input{Li2014}

\bibitem[{{Lis} et~al.(1988){Lis}, {Goldsmith}, {Dickman}, {Predmore}, {Omont},
  and {Cernicharo}}]{Lis1988}
{Lis}, D.~C., {Goldsmith}, P.~F., {Dickman}, R.~L., {Predmore}, C.~R., {Omont},
  A., and {Cernicharo}, J. (1988).
\newblock {Linear polarization of millimeter-wave emission lines in clouds
  without large velocity gradients}.
\newblock \emph{\apj} 328, 304--314.
\newblock \doi{10.1086/166293}
\input{Lis1988}

\bibitem[{{Liu} et~al.(2018){Liu}, {Hasegawa}, {Ching}, {Lai}, {Hirano}, and
  {Rao}}]{Liu2018}
{Liu}, H.~B., {Hasegawa}, Y., {Ching}, T.-C., {Lai}, S.-P., {Hirano}, N., and
  {Rao}, R. (2018).
\newblock {Detection of 40-48 GHz dust continuum linear polarization towards
  the Class 0 young stellar object IRAS 16293-2422}.
\newblock \emph{\aap} 617, A3.
\newblock \doi{10.1051/0004-6361/201832699}
\input{Liu2018}

\bibitem[{{Liu} et~al.(2016){Liu}, {Lai}, {Hasegawa}, {Hirano}, {Rao}, {Li}
  et~al.}]{Liu2016}
{Liu}, H.~B., {Lai}, S.-P., {Hasegawa}, Y., {Hirano}, N., {Rao}, R., {Li},
  I.-H., et~al. (2016).
\newblock {Detection of Linearly Polarized 6.9 mm Continuum Emission from the
  Class 0 Young Stellar Object NGC 1333 IRAS4A}.
\newblock \emph{\apj} 821, 41.
\newblock \doi{10.3847/0004-637X/821/1/41}
\input{Liu2016}

\bibitem[{{Liu} et~al.(2013){Liu}, {Qiu}, {Zhang}, {Girart}, and
  {Ho}}]{Liu2013}
{Liu}, H.~B., {Qiu}, K., {Zhang}, Q., {Girart}, J.~M., and {Ho}, P.~T.~P.
  (2013).
\newblock {Gas Kinematics and the Dragged Magnetic Field in the High-mass
  Molecular Outflow Source G192.16-3.84: An SMA View}.
\newblock \emph{\apj} 771, 71.
\newblock \doi{10.1088/0004-637X/771/1/71}
\input{Liu2013}

\bibitem[{{Lopez-Rodriguez} et~al.(2018){Lopez-Rodriguez}, {Antonucci},
  {Chary}, and {Kishimoto}}]{LopezRodriguez2018}
{Lopez-Rodriguez}, E., {Antonucci}, R., {Chary}, R.-R., and {Kishimoto}, M.
  (2018).
\newblock {The Highly Polarized Dusty Emission Core of Cygnus A}.
\newblock \emph{\apjl} 861, L23.
\newblock \doi{10.3847/2041-8213/aacff5}
\input{LopezRodriguez2018}

\bibitem[{{Lu} et~al.(2018){Lu}, {Zhang}, {Liu}, {Sanhueza}, {Tatematsu},
  {Feng} et~al.}]{Lu2018}
{Lu}, X., {Zhang}, Q., {Liu}, H.~B., {Sanhueza}, P., {Tatematsu}, K., {Feng},
  S., et~al. (2018).
\newblock {Filamentary Fragmentation and Accretion in High-mass Star-forming
  Molecular Clouds}.
\newblock \emph{\apj} 855, 9.
\newblock \doi{10.3847/1538-4357/aaad11}
\input{Lu2018}

\bibitem[{{Machida} et~al.(2006){Machida}, {Matsumoto}, {Hanawa}, and
  {Tomisaka}}]{Machida2006}
{Machida}, M.~N., {Matsumoto}, T., {Hanawa}, T., and {Tomisaka}, K. (2006).
\newblock {Evolution of Rotating Molecular Cloud Core with Oblique Magnetic
  Field}.
\newblock \emph{\apj} 645, 1227--1245.
\newblock \doi{10.1086/504423}
\input{Machida2006}

\bibitem[{{Masson} et~al.(2016){Masson}, {Chabrier}, {Hennebelle}, {Vaytet},
  and {Commer{\c c}on}}]{Masson2016}
{Masson}, J., {Chabrier}, G., {Hennebelle}, P., {Vaytet}, N., and {Commer{\c
  c}on}, B. (2016).
\newblock {Ambipolar diffusion in low-mass star formation. I. General
  comparison with the ideal magnetohydrodynamic case}.
\newblock \emph{\aap} 587, A32.
\newblock \doi{10.1051/0004-6361/201526371}
\input{Masson2016}

\bibitem[{{Matthews} et~al.(2005){Matthews}, {Lai}, {Crutcher}, and
  {Wilson}}]{Matthews2005}
{Matthews}, B.~C., {Lai}, S.-P., {Crutcher}, R.~M., and {Wilson}, C.~D. (2005).
\newblock {Multiscale Magnetic Fields in Star-forming Regions: Interferometric
  Polarimetry of the MMS 6 Core of OMC-3}.
\newblock \emph{\apj} 626, 959--965.
\newblock \doi{10.1086/430127}
\input{Matthews2005}

\bibitem[{{Matthews} et~al.(2009){Matthews}, {McPhee}, {Fissel}, and
  {Curran}}]{Matthews2009}
{Matthews}, B.~C., {McPhee}, C.~A., {Fissel}, L.~M., and {Curran}, R.~L.
  (2009).
\newblock {The Legacy of SCUPOL: 850 {$\mu$}m Imaging Polarimetry from 1997 to
  2005}.
\newblock \emph{\apjs} 182, 143--204.
\newblock \doi{10.1088/0067-0049/182/1/143}
\input{Matthews2009}

\bibitem[{{Maury} et~al.(2018){Maury}, {Girart}, {Zhang}, {Hennebelle}, {Keto},
  {Rao} et~al.}]{Maury2018}
{Maury}, A.~J., {Girart}, J.~M., {Zhang}, Q., {Hennebelle}, P., {Keto}, E.,
  {Rao}, R., et~al. (2018).
\newblock {Magnetically regulated collapse in the B335 protostar? I. ALMA
  observations of the polarized dust emission}.
\newblock \emph{\mnras} 477, 2760--2765.
\newblock \doi{10.1093/mnras/sty574}
\input{Maury2018}

\bibitem[{{McKee} and {Tan}(2003)}]{McKee2003}
{McKee}, C.~F. and {Tan}, J.~C. (2003).
\newblock {The Formation of Massive Stars from Turbulent Cores}.
\newblock \emph{\apj} 585, 850--871.
\newblock \doi{10.1086/346149}
\input{McKee2003}

\bibitem[{{McMullin} et~al.(2007){McMullin}, {Waters}, {Schiebel}, {Young}, and
  {Golap}}]{McMullin2007}
{McMullin}, J.~P., {Waters}, B., {Schiebel}, D., {Young}, W., and {Golap}, K.
  (2007).
\newblock {CASA Architecture and Applications}.
\newblock In \emph{Astronomical Data Analysis Software and Systems XVI}, eds.
  R.~A. {Shaw}, F.~{Hill}, and D.~J. {Bell}. vol. 376 of \emph{Astronomical
  Society of the Pacific Conference Series}, 127
\input{McMullin2007}

\bibitem[{{Mellon} and {Li}(2008)}]{Mellon2008}
{Mellon}, R.~R. and {Li}, Z.-Y. (2008).
\newblock {Magnetic Braking and Protostellar Disk Formation: The Ideal MHD
  Limit}.
\newblock \emph{\apj} 681, 1356--1376.
\newblock \doi{10.1086/587542}
\input{Mellon2008}

\bibitem[{{M{\'e}nard} and {Duch{\^e}ne}(2004)}]{Menard2004}
{M{\'e}nard}, F. and {Duch{\^e}ne}, G. (2004).
\newblock {On the alignment of Classical T Tauri stars with the magnetic field
  in the Taurus-Auriga molecular cloud}.
\newblock \emph{\aap} 425, 973--980.
\newblock \doi{10.1051/0004-6361:20041338}
\input{Menard2004}

\bibitem[{{Mestel} and {Spitzer}(1956)}]{Mestel1956}
{Mestel}, L. and {Spitzer}, L., Jr. (1956).
\newblock {Star formation in magnetic dust clouds}.
\newblock \emph{\mnras} 116, 503.
\newblock \doi{10.1093/mnras/116.5.503}
\input{Mestel1956}

\bibitem[{{Mocz} et~al.(2017){Mocz}, {Burkhart}, {Hernquist}, {McKee}, and
  {Springel}}]{Mocz2017}
{Mocz}, P., {Burkhart}, B., {Hernquist}, L., {McKee}, C.~F., and {Springel}, V.
  (2017).
\newblock {Moving-mesh Simulations of Star-forming Cores in
  Magneto-gravo-turbulence}.
\newblock \emph{\apj} 838, 40.
\newblock \doi{10.3847/1538-4357/aa6475}
\input{Mocz2017}

\bibitem[{{Motte} et~al.(2018){Motte}, {Bontemps}, and {Louvet}}]{Motte2018}
{Motte}, F., {Bontemps}, S., and {Louvet}, F. (2018).
\newblock {High-Mass Star and Massive Cluster Formation in the Milky Way}.
\newblock \emph{\araa} 56, 41--82.
\newblock \doi{10.1146/annurev-astro-091916-055235}
\input{Motte2018}

\bibitem[{{Mouschovias}(1976{\natexlab{a}})}]{Mouschovias1976a}
{Mouschovias}, T.~C. (1976{\natexlab{a}}).
\newblock {Nonhomologous contraction and equilibria of self-gravitating,
  magnetic interstellar clouds embedded in an intercloud medium: Star
  formation. I Formulation of the problem and method of solution}.
\newblock \emph{\apj} 206, 753--767.
\newblock \doi{10.1086/154436}
\input{Mouschovias1976a}

\bibitem[{{Mouschovias}(1976{\natexlab{b}})}]{Mouschovias1976b}
{Mouschovias}, T.~C. (1976{\natexlab{b}}).
\newblock {Nonhomologous contraction and equilibria of self-gravitating,
  magnetic interstellar clouds embedded in an intercloud medium: Star
  formation. II - Results}.
\newblock \emph{\apj} 207, 141--158.
\newblock \doi{10.1086/154478}
\input{Mouschovias1976b}

\bibitem[{{Mouschovias}(1991)}]{Mouschovias1991}
{Mouschovias}, T.~C. (1991).
\newblock {Magnetic braking, ambipolar diffusion, cloud cores, and star
  formation - Natural length scales and protostellar masses}.
\newblock \emph{\apj} 373, 169--186.
\newblock \doi{10.1086/170035}
\input{Mouschovias1991}

\bibitem[{{Mouschovias} and {Ciolek}(1999)}]{Mouschovias1999}
{Mouschovias}, T.~C. and {Ciolek}, G.~E. (1999).
\newblock {Magnetic Fields and Star Formation: A Theory Reaching Adulthood}.
\newblock In \emph{NATO Advanced Science Institutes (ASI) Series C}, eds. C.~J.
  {Lada} and N.~D. {Kylafis}. vol. 540 of \emph{NATO Advanced Science
  Institutes (ASI) Series C}, 305
\input{Mouschovias1999}

\bibitem[{{Offner} and {Chaban}(2017)}]{Offner2017}
{Offner}, S.~S.~R. and {Chaban}, J. (2017).
\newblock {Impact of Protostellar Outflows on Turbulence and Star Formation
  Efficiency in Magnetized Dense Cores}.
\newblock \emph{\apj} 847, 104.
\newblock \doi{10.3847/1538-4357/aa8996}
\input{Offner2017}

\bibitem[{{Offner} et~al.(2016){Offner}, {Dunham}, {Lee}, {Arce}, and
  {Fielding}}]{Offner2016}
{Offner}, S.~S.~R., {Dunham}, M.~M., {Lee}, K.~I., {Arce}, H.~G., and
  {Fielding}, D.~B. (2016).
\newblock {The Turbulent Origin of Outflow and Spin Misalignment in Multiple
  Star Systems}.
\newblock \emph{\apjl} 827, L11.
\newblock \doi{10.3847/2041-8205/827/1/L11}
\input{Offner2016}

\bibitem[{{Ohashi} et~al.(2014){Ohashi}, {Saigo}, {Aso}, {Aikawa}, {Koyamatsu},
  {Machida} et~al.}]{Ohashi2014}
{Ohashi}, N., {Saigo}, K., {Aso}, Y., {Aikawa}, Y., {Koyamatsu}, S., {Machida},
  M.~N., et~al. (2014).
\newblock {Formation of a Keplerian Disk in the Infalling Envelope around L1527
  IRS: Transformation from Infalling Motions to Kepler Motions}.
\newblock \emph{\apj} 796, 131.
\newblock \doi{10.1088/0004-637X/796/2/131}
\input{Ohashi2014}

\bibitem[{{Ohashi} et~al.(2018){Ohashi}, {Kataoka}, {Nagai}, {Momose}, {Muto},
  {Hanawa} et~al.}]{SOhashi2018}
{Ohashi}, S., {Kataoka}, A., {Nagai}, H., {Momose}, M., {Muto}, T., {Hanawa},
  T., et~al. (2018).
\newblock {Two Different Grain Size Distributions within the Protoplanetary
  Disk around HD 142527 Revealed by ALMA Polarization Observation}.
\newblock \emph{\apj} 864, 81.
\newblock \doi{10.3847/1538-4357/aad632}
\input{SOhashi2018}

\bibitem[{{Ohashi} et~al.(2016){Ohashi}, {Sanhueza}, {Chen}, {Zhang},
  {Busquet}, {Nakamura} et~al.}]{Ohashi2016}
{Ohashi}, S., {Sanhueza}, P., {Chen}, H.-R.~V., {Zhang}, Q., {Busquet}, G.,
  {Nakamura}, F., et~al. (2016).
\newblock {Dense Core Properties in the Infrared Dark Cloud G14.225-0.506
  Revealed by ALMA}.
\newblock \emph{\apj} 833, 209.
\newblock \doi{10.3847/1538-4357/833/2/209}
\input{Ohashi2016}

\bibitem[{{Ossenkopf} and {Henning}(1994)}]{Ossenkopf1994}
{Ossenkopf}, V. and {Henning}, T. (1994).
\newblock {Dust opacities for protostellar cores}.
\newblock \emph{\aap} 291, 943--959
\input{Ossenkopf1994}

\bibitem[{{Ostriker} et~al.(2001){Ostriker}, {Stone}, and
  {Gammie}}]{Ostriker2001}
{Ostriker}, E.~C., {Stone}, J.~M., and {Gammie}, C.~F. (2001).
\newblock {Density, Velocity, and Magnetic Field Structure in Turbulent
  Molecular Cloud Models}.
\newblock \emph{\apj} 546, 980--1005.
\newblock \doi{10.1086/318290}
\input{Ostriker2001}

\bibitem[{{Padovani} et~al.(2012){Padovani}, {Brinch}, {Girart},
  {J{\o}rgensen}, {Frau}, {Hennebelle} et~al.}]{Padovani2012}
{Padovani}, M., {Brinch}, C., {Girart}, J.~M., {J{\o}rgensen}, J.~K., {Frau},
  P., {Hennebelle}, P., et~al. (2012).
\newblock {Adaptable radiative transfer innovations for submillimetre
  telescopes (ARTIST). Dust polarisation module (DustPol)}.
\newblock \emph{\aap} 543, A16.
\newblock \doi{10.1051/0004-6361/201219028}
\input{Padovani2012}

\bibitem[{{Pattle} and {Fissel}(2019)}]{PattleFissel2019}
{Pattle}, K. and {Fissel}, L.~M. (2019).
\newblock Submillimeter and far-infrared polarimetric observations of magnetic
  fields in star-forming regions.
\newblock \emph{{Frontiers in Astronomy and Space Sciences}} under review
\input{PattleFissel2019}

\bibitem[{{Pattle} et~al.(2017){Pattle}, {Ward-Thompson}, {Berry}, {Hatchell},
  {Chen}, {Pon} et~al.}]{Pattle2017}
{Pattle}, K., {Ward-Thompson}, D., {Berry}, D., {Hatchell}, J., {Chen}, H.-R.,
  {Pon}, A., et~al. (2017).
\newblock {The JCMT BISTRO Survey: The Magnetic Field Strength in the Orion A
  Filament}.
\newblock \emph{\apj} 846, 122.
\newblock \doi{10.3847/1538-4357/aa80e5}
\input{Pattle2017}

\bibitem[{{Pattle} et~al.(2018){Pattle}, {Ward-Thompson}, {Hasegawa},
  {Bastien}, {Kwon}, {Lai} et~al.}]{Pattle2018}
{Pattle}, K., {Ward-Thompson}, D., {Hasegawa}, T., {Bastien}, P., {Kwon}, W.,
  {Lai}, S.-P., et~al. (2018).
\newblock {First Observations of the Magnetic Field inside the Pillars of
  Creation: Results from the BISTRO Survey}.
\newblock \emph{\apjl} 860, L6.
\newblock \doi{10.3847/2041-8213/aac771}
\input{Pattle2018}

\bibitem[{{Pillai} et~al.(2011){Pillai}, {Kauffmann}, {Wyrowski}, {Hatchell},
  {Gibb}, and {Thompson}}]{Pillai2011}
{Pillai}, T., {Kauffmann}, J., {Wyrowski}, F., {Hatchell}, J., {Gibb}, A.~G.,
  and {Thompson}, M.~A. (2011).
\newblock {Probing the initial conditions of high-mass star formation. II.
  Fragmentation, stability, and chemistry towards high-mass star-forming
  regions G29.96-0.02 and G35.20-1.74}.
\newblock \emph{\aap} 530, A118.
\newblock \doi{10.1051/0004-6361/201015899}
\input{Pillai2011}

\bibitem[{{Plambeck} et~al.(2003){Plambeck}, {Wright}, and
  {Rao}}]{Plambeck2003}
{Plambeck}, R.~L., {Wright}, M.~C.~H., and {Rao}, R. (2003).
\newblock {Magnetic Field Morphology of Orion IRc2 from 86 GHz SiO Maser
  Polarization Images}.
\newblock \emph{\apj} 594, 911--918.
\newblock \doi{10.1086/377097}
\input{Plambeck2003}

\bibitem[{{Planck Collaboration} et~al.(2016{\natexlab{a}}){Planck
  Collaboration}, {Adam}, {Ade}, {Aghanim}, {Alves}, and et~al.}]{PlanckXXXII}
{Planck Collaboration}, {Adam}, R., {Ade}, P.~A.~R., {Aghanim}, N., {Alves},
  M.~I.~R., and et~al. (2016{\natexlab{a}}).
\newblock {Planck intermediate results. XXXII. The relative orientation between
  the magnetic field and structures traced by interstellar dust}.
\newblock \emph{\aap} 586, A135.
\newblock \doi{10.1051/0004-6361/201425044}
\input{PlanckXXXII}

\bibitem[{{Planck Collaboration} et~al.(2015{\natexlab{a}}){Planck
  Collaboration}, {Ade}, {Aghanim}, {Alina}, {Alves}, {Aniano}
  et~al.}]{PlanckXX}
{Planck Collaboration}, {Ade}, P.~A.~R., {Aghanim}, N., {Alina}, D., {Alves},
  M.~I.~R., {Aniano}, G., et~al. (2015{\natexlab{a}}).
\newblock {Planck intermediate results. XX. Comparison of polarized thermal
  emission from Galactic dust with simulations of MHD turbulence}.
\newblock \emph{\aap} 576, A105.
\newblock \doi{10.1051/0004-6361/201424086}
\input{PlanckXX}

\bibitem[{{Planck Collaboration} et~al.(2015{\natexlab{b}}){Planck
  Collaboration}, {Ade}, {Aghanim}, {Alina}, {Alves}, and et~al.}]{PlanckXIX}
{Planck Collaboration}, {Ade}, P.~A.~R., {Aghanim}, N., {Alina}, D., {Alves},
  M.~I.~R., and et~al. (2015{\natexlab{b}}).
\newblock {Planck intermediate results. XIX. An overview of the polarized
  thermal emission from Galactic dust}.
\newblock \emph{\aap} 576, A104.
\newblock \doi{10.1051/0004-6361/201424082}
\input{PlanckXIX}

\bibitem[{{Planck Collaboration} et~al.(2015{\natexlab{c}}){Planck
  Collaboration}, {Ade}, {Aghanim}, {Alina}, {Aniano}, {Armitage-Caplan}
  et~al.}]{PlanckXXI}
{Planck Collaboration}, {Ade}, P.~A.~R., {Aghanim}, N., {Alina}, D., {Aniano},
  G., {Armitage-Caplan}, C., et~al. (2015{\natexlab{c}}).
\newblock {Planck intermediate results. XXI. Comparison of polarized thermal
  emission from Galactic dust at 353 GHz with interstellar polarization in the
  visible}.
\newblock \emph{\aap} 576, A106.
\newblock \doi{10.1051/0004-6361/201424087}
\input{PlanckXXI}

\bibitem[{{Planck Collaboration} et~al.(2016{\natexlab{b}}){Planck
  Collaboration}, {Ade}, {Aghanim}, {Alves}, {Arnaud}, {Arzoumanian}
  et~al.}]{PlanckXXXIII}
{Planck Collaboration}, {Ade}, P.~A.~R., {Aghanim}, N., {Alves}, M.~I.~R.,
  {Arnaud}, M., {Arzoumanian}, D., et~al. (2016{\natexlab{b}}).
\newblock {Planck intermediate results. XXXIII. Signature of the magnetic field
  geometry of interstellar filaments in dust polarization maps}.
\newblock \emph{\aap} 586, A136.
\newblock \doi{10.1051/0004-6361/201425305}
\input{PlanckXXXIII}

\bibitem[{{Planck Collaboration} et~al.(2016{\natexlab{c}}){Planck
  Collaboration}, {Ade}, {Aghanim}, {Alves}, {Arnaud}, {Arzoumanian}
  et~al.}]{PlanckXXXV}
{Planck Collaboration}, {Ade}, P.~A.~R., {Aghanim}, N., {Alves}, M.~I.~R.,
  {Arnaud}, M., {Arzoumanian}, D., et~al. (2016{\natexlab{c}}).
\newblock {Planck intermediate results. XXXV. Probing the role of the magnetic
  field in the formation of structure in molecular clouds}.
\newblock \emph{\aap} 586, A138.
\newblock \doi{10.1051/0004-6361/201525896}
\input{PlanckXXXV}

\bibitem[{{Pohl} et~al.(2016){Pohl}, {Kataoka}, {Pinilla}, {Dullemond},
  {Henning}, and {Birnstiel}}]{Pohl2016}
{Pohl}, A., {Kataoka}, A., {Pinilla}, P., {Dullemond}, C.~P., {Henning}, T.,
  and {Birnstiel}, T. (2016).
\newblock {Investigating dust trapping in transition disks with millimeter-wave
  polarization}.
\newblock \emph{\aap} 593, A12.
\newblock \doi{10.1051/0004-6361/201628637}
\input{Pohl2016}

\bibitem[{{Pudritz} and {Ray}(2019)}]{PudritzRay2019}
{Pudritz}, R.~E. and {Ray}, T.~P. (2019).
\newblock The role of magnetic fields in protostellar outflows and star
  formation.
\newblock \emph{{Frontiers in Astronomy and Space Sciences}} under review
\input{PudritzRay2019}

\bibitem[{{Qiu} et~al.(2013){Qiu}, {Zhang}, {Menten}, {Liu}, and
  {Tang}}]{Qiu2013}
{Qiu}, K., {Zhang}, Q., {Menten}, K.~M., {Liu}, H.~B., and {Tang}, Y.-W.
  (2013).
\newblock {From Poloidal to Toroidal: Detection of a Well-ordered Magnetic
  Field in the High-mass Protocluster G35.2-0.74 N}.
\newblock \emph{\apj} 779, 182.
\newblock \doi{10.1088/0004-637X/779/2/182}
\input{Qiu2013}

\bibitem[{{Qiu} et~al.(2014){Qiu}, {Zhang}, {Menten}, {Liu}, {Tang}, and
  {Girart}}]{Qiu2014}
{Qiu}, K., {Zhang}, Q., {Menten}, K.~M., {Liu}, H.~B., {Tang}, Y.-W., and
  {Girart}, J.~M. (2014).
\newblock {Submillimeter Array Observations of Magnetic Fields in G240.31+0.07:
  An Hourglass in a Massive Cluster-forming Core}.
\newblock \emph{\apjl} 794, L18.
\newblock \doi{10.1088/2041-8205/794/1/L18}
\input{Qiu2014}

\bibitem[{{Qiu} et~al.(2009){Qiu}, {Zhang}, {Wu}, and {Chen}}]{Qiu2009}
{Qiu}, K., {Zhang}, Q., {Wu}, J., and {Chen}, H.-R. (2009).
\newblock {Submillimeter Array Observations of the Molecular Outflow in
  High-Mass Star-Forming Region G240.31+0.07}.
\newblock \emph{\apj} 696, 66--74.
\newblock \doi{10.1088/0004-637X/696/1/66}
\input{Qiu2009}

\bibitem[{{Rao} et~al.(1998){Rao}, {Crutcher}, {Plambeck}, and
  {Wright}}]{Rao1998}
{Rao}, R., {Crutcher}, R.~M., {Plambeck}, R.~L., and {Wright}, M.~C.~H. (1998).
\newblock {High-Resolution Millimeter-Wave Mapping of Linearly Polarized Dust
  Emission: Magnetic Field Structure in Orion}.
\newblock \emph{\apjl} 502, L75+.
\newblock \doi{10.1086/311485}
\input{Rao1998}

\bibitem[{{Rao} et~al.(2014){Rao}, {Girart}, {Lai}, and {Marrone}}]{Rao2014}
{Rao}, R., {Girart}, J.~M., {Lai}, S.-P., and {Marrone}, D.~P. (2014).
\newblock {Detection of a Magnetized Disk around a Very Young Protostar}.
\newblock \emph{\apjl} 780, L6.
\newblock \doi{10.1088/2041-8205/780/1/L6}
\input{Rao2014}

\bibitem[{{Rao} et~al.(2009){Rao}, {Girart}, {Marrone}, {Lai}, and
  {Schnee}}]{Rao2009}
{Rao}, R., {Girart}, J.~M., {Marrone}, D.~P., {Lai}, S.-P., and {Schnee}, S.
  (2009).
\newblock {IRAS 16293: A ''Magnetic'' Tale of Two Cores}.
\newblock \emph{\apj} 707, 921--935.
\newblock \doi{10.1088/0004-637X/707/2/921}
\input{Rao2009}

\bibitem[{{Renbarger} et~al.(2004){Renbarger}, {Chuss}, {Dotson}, {Griffin},
  {Hanna}, {Loewenstein} et~al.}]{Renbarger2004}
{Renbarger}, T., {Chuss}, D.~T., {Dotson}, J.~L., {Griffin}, G.~S., {Hanna},
  J.~L., {Loewenstein}, R.~F., et~al. (2004).
\newblock {Early Results from SPARO: Instrument Characterization and
  Polarimetry of NGC 6334}.
\newblock \emph{\pasp} 116, 415--424.
\newblock \doi{10.1086/383623}
\input{Renbarger2004}

\bibitem[{{Ritacco} et~al.(2017){Ritacco}, {Ponthieu}, {Catalano}, {Adam},
  {Ade}, {Andr{\'e}} et~al.}]{Ritacco2017}
{Ritacco}, A., {Ponthieu}, N., {Catalano}, A., {Adam}, R., {Ade}, P.,
  {Andr{\'e}}, P., et~al. (2017).
\newblock {Polarimetry at millimeter wavelengths with the NIKA camera:
  calibration and performance}.
\newblock \emph{\aap} 599, A34.
\newblock \doi{10.1051/0004-6361/201629666}
\input{Ritacco2017}

\bibitem[{{Robitaille} and {Bressert}(2012)}]{Robitaille2012}
[Dataset] {Robitaille}, T. and {Bressert}, E. (2012).
\newblock {APLpy: Astronomical Plotting Library in Python}.
\newblock Astrophysics Source Code Library
\input{Robitaille2012}

\bibitem[{{Roy} et~al.(2011){Roy}, {Ade}, {Bock}, {Chapin}, {Devlin}, {Dicker}
  et~al.}]{Roy2011}
{Roy}, A., {Ade}, P.~A.~R., {Bock}, J.~J., {Chapin}, E.~L., {Devlin}, M.~J.,
  {Dicker}, S.~R., et~al. (2011).
\newblock {The Balloon-borne Large Aperture Submillimeter Telescope (BLAST)
  2005: A 10 deg$^{2}$ Survey of Star Formation in Cygnus X}.
\newblock \emph{\apj} 727, 114.
\newblock \doi{10.1088/0004-637X/727/2/114}
\input{Roy2011}

\bibitem[{{Sadavoy} et~al.(2018{\natexlab{a}}){Sadavoy}, {Myers}, {Stephens},
  {Tobin}, {Commer{\c c}on}, {Henning} et~al.}]{Sadavoy2018a}
{Sadavoy}, S.~I., {Myers}, P.~C., {Stephens}, I.~W., {Tobin}, J., {Commer{\c
  c}on}, B., {Henning}, T., et~al. (2018{\natexlab{a}}).
\newblock {Dust Polarization toward Embedded Protostars in Ophiuchus with ALMA.
  I. VLA 1623}.
\newblock \emph{\apj} 859, 165.
\newblock \doi{10.3847/1538-4357/aac21a}
\input{Sadavoy2018a}

\bibitem[{{Sadavoy} et~al.(2018{\natexlab{b}}){Sadavoy}, {Myers}, {Stephens},
  {Tobin}, {Kwon}, {Segura-Cox} et~al.}]{Sadavoy2018c}
{Sadavoy}, S.~I., {Myers}, P.~C., {Stephens}, I.~W., {Tobin}, J., {Kwon}, W.,
  {Segura-Cox}, D., et~al. (2018{\natexlab{b}}).
\newblock {Dust Polarization toward Embedded Protostars in Ophiuchus with ALMA.
  II. IRAS 16293-2422}.
\newblock \emph{\apj} 869, 115.
\newblock \doi{10.3847/1538-4357/aaef81}
\input{Sadavoy2018c}

\bibitem[{{Sanhueza} et~al.(2017){Sanhueza}, {Jackson}, {Zhang}, {Guzm{\'a}n},
  {Lu}, {Stephens} et~al.}]{Sanhueza2017}
{Sanhueza}, P., {Jackson}, J.~M., {Zhang}, Q., {Guzm{\'a}n}, A.~E., {Lu}, X.,
  {Stephens}, I.~W., et~al. (2017).
\newblock {A Massive Prestellar Clump Hosting No High-mass Cores}.
\newblock \emph{\apj} 841, 97.
\newblock \doi{10.3847/1538-4357/aa6ff8}
\input{Sanhueza2017}

\bibitem[{{Santos-Lima} et~al.(2012){Santos-Lima}, {de Gouveia Dal Pino}, and
  {Lazarian}}]{SantosLima2012}
{Santos-Lima}, R., {de Gouveia Dal Pino}, E.~M., and {Lazarian}, A. (2012).
\newblock {The Role of Turbulent Magnetic Reconnection in the Formation of
  Rotationally Supported Protostellar Disks}.
\newblock \emph{\apj} 747, 21.
\newblock \doi{10.1088/0004-637X/747/1/21}
\input{SantosLima2012}

\bibitem[{{Saral} et~al.(2017){Saral}, {Hora}, {Audard}, {Koenig},
  {Mart{\'{\i}}nez-Galarza}, {Motte} et~al.}]{Saral2017}
{Saral}, G., {Hora}, J.~L., {Audard}, M., {Koenig}, X.~P.,
  {Mart{\'{\i}}nez-Galarza}, J.~R., {Motte}, F., et~al. (2017).
\newblock {Young Stellar Objects in the Massive Star-forming Regions W51 and
  W43}.
\newblock \emph{\apj} 839, 108.
\newblock \doi{10.3847/1538-4357/aa6575}
\input{Saral2017}

\bibitem[{{Schleuning}(1998)}]{Schleuning1998}
{Schleuning}, D.~A. (1998).
\newblock {Far-Infrared and Submillimeter Polarization of OMC-1: Evidence for
  Magnetically Regulated Star Formation}.
\newblock \emph{\apj} 493, 811.
\newblock \doi{10.1086/305139}
\input{Schleuning1998}

\bibitem[{{Segura-Cox} et~al.(2015){Segura-Cox}, {Looney}, {Stephens},
  {Fern{\'a}ndez-L{\'o}pez}, {Kwon}, {Tobin} et~al.}]{SeguraCox2015}
{Segura-Cox}, D.~M., {Looney}, L.~W., {Stephens}, I.~W.,
  {Fern{\'a}ndez-L{\'o}pez}, M., {Kwon}, W., {Tobin}, J.~J., et~al. (2015).
\newblock {The Magnetic Field in the Class 0 Protostellar Disk of L1527}.
\newblock \emph{\apjl} 798, L2.
\newblock \doi{10.1088/2041-8205/798/1/L2}
\input{SeguraCox2015}

\bibitem[{{Segura-Cox} et~al.(2018){Segura-Cox}, {Looney}, {Tobin}, {Li},
  {Harris}, {Sadavoy} et~al.}]{SeguraCox2018}
{Segura-Cox}, D.~M., {Looney}, L.~W., {Tobin}, J.~J., {Li}, Z.-Y., {Harris},
  R.~J., {Sadavoy}, S., et~al. (2018).
\newblock {The VLA Nascent Disk and Multiplicity Survey of Perseus Protostars
  (VANDAM). V. 18 Candidate Disks around Class 0 and I Protostars in the
  Perseus Molecular Cloud}.
\newblock \emph{\apj} 866, 161.
\newblock \doi{10.3847/1538-4357/aaddf3}
\input{SeguraCox2018}

\bibitem[{{Seifried} et~al.(2015){Seifried}, {Banerjee}, {Pudritz}, and
  {Klessen}}]{Seifried2015}
{Seifried}, D., {Banerjee}, R., {Pudritz}, R.~E., and {Klessen}, R.~S. (2015).
\newblock {Accretion and magnetic field morphology around Class 0 stage
  protostellar discs}.
\newblock \emph{\mnras} 446, 2776--2788.
\newblock \doi{10.1093/mnras/stu2282}
\input{Seifried2015}

\bibitem[{{Shu} et~al.(1987){Shu}, {Adams}, and {Lizano}}]{Shu1987}
{Shu}, F.~H., {Adams}, F.~C., and {Lizano}, S. (1987).
\newblock {Star formation in molecular clouds - Observation and theory}.
\newblock \emph{\araa} 25, 23--81.
\newblock \doi{10.1146/annurev.aa.25.090187.000323}
\input{Shu1987}

\bibitem[{{Shu} et~al.(2000){Shu}, {Najita}, {Shang}, and {Li}}]{Shu2000}
{Shu}, F.~H., {Najita}, J.~R., {Shang}, H., and {Li}, Z.-Y. (2000).
\newblock {X-Winds Theory and Observations}.
\newblock \emph{Protostars and Planets IV} , 789
\input{Shu2000}

\bibitem[{{Siringo} et~al.(2012){Siringo}, {Kov{\'a}cs}, {Kreysa}, {Schuller},
  {Weiss}, {Guesten} et~al.}]{Siringo2012}
{Siringo}, G., {Kov{\'a}cs}, A., {Kreysa}, E., {Schuller}, F., {Weiss}, A.,
  {Guesten}, R., et~al. (2012).
\newblock {First results of the polarimeter for the Large APEX Bolometer Camera
  (LABOCA)}.
\newblock In \emph{Millimeter, Submillimeter, and Far-Infrared Detectors and
  Instrumentation for Astronomy VI}. vol. 8452 of \emph{\procspie}, 845206.
\newblock \doi{10.1117/12.925697}
\input{Siringo2012}

\bibitem[{{Siringo} et~al.(2004){Siringo}, {Kreysa}, {Reichertz}, and
  {Menten}}]{Siringo2004}
{Siringo}, G., {Kreysa}, E., {Reichertz}, L.~A., and {Menten}, K.~M. (2004).
\newblock {A new polarimeter for (sub)millimeter bolometer arrays}.
\newblock \emph{\aap} 422, 751--760.
\newblock \doi{10.1051/0004-6361:20035832}
\input{Siringo2004}

\bibitem[{{Soam} et~al.(2018){Soam}, {Pattle}, {Ward-Thompson}, {Lee},
  {Sadavoy}, {Koch} et~al.}]{Soam2018}
{Soam}, A., {Pattle}, K., {Ward-Thompson}, D., {Lee}, C.~W., {Sadavoy}, S.,
  {Koch}, P.~M., et~al. (2018).
\newblock {Magnetic Fields toward Ophiuchus-B Derived from SCUBA-2 Polarization
  Measurements}.
\newblock \emph{\apj} 861, 65.
\newblock \doi{10.3847/1538-4357/aac4a6}
\input{Soam2018}

\bibitem[{{Sridharan} et~al.(2014){Sridharan}, {Rao}, {Qiu}, {Cortes}, {Li},
  {Pillai} et~al.}]{Sridharan2014}
{Sridharan}, T.~K., {Rao}, R., {Qiu}, K., {Cortes}, P., {Li}, H., {Pillai}, T.,
  et~al. (2014).
\newblock {Hot Core, Outflows, and Magnetic Fields in W43-MM1 (G30.79 FIR 10)}.
\newblock \emph{\apjl} 783, L31.
\newblock \doi{10.1088/2041-8205/783/2/L31}
\input{Sridharan2014}

\bibitem[{{Stephens} et~al.(2018){Stephens}, {Dunham}, {Myers}, {Pokhrel},
  {Bourke}, {Vorobyov} et~al.}]{Stephens2018}
{Stephens}, I.~W., {Dunham}, M.~M., {Myers}, P.~C., {Pokhrel}, R., {Bourke},
  T.~L., {Vorobyov}, E.~I., et~al. (2018).
\newblock {Mass Assembly of Stellar Systems and Their Evolution with the SMA
  (MASSES) --- 1.3 mm Subcompact Data Release}.
\newblock \emph{\apjs} 237, 22.
\newblock \doi{10.3847/1538-4365/aacda9}
\input{Stephens2018}

\bibitem[{{Stephens} et~al.(2017{\natexlab{a}}){Stephens}, {Dunham}, {Myers},
  {Pokhrel}, {Sadavoy}, {Vorobyov} et~al.}]{Stephens2017a}
{Stephens}, I.~W., {Dunham}, M.~M., {Myers}, P.~C., {Pokhrel}, R., {Sadavoy},
  S.~I., {Vorobyov}, E.~I., et~al. (2017{\natexlab{a}}).
\newblock {Alignment between Protostellar Outflows and Filamentary Structure}.
\newblock \emph{\apj} 846, 16.
\newblock \doi{10.3847/1538-4357/aa8262}
\input{Stephens2017a}

\bibitem[{{Stephens} et~al.(2011){Stephens}, {Looney}, {Dowell},
  {Vaillancourt}, and {Tassis}}]{Stephens2011}
{Stephens}, I.~W., {Looney}, L.~W., {Dowell}, C.~D., {Vaillancourt}, J.~E., and
  {Tassis}, K. (2011).
\newblock {The Galactic Magnetic Field's Effect in Star-forming Regions}.
\newblock \emph{\apj} 728, 99.
\newblock \doi{10.1088/0004-637X/728/2/99}
\input{Stephens2011}

\bibitem[{{Stephens} et~al.(2014){Stephens}, {Looney}, {Kwon},
  {Fern{\'a}ndez-L{\'o}pez}, {Hughes}, {Mundy} et~al.}]{Stephens2014}
{Stephens}, I.~W., {Looney}, L.~W., {Kwon}, W., {Fern{\'a}ndez-L{\'o}pez}, M.,
  {Hughes}, A.~M., {Mundy}, L.~G., et~al. (2014).
\newblock {Spatially resolved magnetic field structure in the disk of a T Tauri
  star}.
\newblock \emph{\nat} 514, 597--599.
\newblock \doi{10.1038/nature13850}
\input{Stephens2014}

\bibitem[{{Stephens} et~al.(2013){Stephens}, {Looney}, {Kwon}, {Hull},
  {Plambeck}, {Crutcher} et~al.}]{Stephens2013}
{Stephens}, I.~W., {Looney}, L.~W., {Kwon}, W., {Hull}, C.~L.~H., {Plambeck},
  R.~L., {Crutcher}, R.~M., et~al. (2013).
\newblock {The Magnetic Field Morphology of the Class 0 Protostar L1157-mm}.
\newblock \emph{\apjl} 769, L15.
\newblock \doi{10.1088/2041-8205/769/1/L15}
\input{Stephens2013}

\bibitem[{{Stephens} et~al.(2017{\natexlab{b}}){Stephens}, {Yang}, {Li},
  {Looney}, {Kataoka}, {Kwon} et~al.}]{Stephens2017b}
{Stephens}, I.~W., {Yang}, H., {Li}, Z.-Y., {Looney}, L.~W., {Kataoka}, A.,
  {Kwon}, W., et~al. (2017{\natexlab{b}}).
\newblock {ALMA Reveals Transition of Polarization Pattern with Wavelength in
  HL Tau's Disk}.
\newblock \emph{\apj} 851, 55.
\newblock \doi{10.3847/1538-4357/aa998b}
\input{Stephens2017b}

\bibitem[{{Takahashi} et~al.(2018){Takahashi}, {Machida}, {Tomisaka}, {Ho},
  {Fomalont}, {Nakanishi} et~al.}]{Takahashi2018}
{Takahashi}, S., {Machida}, M.~N., {Tomisaka}, K., {Ho}, P. T.~P., {Fomalont},
  E.~B., {Nakanishi}, K., et~al. (2018).
\newblock {ALMA High Angular Resolution Polarization Study; An Extremely Young
  Class 0 Source, OMC-3/MMS 6}.
\newblock \emph{arXiv e-prints} , arXiv:1812.03189
\input{Takahashi2018}

\bibitem[{{Takahashi} et~al.(2006){Takahashi}, {Saito}, {Takakuwa}, and
  {Kawabe}}]{Takahashi2006}
{Takahashi}, S., {Saito}, M., {Takakuwa}, S., and {Kawabe}, R. (2006).
\newblock {Millimeter- and Submillimeter-Wave Observations of the OMC-2/3
  Region. I. Dispersing and Rotating Core around the Intermediate-Mass
  Protostar MMS 7}.
\newblock \emph{\apj} 651, 933--944.
\newblock \doi{10.1086/507482}
\input{Takahashi2006}

\bibitem[{{Tang} et~al.(2009{\natexlab{a}}){Tang}, {Ho}, {Girart}, {Rao},
  {Koch}, and {Lai}}]{Tang2009a}
{Tang}, Y.-W., {Ho}, P.~T.~P., {Girart}, J.~M., {Rao}, R., {Koch}, P., and
  {Lai}, S.-P. (2009{\natexlab{a}}).
\newblock {Evolution of Magnetic Fields in High Mass Star Formation:
  Submillimeter Array Dust Polarization Image of the Ultracompact H II Region
  G5.89-0.39}.
\newblock \emph{\apj} 695, 1399--1412.
\newblock \doi{10.1088/0004-637X/695/2/1399}
\input{Tang2009a}

\bibitem[{{Tang} et~al.(2009{\natexlab{b}}){Tang}, {Ho}, {Koch}, {Girart},
  {Lai}, and {Rao}}]{Tang2009b}
{Tang}, Y.-W., {Ho}, P.~T.~P., {Koch}, P.~M., {Girart}, J.~M., {Lai}, S.-P.,
  and {Rao}, R. (2009{\natexlab{b}}).
\newblock {Evolution of Magnetic Fields in High-Mass Star Formation: Linking
  Field Geometry and Collapse for the W51 e2/e8 Cores}.
\newblock \emph{\apj} 700, 251--261.
\newblock \doi{10.1088/0004-637X/700/1/251}
\input{Tang2009b}

\bibitem[{{Tang} et~al.(2010){Tang}, {Ho}, {Koch}, and {Rao}}]{Tang2010}
{Tang}, Y.-W., {Ho}, P.~T.~P., {Koch}, P.~M., and {Rao}, R. (2010).
\newblock {High-angular Resolution Dust Polarization Measurements: Shaped
  B-field Lines in the Massive Star-forming Region Orion BN/KL}.
\newblock \emph{\apj} 717, 1262--1273.
\newblock \doi{10.1088/0004-637X/717/2/1262}
\input{Tang2010}

\bibitem[{{Targon} et~al.(2011){Targon}, {Rodrigues}, {Cerqueira}, and
  {Hickel}}]{Targon2011}
{Targon}, C.~G., {Rodrigues}, C.~V., {Cerqueira}, A.~H., and {Hickel}, G.~R.
  (2011).
\newblock {Correlating the Interstellar Magnetic Field with Protostellar Jets
  and Its Sources}.
\newblock \emph{\apj} 743, 54.
\newblock \doi{10.1088/0004-637X/743/1/54}
\input{Targon2011}

\bibitem[{{Tazaki} et~al.(2017){Tazaki}, {Lazarian}, and {Nomura}}]{Tazaki2017}
{Tazaki}, R., {Lazarian}, A., and {Nomura}, H. (2017).
\newblock {Radiative Grain Alignment In Protoplanetary Disks: Implications for
  Polarimetric Observations}.
\newblock \emph{\apj} 839, 56.
\newblock \doi{10.3847/1538-4357/839/1/56}
\input{Tazaki2017}

\bibitem[{{Teyssier} and {Commer{\c{c}}on}(2019)}]{TeyssierCommercon2019}
{Teyssier}, R. and {Commer{\c{c}}on}, B. (2019).
\newblock Numerical methods for simulating star formation.
\newblock \emph{{Frontiers in Astronomy and Space Sciences}} under review
\input{TeyssierCommercon2019}

\bibitem[{{Tobin} et~al.(2012){Tobin}, {Hartmann}, {Chiang}, {Wilner},
  {Looney}, {Loinard} et~al.}]{Tobin2012}
{Tobin}, J.~J., {Hartmann}, L., {Chiang}, H.-F., {Wilner}, D.~J., {Looney},
  L.~W., {Loinard}, L., et~al. (2012).
\newblock {A \~{}0.2-solar-mass protostar with a Keplerian disk in the very
  young L1527 IRS system}.
\newblock \emph{\nat} 492, 83--85.
\newblock \doi{10.1038/nature11610}
\input{Tobin2012}

\bibitem[{{Tobin} et~al.(2016{\natexlab{a}}){Tobin}, {Kratter}, {Persson},
  {Looney}, {Dunham}, {Segura-Cox} et~al.}]{Tobin2016b}
{Tobin}, J.~J., {Kratter}, K.~M., {Persson}, M.~V., {Looney}, L.~W., {Dunham},
  M.~M., {Segura-Cox}, D., et~al. (2016{\natexlab{a}}).
\newblock {A triple protostar system formed via fragmentation of a
  gravitationally unstable disk}.
\newblock \emph{\nat} 538, 483--486.
\newblock \doi{10.1038/nature20094}
\input{Tobin2016b}

\bibitem[{{Tobin} et~al.(2016{\natexlab{b}}){Tobin}, {Looney}, {Li},
  {Chandler}, {Dunham}, {Segura-Cox} et~al.}]{Tobin2016a}
{Tobin}, J.~J., {Looney}, L.~W., {Li}, Z.-Y., {Chandler}, C.~J., {Dunham},
  M.~M., {Segura-Cox}, D., et~al. (2016{\natexlab{b}}).
\newblock {The VLA Nascent Disk and Multiplicity Survey of Perseus Protostars
  (VANDAM). II. Multiplicity of Protostars in the Perseus Molecular Cloud}.
\newblock \emph{\apj} 818, 73.
\newblock \doi{10.3847/0004-637X/818/1/73}
\input{Tobin2016a}

\bibitem[{{Tobin} et~al.(2018){Tobin}, {Looney}, {Li}, {Sadavoy}, {Dunham},
  {Segura-Cox} et~al.}]{Tobin2018}
{Tobin}, J.~J., {Looney}, L.~W., {Li}, Z.-Y., {Sadavoy}, S.~I., {Dunham},
  M.~M., {Segura-Cox}, D., et~al. (2018).
\newblock {The VLA/ALMA Nascent Disk and Multiplicity (VANDAM) Survey of
  Perseus Protostars. VI. Characterizing the Formation Mechanism for Close
  Multiple Systems}.
\newblock \emph{\apj} 867, 43.
\newblock \doi{10.3847/1538-4357/aae1f7}
\input{Tobin2018}

\bibitem[{{Tomida} et~al.(2015){Tomida}, {Okuzumi}, and {Machida}}]{Tomida2015}
{Tomida}, K., {Okuzumi}, S., and {Machida}, M.~N. (2015).
\newblock {Radiation Magnetohydrodynamic Simulations of Protostellar Collapse:
  Nonideal Magnetohydrodynamic Effects and Early Formation of Circumstellar
  Disks}.
\newblock \emph{\apj} 801, 117.
\newblock \doi{10.1088/0004-637X/801/2/117}
\input{Tomida2015}

\bibitem[{{Traficante} et~al.(2018){Traficante}, {Lee}, {Hennebelle},
  {Molinari}, {Kauffmann}, and {Pillai}}]{Traficante2018}
{Traficante}, A., {Lee}, Y.-N., {Hennebelle}, P., {Molinari}, S., {Kauffmann},
  J., and {Pillai}, T. (2018).
\newblock {A possible observational bias in the estimation of the virial
  parameter in virialized clumps}.
\newblock \emph{\aap} 619, L7.
\newblock \doi{10.1051/0004-6361/201833513}
\input{Traficante2018}

\bibitem[{{Troland} and {Heiles}(1986)}]{Troland1986}
{Troland}, T.~H. and {Heiles}, C. (1986).
\newblock {Interstellar magnetic field strengths and gas densities
  Observational and theoretical perspectives}.
\newblock \emph{\apj} 301, 339--345.
\newblock \doi{10.1086/163904}
\input{Troland1986}

\bibitem[{{Tsukamoto} et~al.(2015{\natexlab{a}}){Tsukamoto}, {Iwasaki},
  {Okuzumi}, {Machida}, and {Inutsuka}}]{Tsukamoto2015a}
{Tsukamoto}, Y., {Iwasaki}, K., {Okuzumi}, S., {Machida}, M.~N., and
  {Inutsuka}, S. (2015{\natexlab{a}}).
\newblock {Bimodality of Circumstellar Disk Evolution Induced by the Hall
  Current}.
\newblock \emph{\apjl} 810, L26.
\newblock \doi{10.1088/2041-8205/810/2/L26}
\input{Tsukamoto2015a}

\bibitem[{{Tsukamoto} et~al.(2015{\natexlab{b}}){Tsukamoto}, {Iwasaki},
  {Okuzumi}, {Machida}, and {Inutsuka}}]{Tsukamoto2015b}
{Tsukamoto}, Y., {Iwasaki}, K., {Okuzumi}, S., {Machida}, M.~N., and
  {Inutsuka}, S. (2015{\natexlab{b}}).
\newblock {Bimodality of Circumstellar Disk Evolution Induced by the Hall
  Current}.
\newblock \emph{\apjl} 810, L26.
\newblock \doi{10.1088/2041-8205/810/2/L26}
\input{Tsukamoto2015b}

\bibitem[{{Tsukamoto} et~al.(2018){Tsukamoto}, {Okuzumi}, {Iwasaki}, {Machida},
  and {Inutsuka}}]{Tsukamoto2018}
{Tsukamoto}, Y., {Okuzumi}, S., {Iwasaki}, K., {Machida}, M.~N., and
  {Inutsuka}, S. (2018).
\newblock {Does Misalignment between Magnetic Field and Angular Momentum
  Enhance or Suppress Circumstellar Disk Formation?}
\newblock \emph{\apj} 868, 22.
\newblock \doi{10.3847/1538-4357/aae4dc}
\input{Tsukamoto2018}

\bibitem[{{Tsukamoto} et~al.(2017){Tsukamoto}, {Okuzumi}, {Iwasaki}, {Machida},
  and {Inutsuka}}]{Tsukamoto2017}
{Tsukamoto}, Y., {Okuzumi}, S., {Iwasaki}, K., {Machida}, M.~N., and
  {Inutsuka}, S.-i. (2017).
\newblock {The impact of the Hall effect during cloud core collapse:
  Implications for circumstellar disk evolution}.
\newblock \emph{\pasj} 69, 95.
\newblock \doi{10.1093/pasj/psx113}
\input{Tsukamoto2017}

\bibitem[{{Vaillancourt} et~al.(2007){Vaillancourt}, {Chuss}, {Crutcher},
  {Dotson}, {Dowell}, {Harper} et~al.}]{Vaillancourt2007}
{Vaillancourt}, J.~E., {Chuss}, D.~T., {Crutcher}, R.~M., {Dotson}, J.~L.,
  {Dowell}, C.~D., {Harper}, D.~A., et~al. (2007).
\newblock {Far-infrared polarimetry from the Stratospheric Observatory for
  Infrared Astronomy}.
\newblock In \emph{Infrared Spaceborne Remote Sensing and Instrumentation XV}.
  vol. 6678 of \emph{\procspie}, 66780D.
\newblock \doi{10.1117/12.730922}
\input{Vaillancourt2007}

\bibitem[{{Vall{\'e}e} and {Fiege}(2006)}]{Vallee2006}
{Vall{\'e}e}, J.~P. and {Fiege}, J.~D. (2006).
\newblock {A Cool Filament Crossing the Warm Protostar DR 21(OH): Geometry,
  Kinematics, Magnetic Vectors, and Pressure Balance}.
\newblock \emph{\apj} 636, 332--347.
\newblock \doi{10.1086/497957}
\input{Vallee2006}

\bibitem[{{van Kempen} et~al.(2016){van Kempen}, {Hogerheijde}, {van Dishoeck},
  {Kristensen}, {Belloche}, {Klaassen} et~al.}]{vanKempen2016}
{van Kempen}, T.~A., {Hogerheijde}, M.~R., {van Dishoeck}, E.~F., {Kristensen},
  L.~E., {Belloche}, A., {Klaassen}, P.~D., et~al. (2016).
\newblock {Outflow forces in intermediate-mass star formation}.
\newblock \emph{\aap} 587, A17.
\newblock \doi{10.1051/0004-6361/201424725}
\input{vanKempen2016}

\bibitem[{{Ward-Thompson} et~al.(2017){Ward-Thompson}, {Pattle}, {Bastien},
  {Furuya}, {Kwon}, {Lai} et~al.}]{WardThompson2017}
{Ward-Thompson}, D., {Pattle}, K., {Bastien}, P., {Furuya}, R.~S., {Kwon}, W.,
  {Lai}, S.-P., et~al. (2017).
\newblock {First Results from BISTRO: A SCUBA-2 Polarimeter Survey of the Gould
  Belt}.
\newblock \emph{\apj} 842, 66.
\newblock \doi{10.3847/1538-4357/aa70a0}
\input{WardThompson2017}

\bibitem[{{Wiesemeyer} et~al.(2014){Wiesemeyer}, {Hezareh}, {Kreysa}, {Weiss},
  {G{\"u}sten}, {Menten} et~al.}]{Wiesemeyer2014}
{Wiesemeyer}, H., {Hezareh}, T., {Kreysa}, E., {Weiss}, A., {G{\"u}sten}, R.,
  {Menten}, K.~M., et~al. (2014).
\newblock {Submillimeter Polarimetry with PolKa, a Reflection-Type Modulator
  for the APEX Telescope}.
\newblock \emph{\pasp} 126, 1027.
\newblock \doi{10.1086/679002}
\input{Wiesemeyer2014}

\bibitem[{{Wright} et~al.(2014){Wright}, {Hull}, {Pillai}, {Zhao}, and
  {Sandell}}]{Wright2014}
{Wright}, M.~C.~H., {Hull}, C.~L.~H., {Pillai}, T., {Zhao}, J.-H., and
  {Sandell}, G. (2014).
\newblock {NGC 7538 IRS 1: Interaction of a Polarized Dust Spiral and a
  Molecular Outflow}.
\newblock \emph{\apj} 796, 112.
\newblock \doi{10.1088/0004-637X/796/2/112}
\input{Wright2014}

\bibitem[{{Wurster} et~al.(2018){Wurster}, {Bate}, and {Price}}]{Wurster2018}
{Wurster}, J., {Bate}, M.~R., and {Price}, D.~J. (2018).
\newblock {Hall effect-driven formation of gravitationally unstable discs in
  magnetized molecular cloud cores}.
\newblock \emph{\mnras} 480, 4434--4442.
\newblock \doi{10.1093/mnras/sty2212}
\input{Wurster2018}

\bibitem[{Wurster and Li(2018)}]{WursterLi2018}
Wurster, J. and Li, Z.-Y. (2018).
\newblock The role of magnetic fields in the formation of protostellar discs.
\newblock \emph{Frontiers in Astronomy and Space Sciences} 5, 39.
\newblock \doi{10.3389/fspas.2018.00039}
\input{WursterLi2018}

\bibitem[{{Yang} et~al.(2016{\natexlab{a}}){Yang}, {Li}, {Looney}, and
  {Stephens}}]{Yang2016a}
{Yang}, H., {Li}, Z.-Y., {Looney}, L., and {Stephens}, I. (2016{\natexlab{a}}).
\newblock {Inclination-induced polarization of scattered millimetre radiation
  from protoplanetary discs: the case of HL Tau}.
\newblock \emph{\mnras} 456, 2794--2805.
\newblock \doi{10.1093/mnras/stv2633}
\input{Yang2016a}

\bibitem[{{Yang} et~al.(2016{\natexlab{b}}){Yang}, {Li}, {Looney}, {Cox},
  {Tobin}, {Stephens} et~al.}]{Yang2016b}
{Yang}, H., {Li}, Z.-Y., {Looney}, L.~W., {Cox}, E.~G., {Tobin}, J.,
  {Stephens}, I.~W., et~al. (2016{\natexlab{b}}).
\newblock {Disc polarization from both emission and scattering of magnetically
  aligned grains: the case of NGC 1333 IRAS 4A1}.
\newblock \emph{\mnras} 460, 4109--4121.
\newblock \doi{10.1093/mnras/stw1253}
\input{Yang2016b}

\bibitem[{{Yang} et~al.(2019){Yang}, {Li}, {Stephens}, {Kataoka}, and
  {Looney}}]{Yang2019}
{Yang}, H., {Li}, Z.-Y., {Stephens}, I.~W., {Kataoka}, A., and {Looney}, L.
  (2019).
\newblock {Does HL Tau disc polarization in ALMA band 3 come from radiatively
  aligned grains?}
\newblock \emph{\mnras} 483, 2371--2381.
\newblock \doi{10.1093/mnras/sty3263}
\input{Yang2019}

\bibitem[{{Yuen} and {Lazarian}(2017{\natexlab{a}})}]{YuenLazarian2017b}
{Yuen}, K.~H. and {Lazarian}, A. (2017{\natexlab{a}}).
\newblock {Tracing interstellar magnetic field using the velocity gradient
  technique in shock and self-gravitating media}.
\newblock \emph{ArXiv e-prints}
\input{YuenLazarian2017b}

\bibitem[{{Yuen} and {Lazarian}(2017{\natexlab{b}})}]{YuenLazarian2017a}
{Yuen}, K.~H. and {Lazarian}, A. (2017{\natexlab{b}}).
\newblock {Tracing Interstellar Magnetic Field Using Velocity Gradient
  Technique: Application to Atomic Hydrogen Data}.
\newblock \emph{\apjl} 837, L24.
\newblock \doi{10.3847/2041-8213/aa6255}
\input{YuenLazarian2017a}

\bibitem[{{Zhang} and {Ho}(1997)}]{Zhang1997}
{Zhang}, Q. and {Ho}, P.~T.~P. (1997).
\newblock {Dynamical Collapse in W51 Massive Cores: NH$_3$ Observations}.
\newblock \emph{\apj} 488, 241--+.
\newblock \doi{10.1086/304667}
\input{Zhang1997}

\bibitem[{Zhang et~al.(2014)Zhang, Qiu, Girart, Liu, Tang, Koch
  et~al.}]{Zhang2014}
Zhang, Q., Qiu, K., Girart, J.~M., Liu, H.~B., Tang, Y.-W., Koch, P.~M., et~al.
  (2014).
\newblock Magnetic fields and massive star formation.
\newblock \emph{The Astrophysical Journal} 792, 116
\input{Zhang2014}

\bibitem[{{Zhang} et~al.(2015){Zhang}, {Wang}, {Lu}, and
  {Jim{\'e}nez-Serra}}]{Zhang2015}
{Zhang}, Q., {Wang}, K., {Lu}, X., and {Jim{\'e}nez-Serra}, I. (2015).
\newblock {Fragmentation of Molecular Clumps and Formation of a Protocluster}.
\newblock \emph{\apj} 804, 141.
\newblock \doi{10.1088/0004-637X/804/2/141}
\input{Zhang2015}

\bibitem[{{Zhang} et~al.(2009){Zhang}, {Wang}, {Pillai}, and
  {Rathborne}}]{zhang2009}
{Zhang}, Q., {Wang}, Y., {Pillai}, T., and {Rathborne}, J. (2009).
\newblock {Fragmentation at the Earliest Phase of Massive Star Formation}.
\newblock \emph{\apj} 696, 268--273.
\newblock \doi{10.1088/0004-637X/696/1/268}
\input{zhang2009}

\end{thebibliography}


\end{document}